\documentclass[aip,jcp,reprint,noshowkeys,superscriptaddress]{revtex4-1}
\usepackage{graphicx,dcolumn,bm,xcolor,microtype,multirow,amscd,amsmath,amssymb,amsfonts,physics,wrapfig,txfonts,siunitx,longtable,pifont,soul}
\usepackage[version=4]{mhchem}

\newcommand{\ie}{\textit{i.e.}}

\newcommand{\alert}[1]{\textcolor{black}{#1}}
\usepackage[normalem]{ulem}

\newcommand{\SupMat}{\textcolor{blue}{supplementary material}}

\newcommand{\fnm}{\footnotemark}
\newcommand{\fnt}{\footnotetext}

\newcommand{\sig}{\sigma}

\newcommand{\pis}{{\pi^*}}

\newcommand{\sigCC}{\sig_{\ce{CC}}}
\newcommand{\sigsCC}{\sig_{\ce{CC}}^*}

\newcommand{\sigCN}{\sig_{\ce{CN}}}
\newcommand{\sigsCN}{\sig_{\ce{CN}}^*}

\usepackage[
	colorlinks=true,
    citecolor=blue,
    breaklinks=true
	]{hyperref}
\newcommand{\cmark}{\ding{51}}%
\newcommand{\xmark}{\ding{55}}%
\newcommand{\Y}{\textbf{\textcolor{green}{\cmark}}}
\newcommand{\N}{\textbf{\textcolor{red}{\xmark}}}

\begin{document}

\newcommand{\LCPQ}{Laboratoire de Chimie et Physique Quantiques (UMR 5626), Universit\'e de Toulouse, CNRS, UPS, France}
\newcommand{\CEISAM}{Nantes Universit\'e, CNRS,  CEISAM UMR 6230, F-44000 Nantes, France}

\title{Benchmarking CASPT3 Vertical Excitation Energies}

\author{Martial \surname{Boggio-Pasqua}}
	\email{martial.boggio@irsamc.ups-tlse.fr}
	\affiliation{\LCPQ}
\author{Denis \surname{Jacquemin}}
	\email{Denis.Jacquemin@univ-nantes.fr}
	\affiliation{\CEISAM}
\author{Pierre-Fran\c{c}ois \surname{Loos}}
	\email{loos@irsamc.ups-tlse.fr}
	\affiliation{\LCPQ}

\begin{abstract}
Based on 280 reference vertical transition energies of various natures (singlet, triplet, valence, Rydberg, $n\to\pi^*$, $\pi\to\pi^*$, and double excitations) extracted from the QUEST database, we assess the accuracy of third-order multireference perturbation theory, CASPT3, in the context of molecular excited states.
When one applies the disputable ionization-potential-electron-affinity (IPEA) shift, we show that CASPT3 provides a similar accuracy as its second-order counterpart, CASPT2, with the same mean absolute error of $0.11$ eV.
However, as already reported, we also observe that the accuracy of CASPT3 is almost insensitive to the IPEA shift, irrespective of the transition type and system size, with a small reduction of the mean absolute error to $0.09$ eV when the IPEA shift is switched off.
\end{abstract}

\maketitle

\section{Introduction}
\label{sec:intro}

Perturbation theory is a relatively inexpensive route towards the exact solution of the Schr\"odinger equation. 
However, it rarely works this way in practice as the perturbative series may exhibit quite a large spectrum of behaviors. \cite{Olsen_1996,Christiansen_1996,Cremer_1996,Olsen_2000,Olsen_2019,Stillinger_2000,Goodson_2000a,Goodson_2000b,Goodson_2004,Sergeev_2005,Sergeev_2006,Goodson_2011}
For example, in single-reference M{\o}ller-Plesset (MP) perturbation theory, \cite{Moller_1934} erratic, slowly convergent, and divergent behaviors have been observed. \cite{Laidig_1985,Knowles_1985,Handy_1985,Gill_1986,Laidig_1987,Nobes_1987,Gill_1988,Gill_1988a,Lepetit_1988,Leininger_2000a,Malrieu_2003,Damour_2021}
Systematic improvement is thus difficult to achieve and it is extremely challenging to predict, \textit{a priori}, the evolution of the series when ramping up the perturbation order. \cite{Marie_2021a} 
This has led, in certain specific contexts, to the development of empirical strategy like MP2.5 where one averages the second-order (MP2) and third-order (MP3) total energies, to obtain more accurate values. \cite{Pitonak_2009}

Extension of single-reference perturbation theory to electronic excited states is far from being trivial, and the algebraic diagrammatic
construction (ADC) approximation of the polarization propagator is probably the most natural. \cite{Schirmer_1982,Schirmer_1991,Barth_1995,Schirmer_2004,Schirmer_2018,Trofimov_1997,Trofimov_1997b,Trofimov_2002,Trofimov_2005,Trofimov_2006,Harbach_2014,Dreuw_2015}
However, the ADC series naturally inherits some of the drawbacks of its MP parent and it has been shown to be rather slowly convergent in the context of vertical excitation energies. \cite{Loos_2018a,Loos_2020a,Veril_2021}
This has led some of us to recently propose the ADC(2.5) composite approach, where, in the same spirit as MP2.5, one averages the second-order [ADC(2)] and third-order [ADC(3)] vertical transition energies. \cite{Loos_2020d}

Multi-reference perturbation theory is somewhat easier to generalize to excited states as one has the freedom to select the states of interest to include in the reference (zeroth-order) space via the so-called complete-active-space self-consistent field (CASSCF) formalism, hence effectively catching static correlation in the zeroth-order model space.
The missing dynamical correlation can then be recovered in the (first-order) outer space via low-order perturbation theory, as performed in the complete-active-space second-order perturbation theory (CASPT2) of Roos and coworkers, \cite{Andersson_1990,Andersson_1992,Roos_1995a} the multireference MP2 approach of Hirao, \cite{Hirao_1992} or the $N$-electron valence state second-order perturbation theory (NEVPT2) developed by Angeli, Malrieu, and coworkers. \cite{Angeli_2001a,Angeli_2001b,Angeli_2002,Angeli_2006}
However, these multi-reference formalisms and their implementation are much more involved and costly than their single-reference counterparts.
Although it has well-documented weaknesses, CASPT2 is indisputably the most popular of the three approaches mentioned above. 
As such, it has been employed in countless computational studies involving electronic excited states. 

In the context of excited states, its most severe drawback is certainly the intruder state problem (which is, by construction, absent in NEVPT2) that describes a situation where one or several determinants of the outer (first-order) space, known as perturbers, have an energy close to the zeroth-order CASSCF wave function, hence producing divergences in the denominators of the second-order perturbative energy.
One can introduce a shift in the denominators to avoid such situations, and correcting afterwards the second-order energy for the use of this shift.
The use of real-valued \cite{Roos_1995b,Roos_1996} or imaginary \cite{Forsberg_1997} level shifts has been successfully tested and is now routine in excited-state calculations. \cite{Schapiro_2013,Zobel_2017,Sarkar_2022}

A second pitfall was brought to light by Andersson \textit{et al.} \cite{Andersson_1993,Andersson_1995} and explained by the unbalanced treatment in the zeroth-order Hamiltonian of the open- and closed-shell electronic configurations. 
A cure was quickly proposed via the introduction of an additional parameter in the zeroth-order Hamiltonian, the ionization-potential-electron-affinity (IPEA) shift. \cite{Ghigo_2004}
Although the introduction of an IPEA shift can provide a better agreement between experiment and theory, \cite{Pierloot_2006,Pierloot_2008,Suaud_2009,Kepenekian_2009,Daku_2012,Rudavskyi_2014,Vela_2016,Wen_2018} it has been shown that its application is not systematically justified and that its impact is significantly basis set dependent. \cite{Zobel_2017}

Very recently, based on the highly accurate vertical excitation energies of the QUEST database, \cite{Loos_2018a,Loos_2019,Loos_2020a,Loos_2020b,Loos_2020c,Veril_2021,Loos_2021c,Loos_2021b} we have reported an exhaustive benchmark of CASPT2 and NEVPT2 for 280 excited states of diverse natures (singlet, triplet, valence, Rydberg, $n\to\pis$, $\pi\to\pis$, and double excitations) computed with a large basis set (aug-cc-pVTZ) in 35 small- and medium-sized organic molecules containing from three to six non-hydrogen atoms. \cite{Sarkar_2022}
Our main take-home message was that both CASPT2 with IPEA shift and the partially-contracted version of NEVPT2 provide fairly reliable vertical transition energy estimates, with slight overestimations and mean absolute errors of \SI{0.11}{} and \SI{0.13}{\eV}, respectively.
Importantly, the introduction of the IPEA shift in CASPT2 was found to be crucial as neglecting it increases the mean absolute error to \SI{0.27}{eV}.

In the electronic structure community, third-order perturbation theory has a fairly bad reputation especially within MP perturbation theory where it is rarely worth its extra computational cost. \cite{Rettig_2020} 
Nonetheless, going against popular beliefs and one step further in the perturbative expansion, we propose here to assess the performance of the complete-active-space third-order perturbation theory (CASPT3) method developed by Werner \cite{Werner_1996} and implemented in MOLPRO \cite{Werner_2020} for the same set of electronic transitions as the one used in Ref.~\onlinecite{Sarkar_2022} 
Although CASPT3 calculations have been reported in the literature, 
\cite{Angeli_2006,Yanai_2007,Grabarek_2016,Li_2017,Li_2018,Li_2021,Bittererova_2001,Bokarev_2009,Frankcombe_2011,Gu_2008,Kerkines_2005,Lampart_2008,Leininger_2000,Maranzana_2020,Papakondylis_1999,Schild_2013,Sun_2018,Takatani_2009,Takatani_2010,Verma_2018,Woywod_2010,Yan_2004,Zhang_2020,Zhu_2005,Zhu_2007,Zhu_2013,Zou_2009}
the present study provides, to the best of our knowledge, the first comprehensive benchmark of CASPT3 and allows assessing its accuracy in the framework of electronically excited states.
We underline that, although a third-order version of NEVPT has been developed \cite{Angeli_2006} and has been used in some applications \cite{Pastore_2006a,Pastore_2006b,Pastore_2007,Angeli_2007,Camacho_2010,Angeli_2011,Angeli_2012} by Angeli and coworkers, as far as we are aware of, no NEVPT3 implementation is publicly available.

\alert{Although comparing with experimental values would be interesting on its own right, this involves the computation of 0-0 energies which are much more expensive to determine as they require the equilibrium geometries of the ground and excited states as well as the zero-point vibrational energies for each state.
Moreover, we have recently shown that the accuracy of such quantities are mainly driven by the quality of the (absorption and emission) vertical excitation energies as other geometrical and vibrational effects mostly cancel out. \cite{Loos_2018b,Loos_2019a,Loos_2019b} 
Therefore, it is clear that vertical excitation energies are the key quantities to reproduce.}
\\

\section{Computational details}
\label{sec:compdet}

\begin{figure}
	\includegraphics[width=\linewidth]{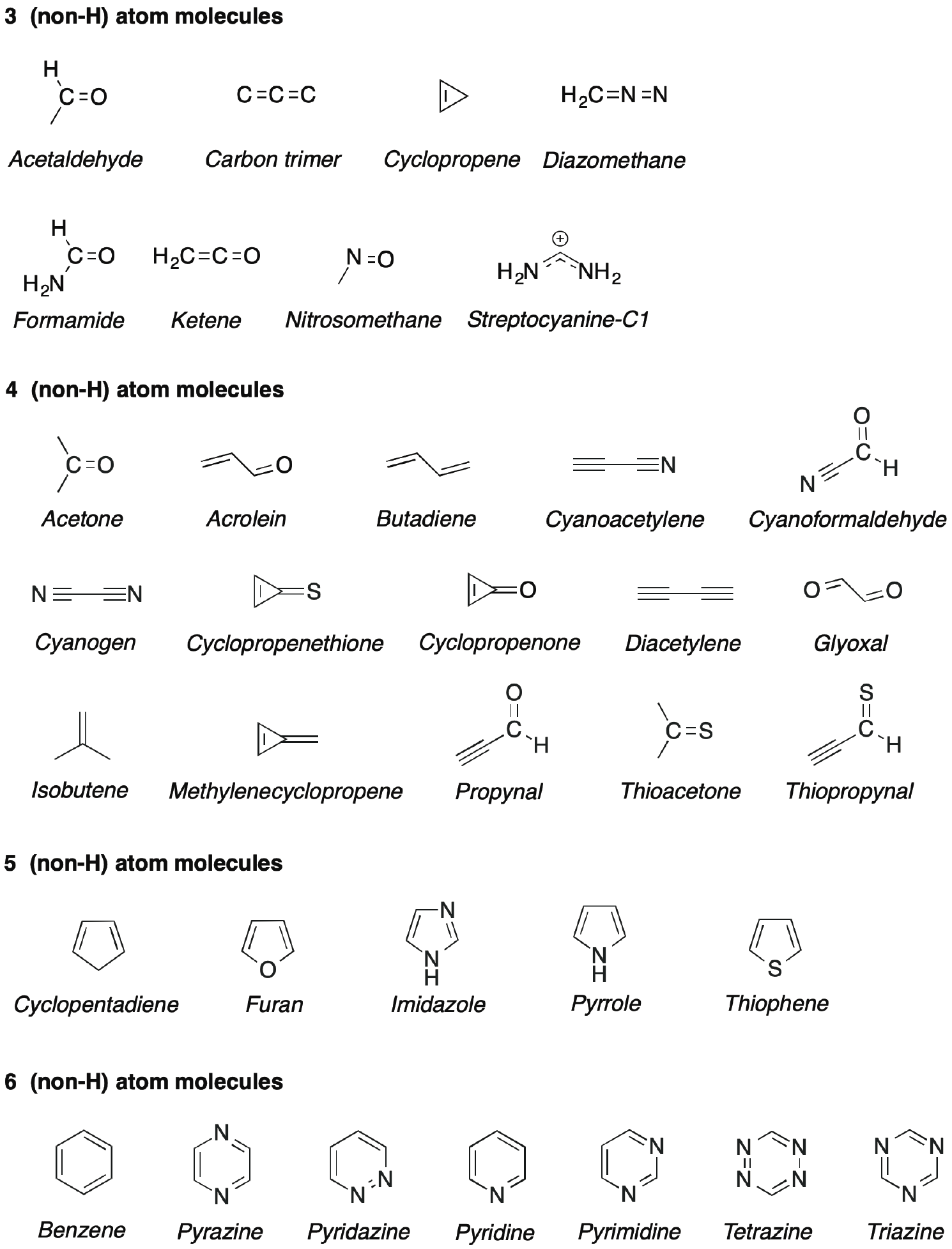}
	\caption{Various molecular systems considered in this study.
	\label{fig:mol}}
\end{figure}

For each compound represented in Fig.~\ref{fig:mol}, we have computed \alert{vertical excitation energies based on single-state CASPT2 and CASPT3 calculations}.
\alert{All calculations reported in the present manuscript have been performed with Dunning's aug-cc-pVTZ basis set.} \cite{Kendall_1992} 
Geometries and reference theoretical best estimates (TBEs) for the vertical excitation energies have been extracted from the QUEST database \cite{Veril_2021} and can be downloaded at \url{https://lcpq.github.io/QUESTDB_website}.

All the CASPT2 and CASPT3 calculations have been carried out in the frozen-core approximation and within the RS2 and RS3 contraction schemes as implemented in MOLPRO and described in Refs.~\onlinecite{Werner_1996} and \onlinecite{Werner_2020}. 
Both methods have been tested with and without IPEA (labeled as NOIPEA).
When an IPEA shift is applied, it is set to the default value of \SI{0.25}{\hartree}, as discussed in Ref.~\onlinecite{Ghigo_2004}.
The MOLPRO implementation of CASPT3 is based on a modification of the multi-reference configuration interaction (MRCI) module. \cite{Werner_1988,Knowles_1988}
For the sake of computational efficiency, the doubly-excited external configurations are internally contracted while the singly-excited internal and semi-internal configurations are left uncontracted. \cite{Werner_1996}
These perturbative calculations have been performed by considering a state-averaged (SA) CASSCF wave function where we have included the ground state and (at least) the excited states of interest.
In several occasions, we have added additional excited states to avoid convergence and/or root-flipping issues.
\alert{Note that the implementation of the IPEA shift is not exactly identical in MOLPRO (used here) and in MOLCAS (used, for example, in Ref.~\onlinecite{Zobel_2017}), since in MOLPRO the singly external configurations are not contracted in the RS2 scheme.}

For each system and transition, we report in the {\SupMat} the exhaustive description of the active spaces for each symmetry representation.
Additionally, for the challenging transitions, we have steadily increased the size of the active space to carefully assess the convergence of the vertical excitation energies of interest.
Note that, compared to our previous CASPT2 benchmark study, \cite{Sarkar_2022} the active spaces of acrolein, pyrimidine, and pyridazine have been slightly reduced in order to make the CASPT3 calculations technically achievable.
In these cases, for the sake of consistency, we have recomputed the CASPT2 values for the same active space.
Although these active space reductions are overall statistically negligible, this explains the small deviations that one may observe between the data reported here and in Ref.~\onlinecite{Sarkar_2022}.
Finally, to alleviate the intruder state problem, a level shift of \SI{0.3}{\hartree} has been systematically applied. \cite{Roos_1995b,Roos_1996}
This value has been slightly increased in particularly difficult cases, and such cases are detailed in {\SupMat}.

\section{Results and discussion}
\label{sec:res}

\begin{figure}
	\includegraphics[width=0.5\linewidth]{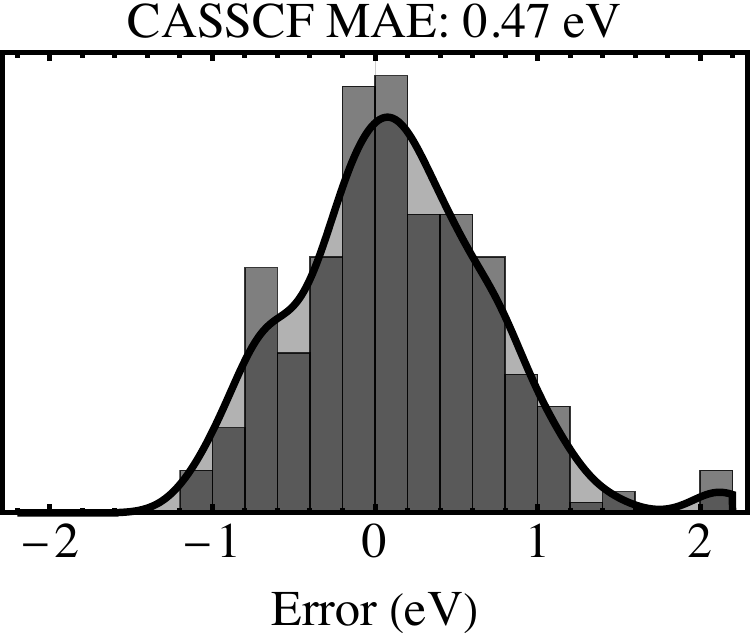}
	\\
	\vspace{0.01\textwidth}
	\includegraphics[width=\linewidth]{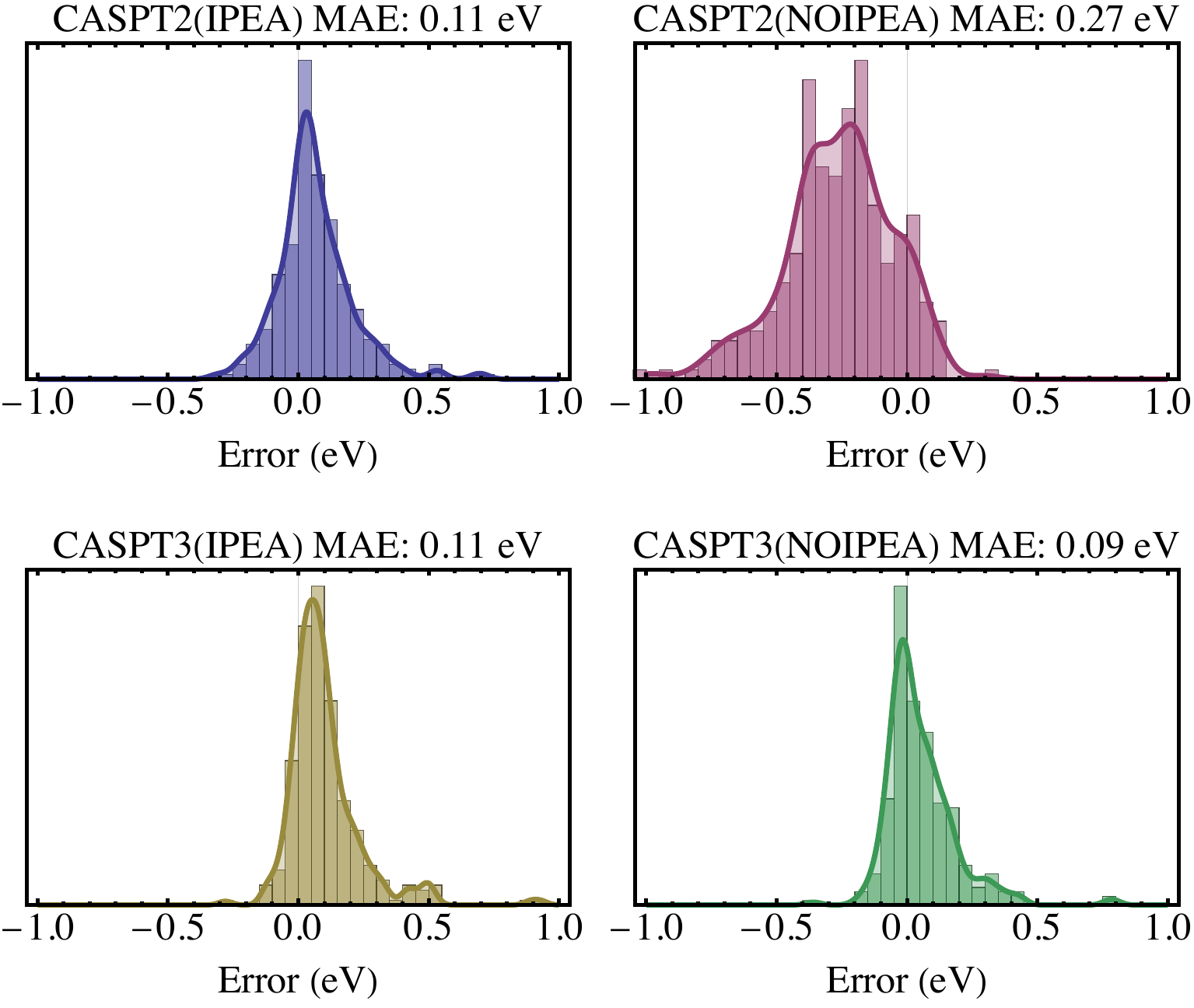}
	\caption{Histograms of the errors (in \si{\eV}) obtained for \alert{CASSCF as well as} CASPT2 and CASPT3 with and without IPEA shift.
	Raw data are given in Table \ref{tab:BigTab}.}
	\label{fig:PT2_vs_PT3}
\end{figure}

A detailed discussion of each individual molecule can be found in Ref.~\onlinecite{Sarkar_2022} and in earlier works, \cite{Loos_2018a,Loos_2020b} where theoretical and experimental literature values are discussed.
We therefore decided to focus on global trends here.
The exhaustive list of \alert{CASSCF}, CASPT2, and CASPT3 transitions can be found in Table \ref{tab:BigTab} and the distribution of the errors are represented in Fig.~\ref{fig:PT2_vs_PT3}.
The usual statistical indicators are used in the following, namely, the mean signed error (MSE), the mean absolute error (MAE), the root-mean-square error (RMSE), the standard deviation of the errors (SDE), as well as the largest positive and negative deviations [Max($+$) and Max($-$), respectively].
These are given in Table \ref{tab:stat} considering the 265 ``safe'' TBEs (out of 280) for which chemical accuracy is assumed (absolute error below \SI{0.043}{\eV}). 
The MAEs determined for subsets of transitions (singlet, triplet, valence, Rydberg, $n\to\pis$, $\pi\to\pis$, and double excitations) and system sizes (3 non-H atoms, 4 non-H atoms, and 5-6 non-H atoms) can be found in Table \ref{tab:stat_subset}. 
Error patterns for selected subsets are reported in {\SupMat}.

\begin{longtable*}{cllccccccccc}
\caption{Vertical excitation energies (in \si{\eV}) computed with various multi-reference methods and the aug-cc-pVTZ basis.
The reference TBEs of the QUEST database, their percentage of single excitations $\%T_1$ involved in the transition (computed at the CC3 level), their nature 
(V and R stand for valence and Rydberg, respectively) are also reported.
TBEs listed as ``safe'' are assumed to be chemically accurate (\ie, absolute error below \SI{0.043}{\eV}).
[F] indicates a fluorescence transition, \ie, a vertical transition energy computed from an excited-state equilibrium geometry.
\label{tab:BigTab}}
\\
\hline\hline
\#	&Molecule			&Excitation				&Nature	&$\%T_1$	&TBE	&Safe?	&CASSCF	&CASPT2	&CASPT2	&CASPT3	&CASPT3\\
	&					&						&		&			&		&		&		&(IPEA)	&(NOIPEA)&(IPEA)&(NOIPEA)\\
\hline
\endfirsthead
\hline\hline
\#	&Molecule			&Excitation				&Nature	&$\%T_1$	&TBE	&Safe?	&CASSCF	&CASPT2	&CASPT2	&CASPT3	&CASPT3\\
	&					&						&		&			&		&		&		&(IPEA)	&(NOIPEA)&(IPEA)&(NOIPEA)\\
\hline
\endhead
\hline 
\multicolumn{12}{r}{{Continued on next page}} \\
\endfoot
\hline\hline
\endlastfoot
1	&Acetaldehyde		&$^1A''(n,\pis)$			&V	&91.3	&4.31	&\Y	&4.62	&4.35	&4.13	&4.44	&4.41\\
2	&					&$^3A''(n,\pis)$			&V	&97.9	&3.97	&\Y	&4.28	&3.94	&3.71	&4.06	&4.03\\
3	&Acetone			&$^1A_2(n,\pis)$			&V	&91.1	&4.47	&\Y	&4.77	&4.44	&4.19	&4.57	&4.55\\
4	&					&$^1B_2(n,3s)$				&R	&90.5	&6.46	&\Y	&5.50	&6.46	&6.35	&6.64	&6.67\\
5	&					&$^1A_2(n,3p)$				&R	&90.9	&7.47	&\Y	&7.46	&7.80	&7.55	&7.76	&7.68\\
6	&					&$^1A_1(n,3p)$				&R	&90.6	&7.51	&\Y	&7.03	&7.67	&7.46	&7.76	&7.75\\
7	&					&$^1B_2(n,3p)$				&R	&91.2	&7.62	&\Y	&6.44	&7.56	&7.47	&7.73	&7.76\\
8	&					&$^3A_2(n,\pis)$			&V	&97.8	&4.13	&\Y	&4.47	&4.13	&3.89	&4.27	&4.24\\
9	&					&$^3A_1(\pi,\pis)$			&V	&98.7	&6.25	&\Y	&6.22	&6.24	&6.07	&6.26	&6.22\\
10	&Acrolein			&$^1A''(n,\pis)$			&V	&87.6	&3.78	&\Y	&3.48	&3.58	&3.46	&3.66	&3.66\\
11	&					&$^1A'(\pi,\pis)$			&V	&91.2	&6.69	&\Y	&8.84	&6.93	&6.28	&7.18	&7.05\\
12	&					&$^1A''(n,\pis)$			&V	&79.4	&6.72	&\N	&6.76	&6.79	&6.34	&6.88	&6.80\\
13	&					&$^1A'(n,3s)$				&R	&89.4	&7.08	&\Y	&7.20	&7.21	&6.98	&7.20	&7.16\\
14	&					&$^1A'(\pi,\pis)$			&V	&75.0	&7.87	&\Y	&7.91	&8.10	&7.75	&8.02	&7.95\\
15	&					&$^3A''(n,\pis)$			&V	&97.0	&3.51	&\Y	&3.25	&3.28	&3.15	&3.39	&3.40\\
16	&					&$^3A'(\pi,\pis)$			&V	&98.6	&3.94	&\Y	&3.89	&4.01	&3.78	&3.96	&3.91\\
17	&					&$^3A'(\pi,\pis)$			&V	&98.4	&6.18	&\Y	&5.89	&6.20	&5.93	&6.10	&6.02\\
18	&					&$^3A''(n,\pis)$			&V	&92.7	&6.54	&\N	&6.67	&6.65	&6.21	&6.74	&6.66\\
19	&Benzene			&$^1B_{2u}(\pi,\pis)$		&V	&86.3	&5.06	&\Y	&4.98	&5.14	&4.66	&5.09	&5.01\\
20	&					&$^1B_{1u}(\pi,\pis)$		&V	&92.9	&6.45	&\Y	&7.27	&6.65	&6.23	&6.67	&6.58\\
21	&					&$^1E_{1g}(\pi,3s)$			&R	&92.8	&6.52	&\Y	&5.90	&6.70	&6.57	&6.56	&6.51\\
22	&					&$^1A_{2u}(\pi,3p)$			&R	&93.4	&7.08	&\Y	&6.14	&7.21	&7.07	&7.07	&7.02\\
23	&					&$^1E_{2u}(\pi,3p)$			&R	&92.8	&7.15	&\Y	&6.21	&7.26	&7.12	&7.13	&7.08\\
24	&					&$^1E_{2g}(\pi,\pis)$		&V	&73.0	&8.28	&\Y	&8.10	&8.31	&7.82	&8.26	&8.16\\
25	&					&$^3B_{1u}(\pi,\pis)$		&V	&98.6	&4.16	&\Y	&3.85	&4.22	&3.92	&4.14	&4.08\\
26	&					&$^3E_{1u}(\pi,\pis)$		&V	&97.1	&4.85	&\Y	&4.85	&4.89	&4.51	&4.87	&4.80\\
27	&					&$^3B_{2u}(\pi,\pis)$		&V	&98.1	&5.81	&\Y	&6.75	&5.85	&5.40	&5.90	&5.81\\
28	&Butadiene			&$^1B_u(\pi,\pis)$			&V	&93.3	&6.22	&\Y	&6.65	&6.76	&6.52	&6.72	&6.65\\
29	&					&$^1B_g(\pi,3s)$			&R	&94.1	&6.33	&\Y	&5.94	&6.49	&6.32	&6.43	&6.38\\
30	&					&$^1A_g(\pi,\pis)$			&V	&75.1	&6.50	&\Y	&6.99	&6.74	&6.30	&6.73	&6.66\\
31	&					&$^1A_u(\pi,3p)$			&R	&94.1	&6.64	&\Y	&5.95	&6.74	&6.64	&6.70	&6.67\\
32	&					&$^1A_u(\pi,3p)$			&R	&94.1	&6.80	&\Y	&6.12	&6.95	&6.84	&6.90	&6.86\\
33	&					&$^1B_u(\pi,3p)$			&R	&93.8	&7.68	&\Y	&7.93	&7.60	&7.30	&7.62	&7.54\\
34	&					&$^3B_u(\pi,\pis)$			&V	&98.4	&3.36	&\Y	&3.55	&3.40	&3.19	&3.40	&3.35\\
35	&					&$^3A_g(\pi,\pis)$			&V	&98.7	&5.20	&\Y	&5.52	&5.32	&4.93	&5.29	&5.19\\
36	&					&$^3B_g(\pi,3s)$			&R	&97.9	&6.29	&\Y	&5.89	&6.44	&6.27	&6.38	&6.33\\
37	&Carbon trimer		&$^1\Delta_g(\text{double})$&V	&1.0	&5.22	&\Y	&4.98	&5.08	&4.85	&5.20	&5.19\\
38	&					&$^1\Sigma^+_g(\text{double})$&V&1.0	&5.91	&\Y	&5.84	&5.82	&5.58	&5.92	&5.89\\
39	&Cyanoacetylene		&$^1\Sigma^-(\pi,\pis)$		&V	&94.3	&5.80	&\Y	&6.54	&5.85	&5.47	&5.89	&5.81\\
40	&					&$^1\Delta(\pi,\pis)$		&V	&94.0	&6.07	&\Y	&6.80	&6.13	&5.78	&6.17	&6.09\\
41	&					&$^3\Sigma^+(\pi,\pis)$		&V	&98.5	&4.44	&\Y	&4.86	&4.45	&4.04	&4.52	&4.45\\
42	&					&$^3\Delta(\pi,\pis)$		&V	&98.2	&5.21	&\Y	&5.64	&5.21	&4.86	&5.26	&5.19\\
43	&					&$^1A''[F](\pi,\pis)$		&V	&93.6	&3.54	&\Y	&4.30	&3.67	&3.47	&3.64	&3.58\\
44	&Cyanoformaldehyde	&$^1A''(n,\pis)$			&V	&89.8	&3.81	&\Y	&4.02	&3.98	&3.67	&3.94	&3.89\\
45	&					&$^1A''(\pi,\pis)$			&V	&91.9	&6.46	&\Y	&7.61	&6.79	&6.43	&6.77	&6.67\\
46	&					&$^3A''(n,\pis)$			&V	&97.6	&3.44	&\Y	&3.52	&3.46	&3.25	&3.51	&3.50\\
47	&					&$^3A'(\pi,\pis)$			&V	&98.4	&5.01	&\Y	&4.98	&5.25	&5.03	&5.16	&5.12\\
48	&Cyanogen			&$^1\Sigma_u^-(\pi,\pis)$	&V	&94.1	&6.39	&\Y	&7.14	&6.40	&6.03	&6.46	&6.39\\
49	&					&$^1\Delta_u(\pi,\pis)$		&V	&93.4	&6.66	&\Y	&7.46	&6.70	&6.35	&6.75	&6.68\\
50	&					&$^3\Sigma_u^+(\pi,\pis)$	&V	&98.5	&4.91	&\Y	&5.28	&4.85	&4.46	&4.95	&4.89\\
51	&					&$^1\Sigma_u^-[F](\pi,\pis)$&V	&93.4	&5.05	&\Y	&5.68	&5.07	&4.75	&5.11	&5.04\\
52	&Cyclopentadiene	&$^1B_2(\pi,\pis)$			&V	&93.8	&5.56	&\Y	&6.71	&5.96	&5.62	&6.06	&5.99\\
53	&					&$^1A_2(\pi,3s)$			&R	&94.0	&5.78	&\Y	&5.21	&5.88	&5.78	&5.81	&5.77\\
54	&					&$^1B_1(\pi,3p)$			&R	&94.2	&6.41	&\Y	&6.08	&6.59	&6.44	&6.47	&6.41\\
55	&					&$^1A_2(\pi,3p)$			&R	&93.8	&6.46	&\Y	&5.78	&6.55	&6.46	&6.45	&6.41\\
56	&					&$^1B_2(\pi,3p)$			&R	&94.2	&6.56	&\Y	&6.16	&6.72	&6.56	&6.61	&6.54\\
57	&					&$^1A_1(\pi,\pis)$			&V	&78.9	&6.52	&\N	&6.49	&6.63	&6.13	&6.59	&6.50\\
58	&					&$^3B_2(\pi,\pis)$			&V	&98.4	&3.31	&\Y	&3.26	&3.34	&3.09	&3.31	&3.26\\
59	&					&$^3A_1(\pi,\pis)$			&V	&98.6	&5.11	&\Y	&4.92	&5.14	&4.78	&5.10	&5.03\\
60	&					&$^3A_2(\pi,3s)$			&R	&97.9	&5.73	&\Y	&5.53	&5.91	&5.74	&5.81	&5.75\\
61	&					&$^3B_1(\pi,3p)$			&R	&97.9	&6.36	&\Y	&6.05	&6.56	&6.40	&6.43	&6.37\\
62	&Cyclopropene 		&$^1B_1(\sig,\pis)$			&V	&92.8	&6.68	&\Y	&7.48	&6.86	&6.58	&6.85	&6.77\\
63	&					&$^1B_2(\pi,\pis)$			&V	&95.1	&6.79	&\Y	&7.47	&6.89	&6.47	&6.96	&6.87\\
64	&					&$^3B_2(\pi,\pis)$			&V	&98.0	&4.38	&\Y	&4.60	&4.47	&4.27	&4.46	&4.40\\
65	&					&$^3B_1(\sig,\pis)$			&V	&98.9	&6.45	&\Y	&7.08	&6.56	&6.32	&6.55	&6.47\\
66	&Cyclopropenethione	&$^1A_2(n,\pis)$			&V	&89.6	&3.41	&\Y	&3.44	&3.43	&3.14	&3.46	&3.40\\
67	&					&$^1B_1(n,\pis)$			&V	&84.8	&3.45	&\Y	&3.57	&3.45	&3.17	&3.52	&3.46\\
68	&					&$^1B_2(\pi,\pis)$			&V	&83.0	&4.60	&\Y	&4.51	&4.64	&4.35	&4.66	&4.61\\
69	&					&$^1B_2(n,3s)$				&R	&91.8	&5.34	&\Y	&4.59	&5.25	&5.15	&5.25	&5.22\\
70	&					&$^1A_1(\pi,\pis)$			&V	&89.0	&5.46	&\Y	&6.46	&5.84	&5.32	&5.88	&5.75\\
71	&					&$^1B_2(n,3p)$				&R	&91.3	&5.92	&\Y	&5.27	&5.93	&5.86	&5.92	&5.90\\
72	&					&$^3A_2(n,\pis)$			&V	&97.2	&3.28	&\Y	&3.26	&3.28	&3.00	&3.33	&3.28\\
73	&					&$^3B_1(n,\pis)$			&V	&94.5	&3.32	&\Y	&3.51	&3.35	&3.07	&3.42	&3.36\\
74	&					&$^3B_2(\pi,\pis)$			&V	&96.5	&4.01	&\Y	&3.80	&3.97	&3.75	&3.99	&3.95\\
75	&					&$^3A_1(\pi,\pis)$			&V	&98.2	&4.01	&\Y	&3.83	&4.01	&3.77	&4.00	&3.95\\
76	&Cyclopropenone		&$^1B_1(n,\pis)$			&V	&87.7	&4.26	&\Y	&4.92	&4.12	&3.75	&4.40	&4.38\\
77	&					&$^1A_2(n,\pis)$			&V	&91.0	&5.55	&\Y	&5.64	&5.62	&5.31	&5.67	&5.64\\
78	&					&$^1B_2(n,3s)$				&R	&90.8	&6.34	&\Y	&5.68	&6.28	&6.21	&6.41	&6.44\\
79	&					&$^1B_2(\pi,\pis)$			&V	&86.5	&6.54	&\Y	&6.40	&6.54	&6.20	&6.63	&6.62\\
80	&					&$^1B_2(n,3p)$				&R	&91.1	&6.98	&\Y	&6.35	&6.84	&6.70	&6.99	&7.01\\
81	&					&$^1A_1(n,3p)$				&R	&91.2	&7.02	&\Y	&6.84	&7.27	&7.03	&7.26	&7.24\\
82	&					&$^1A_1(\pi,\pis)$			&V	&90.8	&8.28	&\Y	&10.42	&8.96	&8.11	&9.21	&9.07\\
83	&					&$^3B_1(n,\pis)$			&V	&96.0	&3.93	&\Y	&4.72	&3.65	&3.28	&4.00	&3.98\\
84	&					&$^3B_2(\pi,\pis)$			&V	&97.9	&4.88	&\Y	&4.39	&4.76	&4.60	&4.76	&4.74\\
85	&					&$^3A_2(n,\pis)$			&V	&97.5	&5.35	&\Y	&5.40	&5.36	&5.06	&5.44	&5.42\\
86	&					&$^3A_1(\pi,\pis)$			&V	&98.1	&6.79	&\Y	&6.59	&6.93	&6.61	&6.86	&6.82\\
87	&Diacetylene		&$^1\Sigma_u^-(\pi,\pis)$	&V	&94.4	&5.33	&\Y	&6.13	&5.42	&5.01	&5.45	&5.36\\
88	&					&$^1\Delta_u(\pi,\pis)$		&V	&94.1	&5.61	&\Y	&6.39	&5.68	&5.30	&5.72	&5.63\\
89	&					&$^3\Sigma_u^+(\pi,\pis)$	&V	&98.5	&4.10	&\Y	&4.54	&4.11	&3.67	&4.17	&4.09\\
90	&					&$^3\Delta_u(\pi,\pis)$		&V	&98.2	&4.78	&\Y	&5.28	&4.82	&4.45	&4.86	&4.78\\
91	&Diazomethane 		&$^1A_2(\pi,\pis)$			&V	&90.1	&3.14	&\Y	&3.27	&3.13	&2.92	&3.09	&3.04\\
92	&					&$^1B_1(\pi,3s)$			&R	&93.8	&5.54	&\Y	&4.59	&5.50	&5.30	&5.48	&5.45\\
93	&					&$^1A_1(\pi,\pis)$			&V	&91.4	&5.90	&\Y	&5.65	&6.21	&5.92	&6.18	&6.13\\
94	&					&$^3A_2(\pi,\pis)$			&V	&97.7	&2.79	&\Y	&3.02	&2.87	&2.67	&2.84	&2.79\\
95	&					&$^3A_1(\pi,\pis)$			&V	&98.6	&4.05	&\Y	&4.27	&4.10	&3.88	&4.06	&4.01\\
96	&					&$^3B_1(\pi,3s)$			&R	&98.0	&5.35	&\Y	&4.45	&5.34	&5.15	&5.33	&5.30\\
97	&					&$^3A_1(\pi,3p)$			&R	&98.5	&6.82	&\Y	&6.34	&7.00	&6.76	&6.96	&6.91\\
98	&					&$^1A''[F](\pi,\pis)$		&V	&87.4	&0.71	&\Y	&0.72	&0.69	&0.52	&0.66	&0.62\\
99	&Formamide 			&$^1A''(n,\pis)$			&V	&90.8	&5.65	&\Y	&5.95	&5.66	&5.45	&5.71	&5.67\\
100	&					&$^1A'(n,3s)$				&R	&88.6	&6.77	&\Y	&6.17	&6.80	&6.64	&6.82	&6.81\\
101	&					&$^1A'(n,3p)$				&R	&89.6	&7.38	&\N	&6.74	&7.45	&7.32	&7.46	&7.46\\
102	&					&$^1A'(\pi,\pis)$			&V	&89.3	&7.63	&\N	&8.80	&7.88	&7.13	&7.95	&7.78\\
103	&					&$^3A''(n,\pis)$			&V	&97.7	&5.38	&\Y	&5.89	&5.36	&5.16	&5.41	&5.37\\
104	&					&$^3A'(\pi,\pis)$			&V	&98.2	&5.81	&\Y	&6.10	&5.88	&5.62	&5.91	&5.87\\
105	&Furan				&$^1A_2(\pi,3s)$			&R	&93.8	&6.09	&\Y	&5.26	&6.16	&6.04	&6.06	&6.02\\
106	&					&$^1B_2(\pi,\pis)$			&V	&93.0	&6.37	&\Y	&7.78	&6.59	&6.02	&6.80	&6.71\\
107	&					&$^1A_1(\pi,\pis)$			&V	&92.4	&6.56	&\Y	&6.73	&6.66	&6.10	&6.69	&6.62\\
108	&					&$^1B_1(\pi,3p)$			&R	&93.9	&6.64	&\Y	&6.07	&6.79	&6.63	&6.65	&6.60\\
109	&					&$^1A_2(\pi,3p)$			&R	&93.6	&6.81	&\Y	&5.87	&6.87	&6.77	&6.76	&6.72\\
110	&					&$^1B_2(\pi,3p)$			&R	&93.5	&7.24	&\Y	&6.54	&7.11	&6.84	&6.96	&6.88\\
111	&					&$^3B_2(\pi,\pis)$			&V	&98.4	&4.20	&\Y	&3.94	&4.26	&4.01	&4.17	&4.12\\
112	&					&$^3A_1(\pi,\pis)$			&V	&98.1	&5.46	&\Y	&5.41	&5.50	&5.09	&5.47	&5.40\\
113	&					&$^3A_2(\pi,3s)$			&R	&97.9	&6.02	&\Y	&5.57	&6.16	&5.99	&6.05	&5.99\\
114	&					&$^3B_1(\pi,3p)$			&R	&97.9	&6.59	&\Y	&6.04	&6.76	&6.60	&6.62	&6.56\\
115	&Glyoxal			&$^1A_u(n,\pis)$			&V	&91.0	&2.88	&\Y	&3.42	&2.82	&2.51	&2.97	&2.94\\
116	&					&$^1B_g(n,\pis)$			&V	&88.3	&4.24	&\Y	&4.68	&4.21	&3.89	&4.36	&4.31\\
117	&					&$^1A_g(\text{double})$		&V	&0.5	&5.61	&\Y	&5.92	&5.37	&5.21	&5.53	&5.55\\
118	&					&$^1B_g(n,\pis)$			&V	&83.9	&6.57	&\Y	&7.35	&6.52	&5.98	&6.76	&6.72\\
119	&					&$^1B_u(n,3p)$				&R	&91.7	&7.71	&\Y	&7.04	&7.61	&7.34	&7.78	&7.81\\
120	&					&$^3A_u(n,\pis)$			&V	&97.6	&2.49	&\Y	&3.06	&2.41	&2.12	&2.57	&2.55\\
121	&					&$^3B_g(n,\pis)$			&V	&97.4	&3.89	&\Y	&4.61	&3.90	&3.53	&4.04	&4.01\\
122	&					&$^3B_u(\pi,\pis)$			&V	&98.5	&5.15	&\Y	&5.46	&5.14	&4.91	&5.17	&5.14\\
123	&					&$^3A_g(\pi,\pis)$			&V	&98.8	&6.30	&\Y	&6.69	&6.32	&6.02	&6.33	&6.27\\
124	&Imidazole			&$^1A''(\pi,3s)$			&R	&93.0	&5.70	&\Y	&5.04	&5.88	&5.66	&5.74	&5.68\\
125	&					&$^1A'(\pi,3p)$				&R	&90.0	&6.41	&\Y	&6.18	&6.69	&6.45	&6.61	&6.56\\
126	&					&$^1A''(\pi,3p)$			&R	&93.6	&6.50	&\Y	&5.43	&6.57	&6.47	&6.47	&6.44\\
127	&					&$^1A''(n,\pis)$			&V	&89.0	&6.71	&\Y	&7.13	&6.94	&6.57	&6.92	&6.85\\
128	&					&$^1A'(\pi,\pis)$			&V	&88.9	&6.86	&\Y	&6.73	&6.88	&6.46	&6.89	&6.83\\
129	&					&$^1A'(n,3s)$				&R	&89.0	&7.00	&\Y	&6.36	&7.10	&6.91	&7.09	&7.07\\
130	&					&$^3A'(\pi,\pis)$			&V	&98.3	&4.73	&\Y	&4.55	&4.78	&4.53	&4.73	&4.68\\
131	&					&$^3A''(\pi,3s)$			&R	&97.6	&5.66	&\Y	&5.03	&5.86	&5.63	&5.72	&5.66\\
132	&					&$^3A'(\pi,\pis)$			&V	&97.9	&5.74	&\Y	&5.69	&5.85	&5.48	&5.80	&5.72\\
133	&					&$^3A''(n,\pis)$			&V	&97.3	&6.31	&\Y	&6.58	&6.44	&6.10	&6.43	&6.37\\
134	&Isobutene			&$^1B_1(\pi,3s)$			&R	&94.1	&6.46	&\Y	&6.21	&6.74	&6.59	&6.64	&6.57\\
135	&					&$^1A_1(\pi,3p)$			&R	&94.2	&7.01	&\Y	&6.90	&7.32	&7.14	&7.24	&7.18\\
136	&					&$^3A_1(\pi,\pis)$			&V	&98.9	&4.53	&\Y	&4.66	&4.59	&4.41	&4.58	&4.53\\
137	&Ketene 			&$^1A_2(\pi,\pis)$			&V	&91.0	&3.86	&\Y	&3.98	&3.92	&3.70	&3.90	&3.85\\
138	&					&$^1B_1(\pi,3s)$			&R	&93.9	&6.01	&\Y	&5.22	&5.99	&5.79	&6.00	&5.97\\
139	&					&$^1A_2(\pi,3p)$			&R	&94.4	&7.18	&\Y	&6.38	&7.25	&7.05	&7.19	&7.15\\
140	&					&$^3A_2(\pi,\pis)$			&V	&91.0	&3.77	&\Y	&3.92	&3.81	&3.59	&3.79	&3.74\\
141	&					&$^3A_1(\pi,\pis)$			&V	&98.6	&5.61	&\Y	&5.79	&5.65	&5.43	&5.63	&5.59\\
142	&					&$^3B_1(\pi,3s)$			&R	&98.1	&5.79	&\Y	&5.05	&5.79	&5.60	&5.80	&5.77\\
143	&					&$^3A_2(\pi,3p)$			&R	&94.4	&7.12	&\Y	&6.35	&7.22	&7.01	&7.15	&7.11\\
144	&					&$^1A''[F](\pi,\pis)$		&V	&87.9	&1.00	&\Y	&0.95	&1.05	&0.88	&1.00	&0.95\\
145	&Methylenecyclopropene&$^1B_2(\pi,\pis)$		&V	&85.4	&4.28	&\Y	&4.47	&4.40	&4.12	&4.39	&4.33\\
146	&					&$^1B_1(\pi,3s)$			&R	&93.6	&5.44	&\Y	&4.92	&5.57	&5.44	&5.46	&5.41\\
147	&					&$^1A_2(\pi,3p)$			&R	&93.3	&5.96	&\Y	&5.37	&6.09	&5.97	&5.97	&5.92\\
148	&					&$^1A_1(\pi,\pis)$			&V	&92.8	&6.12	&\N	&5.37	&6.26	&6.16	&6.17	&6.13\\
149	&					&$^3B_2(\pi,\pis)$			&V	&97.2	&3.49	&\Y	&3.44	&3.57	&3.34	&3.55	&3.49\\
150	&					&$^3A_1(\pi,\pis)$			&V	&98.6	&4.74	&\Y	&4.60	&4.82	&4.58	&4.77	&4.72\\
151	&Nitrosomethane 	&$^1A''(n,\pis)$			&V	&93.0	&1.96	&\Y	&2.12	&1.84	&1.60	&1.94	&1.91\\
152	&					&$^1A'(\text{double})$		&V	&2.5	&4.76	&\Y	&4.74	&4.69	&4.67	&4.71	&4.71\\
153	&					&$^1A'(n,3s)$				&R	&90.8	&6.29	&\Y	&5.87	&6.32	&6.07	&6.34	&6.31\\
154	&					&$^3A''(n,\pis)$			&V	&98.4	&1.16	&\Y	&1.31	&1.00	&0.75	&1.12	&1.09\\
155	&					&$^3A'(\pi,\pis)$			&V	&98.9	&5.60	&\Y	&5.52	&5.52	&5.37	&5.54	&5.50\\
156	&					&$^1A''[F](n,\pis)$			&V	&92.7	&1.67	&\Y	&1.83	&1.55	&1.32	&1.66	&1.62\\
157	&Propynal			&$^1A''(n,\pis)$			&V	&89.0	&3.80	&\Y	&4.00	&3.92	&3.64	&3.90	&3.86\\
158	&					&$^1A''(\pi,\pis)$			&V	&92.9	&5.54	&\Y	&6.62	&5.82	&5.49	&5.81	&5.72\\
159	&					&$^3A''(n,\pis)$			&V	&97.4	&3.47	&\Y	&3.52	&3.48	&3.26	&3.52	&3.50\\
160	&					&$^3A'(\pi,\pis)$			&V	&98.3	&4.47	&\Y	&4.69	&4.59	&4.30	&4.59	&4.54\\
161	&Pyrazine			&$^1B_{3u}(n,\pis)$			&V	&90.1	&4.15	&\Y	&4.76	&4.09	&3.66	&4.31	&4.30\\
162	&					&$^1A_u(n,\pis)$			&V	&88.6	&4.98	&\Y	&5.90	&4.76	&4.26	&5.10	&5.10\\
163	&					&$^1B_{2u}(\pi,\pis)$		&V	&86.9	&5.02	&\Y	&4.97	&5.13	&4.65	&5.09	&5.03\\
164	&					&$^1B_{2g}(n,\pis)$			&V	&85.6	&5.71	&\Y	&5.80	&5.68	&5.27	&5.73	&5.70\\
165	&					&$^1A_g(n,3s)$				&R	&91.1	&6.65	&\Y	&6.69	&6.66	&6.27	&6.81	&6.80\\
166	&					&$^1B_{1g}(n,\pis)$			&V	&84.2	&6.74	&\Y	&7.16	&6.61	&6.07	&6.78	&6.76\\
167	&					&$^1B_{1u}(\pi,\pis)$		&V	&92.8	&6.88	&\Y	&8.04	&7.14	&6.72	&7.20	&7.12\\
168	&					&$^1B_{1g}(\pi,3s)$			&R	&93.8	&7.21	&\Y	&6.73	&7.41	&7.27	&7.24	&7.18\\
169	&					&$^1B_{2u}(n,3p)$			&R	&90.8	&7.24	&\Y	&7.49	&7.34	&6.93	&7.43	&7.40\\
170	&					&$^1B_{1u}(n,3p)$			&R	&91.4	&7.44	&\Y	&7.83	&7.55	&7.08	&7.64	&7.59\\
171	&					&$^1B_{1u}(\pi,\pis)$		&V	&90.5	&7.98	&\N	&9.65	&8.59	&7.96	&8.68	&8.57\\
172	&					&$^3B_{3u}(n,\pis)$			&V	&97.3	&3.59	&\Y	&4.16	&3.49	&3.08	&3.72	&3.71\\
173	&					&$^3B_{1u}(\pi,\pis)$		&V	&98.5	&4.35	&\Y	&3.98	&4.44	&4.15	&4.34	&4.28\\
174	&					&$^3B_{2u}(\pi,\pis)$		&V	&97.6	&4.39	&\Y	&4.62	&4.44	&4.09	&4.47	&4.41\\
175	&					&$^3A_u(n,\pis)$			&V	&96.1	&4.93	&\Y	&5.85	&4.73	&4.21	&5.07	&5.07\\
176	&					&$^3B_{2g}(n,\pis)$			&V	&97.0	&5.08	&\Y	&5.25	&5.04	&4.66	&5.14	&5.11\\
177	&					&$^3B_{1u}(\pi,\pis)$		&V	&97.0	&5.28	&\Y	&5.15	&5.29	&4.92	&5.25	&5.19\\
178	&Pyridazine			&$^1B_1(n,\pis)$			&V	&89.0	&3.83	&\Y	&4.29	&3.74	&3.36	&3.94	&3.92\\
179	&					&$^1A_2(n,\pis)$			&V	&86.9	&4.37	&\Y	&4.83	&4.29	&3.87	&4.49	&4.48\\
180	&					&$^1A_1(\pi,\pis)$			&V	&85.8	&5.26	&\Y	&5.12	&5.34	&4.87	&5.30	&5.25\\
181	&					&$^1A_2(n,\pis)$			&V	&86.2	&5.72	&\Y	&6.26	&5.73	&5.19	&5.93	&5.89\\
182	&					&$^1B_2(n,3s)$				&R	&88.5	&6.17	&\Y	&5.99	&6.18	&5.90	&6.28	&6.27\\
183	&					&$^1B_1(n,\pis)$			&V	&87.0	&6.37	&\Y	&7.16	&6.50	&5.94	&6.72	&6.67\\
184	&					&$^1B_2(\pi,\pis)$			&V	&90.6	&6.75	&\Y	&7.54	&7.26	&6.82	&7.25	&7.17\\
185	&					&$^3B_1(n,\pis)$			&V	&97.1	&3.19	&\Y	&3.60	&3.08	&2.72	&3.29	&3.28\\
186	&					&$^3A_2(n,\pis)$			&V	&96.1	&4.11	&\Y	&4.49	&4.01	&3.59	&4.20	&4.18\\
187	&					&$^3B_2(\pi,\pis)$			&V	&98.5	&4.34	&\N	&3.93	&4.44	&4.13	&4.30	&4.24\\
188	&					&$^3A_1(\pi,\pis)$			&V	&97.3	&4.82	&\Y	&4.93	&4.87	&4.48	&4.89	&4.83\\
189	&Pyridine			&$^1B_1(n,\pis)$			&V	&88.4	&4.95	&\Y	&5.43	&5.15	&4.81	&5.18	&5.13\\
190	&					&$^1B_2(\pi,\pis)$			&V	&86.5	&5.14	&\Y	&5.03	&5.18	&4.76	&5.15	&5.09\\
191	&					&$^1A_2(n,\pis)$			&V	&87.9	&5.40	&\Y	&6.30	&5.46	&5.03	&5.63	&5.59\\
192	&					&$^1A_1(\pi,\pis)$			&V	&92.1	&6.62	&\Y	&7.90	&6.92	&6.27	&7.04	&6.93\\
193	&					&$^1A_1(n,3s)$				&R	&89.7	&6.76	&\Y	&6.40	&6.90	&6.67	&6.97	&6.96\\
194	&					&$^1A_2(\pi,3s)$			&R	&93.2	&6.82	&\Y	&6.60	&7.08	&6.87	&6.88	&6.80\\
195	&					&$^1B_1(\pi,3p)$			&R	&93.6	&7.38	&\Y	&7.12	&7.70	&7.51	&7.48	&7.40\\
196	&					&$^1A_1(\pi,\pis)$			&V	&90.5	&7.39	&\Y	&9.49	&7.66	&6.63	&7.87	&7.70\\
197	&					&$^1B_2(\pi,\pis)$			&V	&90.0	&7.40	&\N	&7.45	&7.92	&7.67	&7.80	&7.73\\
198	&					&$^3A_1(\pi,\pis)$			&V	&98.5	&4.30	&\Y	&3.98	&4.40	&4.06	&4.29	&4.22\\
199	&					&$^3B_1(n,\pis)$			&V	&97.0	&4.46	&\Y	&4.65	&4.48	&4.21	&4.57	&4.55\\
200	&					&$^3B_2(\pi,\pis)$			&V	&97.3	&4.79	&\Y	&4.83	&4.86	&4.53	&4.81	&4.74\\
201	&					&$^3A_1(\pi,\pis)$			&V	&97.1	&5.04	&\Y	&5.11	&5.09	&4.63	&5.09	&5.02\\
202	&					&$^3A_2(n,\pis)$			&V	&95.8	&5.36	&\Y	&5.94	&5.33	&4.96	&5.53	&5.51\\
203	&					&$^3B_2(\pi,\pis)$			&V	&97.7	&6.24	&\Y	&6.93	&6.40	&5.99	&6.43	&6.35\\
204	&Pyrimidine			&$^1B_1(n,\pis)$			&V	&88.6	&4.44	&\Y	&4.85	&4.44	&4.07	&4.58	&4.55\\
205	&					&$^1A_2(n,\pis)$			&V	&88.5	&4.85	&\Y	&5.52	&4.80	&4.36	&5.02	&5.00\\
206	&					&$^1B_2(\pi,\pis)$			&V	&86.3	&5.38	&\Y	&5.28	&5.42	&4.98	&5.41	&5.36\\
207	&					&$^1A_2(n,\pis)$			&V	&86.7	&5.92	&\Y	&6.70	&5.92	&5.32	&6.16	&6.10\\
208	&					&$^1B_1(n,\pis)$			&V	&86.7	&6.26	&\Y	&7.20	&6.31	&5.65	&6.58	&6.53\\
209	&					&$^1B_2(n,3s)$				&R	&90.3	&6.70	&\Y	&6.86	&6.85	&6.50	&6.89	&6.86\\
210	&					&$^1A_1(\pi,\pis)$			&V	&91.5	&6.88	&\Y	&7.69	&7.31	&6.94	&7.29	&7.22\\
211	&					&$^3B_1(n,\pis)$			&V	&96.8	&4.09	&\Y	&4.45	&4.05	&3.67	&4.20	&4.18\\
212	&					&$^3A_1(\pi,\pis)$			&V	&98.3	&4.51	&\N	&4.22	&4.57	&4.25	&4.51	&4.44\\
213	&					&$^3A_2(n,\pis)$			&V	&96.5	&4.66	&\Y	&5.20	&4.63	&4.16	&4.81	&4.78\\
214	&					&$^3B_2(\pi,\pis)$			&V	&97.4	&4.96	&\Y	&5.10	&5.01	&4.60	&5.03	&4.97\\
215	&Pyrrole			&$^1A_2(\pi,3s)$			&R	&92.9	&5.24	&\Y	&4.49	&5.44	&5.23	&5.28	&5.23\\
216	&					&$^1B_1(\pi,3p)$			&R	&92.4	&6.00	&\Y	&5.22	&6.26	&6.07	&6.08	&6.02\\
217	&					&$^1A_2(\pi,3p)$			&R	&93.0	&6.00	&\Y	&4.89	&6.16	&6.02	&6.01	&5.97\\
218	&					&$^1B_2(\pi,\pis)$			&V	&92.5	&6.26	&\Y	&7.73	&6.62	&6.36	&6.45	&6.38\\
219	&					&$^1A_1(\pi,\pis)$			&V	&86.3	&6.30	&\Y	&6.47	&6.41	&5.84	&6.43	&6.34\\
220	&					&$^1B_2(\pi,3p)$			&R	&92.6	&6.83	&\Y	&5.82	&6.75	&6.11	&6.92	&6.82\\
221	&					&$^3B_2(\pi,\pis)$			&V	&98.3	&4.51	&\Y	&4.24	&4.57	&4.30	&4.49	&4.44\\
222	&					&$^3A_2(\pi,3s)$			&R	&97.6	&5.21	&\Y	&4.47	&5.41	&5.21	&5.26	&5.20\\
223	&					&$^3A_1(\pi,\pis)$			&V	&97.8	&5.45	&\Y	&5.52	&5.50	&5.04	&5.49	&5.40\\
224	&					&$^3B_1(\pi,3p)$			&R	&97.4	&5.91	&\Y	&5.18	&6.22	&6.03	&6.04	&5.98\\
225	&Streptocyanine-C1	&$^1B_2(\pi,\pis)$			&V	&88.7	&7.13	&\Y	&7.82	&7.17	&6.76	&7.28	&7.21\\
226	&					&$^3B_2(\pi,\pis)$			&V	&98.3	&5.52	&\Y	&5.86	&5.49	&5.22	&5.54	&5.49\\
227	&Tetrazine			&$^1B_{3u}(n,\pis)$			&V	&89.8	&2.47	&\Y	&2.99	&2.31	&1.91	&2.54	&2.53\\
228	&					&$^1A_u(n,\pis)$			&V	&87.9	&3.69	&\Y	&4.37	&3.49	&3.00	&3.77	&3.78\\
229	&					&$^1A_g(\text{double})$		&V	&0.7	&4.61	&\N	&5.42	&4.69	&4.48	&4.85	&4.87\\
230	&					&$^1B_{1g}(n,\pis)$			&V	&83.1	&4.93	&\Y	&5.41	&4.83	&4.33	&5.02	&5.00\\
231	&					&$^1B_{2u}(\pi,\pis)$		&V	&85.4	&5.21	&\Y	&5.04	&5.31	&4.84	&5.26	&5.23\\
232	&					&$^1B_{2g}(n,\pis)$			&V	&81.7	&5.45	&\Y	&5.43	&5.38	&4.90	&5.42	&5.38\\
233	&					&$^1A_u(n,\pis)$			&V	&87.7	&5.53	&\Y	&6.37	&5.51	&4.92	&5.80	&5.80\\
234	&					&$^1B_{3g}(\text{double})$	&V	&0.7	&6.15	&\N	&6.59	&5.85	&5.22	&6.20	&6.22\\
235	&					&$^1B_{2g}(n,\pis)$			&V	&80.2	&6.12	&\Y	&6.79	&5.96	&5.18	&6.27	&6.28\\
236	&					&$^1B_{1g}(n,\pis)$			&V	&85.1	&6.91	&\Y	&7.18	&6.59	&5.89	&6.79	&6.72\\
237	&					&$^3B_{3u}(n,\pis)$			&V	&97.1	&1.85	&\Y	&2.38	&1.70	&1.31	&1.94	&1.93\\
238	&					&$^3A_u(n,\pis)$			&V	&96.3	&3.45	&\Y	&4.06	&3.26	&2.78	&3.52	&3.52\\
239	&					&$^3B_{1g}(n,\pis)$			&V	&97.0	&4.20	&\Y	&4.66	&4.10	&3.62	&4.32	&4.30\\
240	&					&$^1B_{1u}(\pi,\pis)$		&V	&98.5	&4.49	&\N	&3.90	&4.55	&4.29	&4.39	&4.34\\
241	&					&$^3B_{2u}(\pi,\pis)$		&V	&97.5	&4.52	&\Y	&4.68	&4.55	&4.20	&4.60	&4.55\\
242	&					&$^3B_{2g}(n,\pis)$			&V	&96.4	&5.04	&\Y	&5.17	&5.02	&4.53	&5.10	&5.07\\
243	&					&$^3A_u(n,\pis)$			&V	&96.6	&5.11	&\Y	&6.12	&5.07	&4.44	&5.41	&5.41\\
244	&					&$^3B_{3g}(\text{double})$	&V	&5.7	&5.51	&\N	&6.56	&5.39	&4.86	&5.83	&5.85\\
245	&					&$^3B_{1u}(\pi,\pis)$		&V	&96.6	&5.42	&\Y	&5.32	&5.46	&5.08	&5.44	&5.39\\
246	&Thioacetone		&$^1A_2(n,\pis)$			&V	&88.9	&2.53	&\Y	&2.72	&2.58	&2.33	&2.60	&2.53\\
247	&					&$^1B_2(n,3s)$				&R	&91.3	&5.56	&\Y	&4.80	&5.60	&5.48	&5.64	&5.61\\
248	&					&$^1A_1(\pi,\pis)$			&V	&90.6	&5.88	&\Y	&6.94	&6.42	&5.98	&6.40	&6.26\\
249	&					&$^1B_2(n,3p)$				&R	&92.4	&6.51	&\Y	&5.57	&6.51	&6.40	&6.53	&6.49\\
250	&					&$^1A_1(n,3p)$				&R	&91.6	&6.61	&\Y	&6.24	&6.66	&6.41	&6.59	&6.50\\
251	&					&$^3A_2(n,\pis)$			&V	&97.4	&2.33	&\Y	&2.52	&2.34	&2.09	&2.38	&2.31\\
252	&					&$^3A_1(\pi,\pis)$			&V	&98.7	&3.45	&\Y	&3.52	&3.48	&3.29	&3.48	&3.43\\
253	&Thiophene			&$^1A_1(\pi,\pis)$			&V	&87.6	&5.64	&\Y	&6.11	&5.84	&5.21	&5.89	&5.79\\
254	&					&$^1B_2(\pi,\pis)$			&V	&91.5	&5.98	&\Y	&6.94	&6.35	&5.89	&6.44	&6.35\\
255	&					&$^1A_2(\pi,3s)$			&R	&92.6	&6.14	&\Y	&5.70	&6.28	&6.07	&6.16	&6.10\\
256	&					&$^1B_1(\pi,3p)$			&R	&90.1	&6.14	&\Y	&6.02	&6.21	&5.90	&6.16	&6.10\\
257	&					&$^1A_2(\pi,3p)$			&R	&91.8	&6.21	&\Y	&6.05	&6.32	&5.98	&6.28	&6.21\\
258	&					&$^1B_1(\pi,3s)$			&R	&92.8	&6.49	&\Y	&5.78	&6.57	&6.28	&6.51	&6.44\\
259	&					&$^1B_2(\pi,3p)$			&R	&92.4	&7.29	&\Y	&6.80	&7.29	&7.03	&7.20	&7.13\\
260	&					&$^1A_1(\pi,\pis)$			&V	&86.5	&7.31	&\N	&8.29	&7.62	&6.85	&7.71	&7.56\\
261	&					&$^3B_2(\pi,\pis)$			&V	&98.2	&3.92	&\Y	&3.68	&3.98	&3.71	&3.90	&3.84\\
262	&					&$^3A_1(\pi,\pis)$			&V	&97.7	&4.76	&\Y	&4.97	&4.85	&4.39	&4.87	&4.79\\
263	&					&$^3B_1(\pi,3p)$			&R	&96.6	&5.93	&\Y	&5.86	&5.97	&5.64	&5.94	&5.88\\
264	&					&$^3A_2(\pi,3s)$			&R	&97.5	&6.08	&\Y	&5.65	&6.22	&6.01	&6.11	&6.04\\
265	&Thiopropynal		&$^1A''(n,\pis)$			&V	&87.5	&2.03	&\Y	&2.06	&2.05	&1.84	&2.05	&2.00\\
266	&					&$^3A''(n,\pis)$			&V	&97.2	&1.80	&\Y	&1.85	&1.81	&1.60	&1.84	&1.79\\
267	&Triazine			&$^1A_1''(n,\pis)$			&V	&88.3	&4.72	&\Y	&5.88	&4.62	&3.90	&5.00	&4.99\\
268	&					&$^1A_2''(n,\pis)$			&V	&88.3	&4.75	&\Y	&5.14	&4.77	&4.39	&4.90	&4.87\\
269	&					&$^1E''(n,\pis)$			&V	&88.3	&4.78	&\Y	&5.51	&4.76	&4.14	&5.01	&4.98\\
270	&					&$^1A_2'(\pi,\pis)$			&V	&85.7	&5.75	&\Y	&5.55	&5.76	&5.32	&5.75	&5.72\\
271	&					&$^1A_1'(\pi,\pis)$			&V	&90.4	&7.24	&\Y	&8.20	&7.43	&6.89	&7.50	&7.41\\
272	&					&$^1E'(n,3s)$				&R	&90.9	&7.32	&\Y	&7.40	&7.48	&7.15	&7.53	&7.49\\
273	&					&$^1E''(n,\pis)$			&V	&82.6	&7.78	&\Y	&8.26	&7.75	&7.04	&7.92	&7.90\\
274	&					&$^1E'(\pi,\pis)$			&V	&90.0	&7.94	&\Y	&10.03	&8.65	&7.70	&8.83	&8.72\\
275	&					&$^3A_2''(n,\pis)$			&V	&96.7	&4.33	&\Y	&4.74	&4.37	&3.99	&4.51	&4.49\\
276	&					&$^3E''(n,\pis)$			&V	&96.6	&4.51	&\Y	&5.14	&4.47	&3.88	&4.71	&4.68\\
277	&					&$^3A_1''(n,\pis)$			&V	&96.2	&4.73	&\Y	&5.88	&4.70	&3.94	&5.06	&5.04\\
278	&					&$^3A_1'(\pi,\pis)$			&V	&98.2	&4.85	&\Y	&4.46	&4.88	&4.55	&4.81	&4.75\\
279	&					&$^3E'(\pi,\pis)$			&V	&96.9	&5.59	&\Y	&5.57	&5.62	&5.20	&5.62	&5.57\\
280	&					&$^3A_2'(\pi,\pis)$			&V	&97.6	&6.62	&\Y	&7.70	&6.62	&6.12	&6.76	&6.68\\
\end{longtable*}

\begin{table*}
	\caption{Statistical quantities (in eV), considering the 265 ``safe'' TBEs (out of 280) as reference, for various multi-reference methods.
	Raw data are given in Table \ref{tab:BigTab}.}
	\label{tab:stat}
	\begin{ruledtabular}
        \begin{tabular}{lrrrrrrr}
			&CASSCF\fnm[1]	&CASPT2\fnm[1]	&CASPT2\fnm[1]	&CASPT3\fnm[1]	&CASPT3\fnm[1]	&SC-NEVPT2\fnm[2]	&PC-NEVPT2\fnm[2]\\
			&				&(IPEA)			&(NOIPEA)		&(IPEA)			&(NOIPEA)\\
			\hline
            MSE			&$0.12$		&$0.06$		&$-0.26$	&$0.10$		&$0.05$		&$0.13$		&$0.09$\\
            SDE			&$0.58$		&$0.14$		&$0.21$		&$0.14$		&$0.13$		&$0.14$		&$0.14$\\
            RMSE		&$0.61$		&$0.16$		&$0.33$		&$0.17$		&$0.14$		&$0.19$		&$0.17$\\
            MAE			&$0.47$		&$0.11$		&$0.27$		&$0.11$		&$0.09$		&$0.15$		&$0.13$\\
            Max($+$)	&$2.15$		&$0.71$		&$0.30$		&$0.93$		&$0.79$		&$0.65$		&$0.46$\\
            Max($-$)	&$-1.18$	&$-0.32$	&$-1.02$	&$-0.28$	&$-0.36$	&$-0.38$	&$-0.57$\\
		\end{tabular}
	\end{ruledtabular}
	\fnt[1]{Values from the present study.}
	\fnt[2]{Values taken from Ref.~\onlinecite{Sarkar_2022}.}
\end{table*}

\begin{table*}
	\caption{MAEs determined for several subsets of transitions and system sizes computed with various multi-reference methods.
	Count is the number of excited states considered in each subset.
	Raw data are given in Table \ref{tab:BigTab}.}
	\label{tab:stat_subset}
	\begin{ruledtabular}
		\begin{tabular}{lrrrrrrrr}
			Transitions		&	Count	&CASSCF\fnm[1]	&CASPT2\fnm[1]	&CASPT2\fnm[1]	&CASPT3\fnm[1]	&CASPT3\fnm[1]	&SC-NEVPT2\fnm[2]	&PC-NEVPT2\fnm[2]\\
							&			&				&(IPEA)			&(NOIPEA)		&(IPEA)			&(NOIPEA)\\
			\hline
 			Singlet			&174	&0.56	&0.14	&0.27	&0.14	&0.12	&0.16	&0.14\\
			Triplet			&110	&0.34	&0.07	&0.29	&0.07	&0.06	&0.13	&0.11\\
			Valence 		&208	&0.44	&0.11	&0.33	&0.13	&0.10	&0.15	&0.12\\
			Rydberg			&76		&0.55	&0.13	&0.13	&0.08	&0.07	&0.15	&0.15\\
			$n \to \pis$	&78		&0.44	&0.08	&0.44	&0.13	&0.10	&0.12	&0.10\\
			$\pi \to \pis$ 	&119	&0.45	&0.12	&0.27	&0.13	&0.10	&0.18	&0.14\\
			Double			&9		&0.46	&0.11	&0.22	&0.12	&0.09	&0.14	&0.13\\
			3 non-H atoms	&39		&0.38	&0.07	&0.21	&0.06	&0.05	&0.10	&0.08\\
			4 non-H	atoms	&94		&0.45	&0.11	&0.22	&0.12	&0.09	&0.14	&0.13\\
			5-6	non-H atoms	&151	&0.51	&0.12	&0.33	&0.13	&0.11	&0.17	&0.15\\	
		\end{tabular}
	\end{ruledtabular}
	\fnt[1]{Values from the present study.}
	\fnt[2]{Values taken from Ref.~\onlinecite{Sarkar_2022}.}
\end{table*}

From the different statistical quantities reported in Table \ref{tab:stat}, one can highlight the following trends.
\alert{First, as expected, CASSCF returns a large MAE of \SI{0.47}{\eV} and a relative small MSE of \SI{0.12}{\eV}, with a better accuracy for triplet states than for singlet states.}
\alert{Second}, as previously reported, \cite{Werner_1996,Grabarek_2016} CASPT3 vertical excitation energies are much less sensitive to the IPEA shift, which drastically alters the accuracy of CASPT2: the mean absolute deviation between the CASPT2(NOIPEA) and CASPT2(IPEA) data is \SI{0.329}{\eV} while it is only \SI{0.051}{\eV} between CASPT3(NOIPEA) and CASPT3(IPEA).
Consequently, the MAEs of CASPT3(IPEA) and CASPT3(NOIPEA) are amazingly close (\SI{0.11}{} and \SI{0.09}{\eV}), while the MAEs of CASPT2(IPEA) and CASPT2(NOIPEA) are remarkably different (\SI{0.11}{} and \SI{0.27}{\eV}).
Likewise, the MSEs of CASPT2(IPEA) and CASPT2(NOIPEA), \SI{0.06}{} and \SI{-0.26}{\eV}, clearly evidence the well-known global underestimation of the CASPT2(NOIPEA) excitation energies in molecular systems when large basis sets are used.  
For CASPT3, the MSE with IPEA shift is only slightly larger without IPEA (\SI{0.10}{} and \SI{0.05}{\eV}, respectively).
Importantly, CASPT3 performs slightly better without IPEA shift, which is a nice outcome that holds for each group of transitions and system size (see the MAEs in  Table \ref{tab:stat_subset}).

Second, CASPT3 (with or without IPEA) has a similar accuracy as CASPT2(IPEA).
Again, this observation stands for each subset of excitations and irrespectively of the system size (see Table \ref{tab:stat_subset}).
Because the relative size of the active space naturally decreases as the number of electrons and orbitals get larger, we observe that the MAEs of each subset increase with the size of the molecules.
Note that combining CASPT2 and CASPT3 via an hybrid protocol such as CASPT2.5, as proposed by Zhang and Truhlar in the context of spin splitting energies in transition metals, \cite{Zhang_2020} is not beneficial in the present situation.

It is worth mentioning that CASPT3(NOIPEA) yields MAEs for each subset that is almost systematically below \SI{0.1}{\eV}, except for the singlet subset which contains some states showing large (positive) deviations at both the CASPT2 and CASPT3 levels.
This is most notably the case for the $^1 B_u(\pi,\pis)$ state of butadiene, the $^1B_2(\pi,\pis)$ state of cyclopentadiene, the $^1A_1(\pi,\pis)$ state of cyclopropenone, the second $^1B_{1u}(\pi,\pis)$ state of pyrazine, the $^1B_2(\pi,\pis)$ state of pyridazine, and the $^1E'(\pi,\pis)$ state of triazine, for which both CASPT2(IPEA) and CASPT3(NOIPEA) overestimate the corresponding vertical transition energies by at least \SI{0.4}{\eV} with respect to the TBEs. 
This can be tracked down to the relatively small active spaces that we have considered here and, more precisely, to the lack of direct $\sig$-$\pi$ coupling in the active space that is known to be important in ionic states, for example. \cite{Davidson_1996,Borden_1996,Boggio-Pasqua_2004,Angeli_2009,Garniron_2018,Tran_2019,BenAmor_2020} 
For this family of states, it is particularly important to describe the dynamic response of the $\sig$-electron framework to the field of the $\pi$-electron system, a phenomenon known as dynamic $\sig$ polarization \alert{(that should not be confused with so-called left-right polarization \cite{Mok_1996})}. 
Because the dynamic $\sig$ polarization is generally more important for the ionic excited state than for the ground state, its contribution is expected to lower the vertical transition energy. 
Furthermore, this part of the dynamic $\sig$-$\pi$ correlation needs to be included at the orbital optimization stage, otherwise the orbitals become too diffuse, resulting in artificial valence-Rydberg mixing which cannot be disentangled using non-degenerate perturbation theory such as the version of CASPT2 and CASPT3 considered here. \cite{Angeli_2009}

As an illustration of this problematic, we have chosen to address the specific case of the second $^1B_{1u}(\pi,\pis)$ state of pyrazine, which is known to exhibit a strong ionic character. \cite{Fulscher_1994}
As shown in Table \ref{tab:BigTab} (\#171), the TBE for the vertical transition energy to this state is \SI{7.98}{\eV}. 
CASPT2(IPEA) and CASPT3(NOIPEA) locate this state at \SI{8.59}{} and \SI{8.57}{\eV}, respectively, providing a large overestimation of \SI{0.6}{\eV}. 
This state was computed using a reference CASSCF wave function averaged over four states [the ground state, two valence $B_{1u}(\pi,\pis)$ states and one Rydberg $B_{1u}(\pi,3p_x)$ state] with an active space comprising the $\pi$ valence and three $3p_x$ orbitals. 
(The $3p_x$ orbitals were included to recover part of the radial correlation.)
However, this strategy leads to a valence-Rydberg mixing due to the fact that the dynamic correlation is not sufficiently described at the CASSCF level. 
The ionic $B_{1u}(\pi,\pis)$ state lies \SI{9.65}{\eV} vertically above the ground state, while the Rydberg $B_{1u}(\pi,3p_x)$ state is \SI{0.2}{\eV} below at the CASSCF level. 
For this reason, the two states are mixed and both CASPT2 and CASPT3 fail to predict accurate transition energies for the ionic state. 
The Rydberg character of the ionic $B_{1u}(\pi,\pis)$ state is evident from the inspection of the CASSCF wave function and also from its value of $\expval*{x^2}$, which measures the spatial extent of the wave function out of the molecular plane (hence characteristic of the size of the $\pi$ orbitals in the considered state). 
The $\expval*{x^2}$ value is \SI{31.9}{\bohr^2} for the ionic $B_{1u}(\pi,\pis)$ state and \SI{51.1}{\bohr^2} for the $B_{1u}(\pi,3p_x)$ Rydberg state, while it is only \SI{26.6}{\bohr^2} for the ground state. 

To remove the artificial valence-Rydberg mixing in the reference CASSCF wave function, we included the dynamic $\sig$ polarization at the orbital optimization stage using a restricted active space self-consistent field (RASSCF) approach. \cite{Olsen_1988} 
We selected the bonding $\sigCC$ and $\sigCN$ orbitals in the RAS1 partition and the corresponding anti-bonding $\sigsCC$ and $\sigsCN$ orbitals in RAS3 allowing a single hole in RAS1 and a single electron in RAS3. 
The six valence $\pi$ orbitals were kept in RAS2 (full CI space). 
In this way, the contraction of the $\pi$ orbitals as a result of the dynamic $\sig$ polarization is ensured and the interference of the Rydberg state is removed allowing to compute the two valence $B_{1u}(\pi,\pis)$ states without including the Rydberg $B_{1u}(\pi,3p_x)$ state in the state-averaging procedure. 
The $\expval*{x^2}$ value associated with the ionic $B_{1u}(\pi,\pis)$ state is reduced to \SI{26.9}{\bohr^2}, providing a spatial extent similar to that of the ground state ($\expval*{x^2} = \SI{27.0}{\bohr^2}$ at the RASSCF level). 
Using the RASSCF orbitals to perform the CASPT2(IPEA) and CASPT3(NOIPEA) calculations using a CAS-CI(6,6) reference, we obtain vertical transition energies of \SI{7.92}{} and \SI{8.10}{\eV}, respectively. 
The agreement with the TBE is now within the expected accuracy of the method with an error of about \SI{0.1}{\eV}. 
To be complete the vertical transition energy to the first $B_{1u}(\pi,\pis)$ state, which also possesses a significant ionic character, is improved too with respect to the TBE at \SI{6.88}{\eV} with transition energies of \SI{6.83}{\eV} and \SI{6.87}{\eV} at the CASPT2(IPEA) and CASPT3(NOIPEA) levels, respectively. 
This represents a significant improvement compared to the \SI{7.14}{} and \SI{7.12}{\eV} values obtained at the same level of theory but using a reference SA4-CASSCF wave function. 
We thus believe that the difficult cases listed above can be handled more rigorously provided that more suitable active spaces are used to describe the reference (zeroth-order) wave function prior to the CASPT2/CASPT3 calculations.

Comparatively, Liang \textit{et al.} have recently shown, for a larger set of transitions, that time-dependent density-functional theory with the best exchange-correlation functionals yield RMSEs of the order of \SI{0.3}{\eV}, \cite{Liang_2022} outperforming (more expensive) wave function methods like CIS(D). \cite{Head-Gordon_1994,Head-Gordon_1995} 
The accuracy of CASPT2(IPEA) and CASPT3 is clearly a step beyond but at a much larger computational cost.
Although, these two methods do not beat the approximate third-order coupled-cluster method, CC3, \cite{Christiansen_1995b,Koch_1997} for transitions with a dominant single excitation character (for which CC3 returns a MAEs below the chemical accuracy threshold of \SI{0.043}{\eV}) \cite{Veril_2021} it has the undisputable advantage to describe with the same accuracy both single and double excitations.
This feature is crucial in the description of some photochemistry mechanisms. \cite{Boggio-Pasqua_2007}
 
\begin{table}
	\caption{Wall times (in seconds) for the computation of the (ground-state) second-order (PT2) and third-order (PT3) energies of benzene.
	Calculations have been performed in the frozen-core approximation and with the aug-cc-pVTZ basis set on an Intel Xeon node (see main text).}
	\label{tab:timings}
    \begin{ruledtabular}
        \begin{tabular}{cccccc}
        	Active	&\# CAS 	&\# contracted	&\# uncontracted 	&$t_\text{PT2}$ 	&$t_\text{PT3}$\\
        	space	&det.		&config. 		&config.			&	&	\\
        	\hline
	(6e,6o)	&104 	&$4.50 \times 10^6$	&$2.29 \times 10^8$	&11	&60\\
	(6e,7o)	&165 	&$7.27 \times 10^6$	&$3.69 \times 10^8$	&39	&249\\
	(6e,8o)	&412 	&$1.59 \times 10^7$	&$8.98 \times 10^8$	&159&1333\\
	(6e,9o)	&1800	&$3.96 \times 10^7$	&$3.53 \times 10^9$	&578&6332\\
        \end{tabular}
    \end{ruledtabular}
\end{table}

\begin{figure}
	\includegraphics[width=\linewidth]{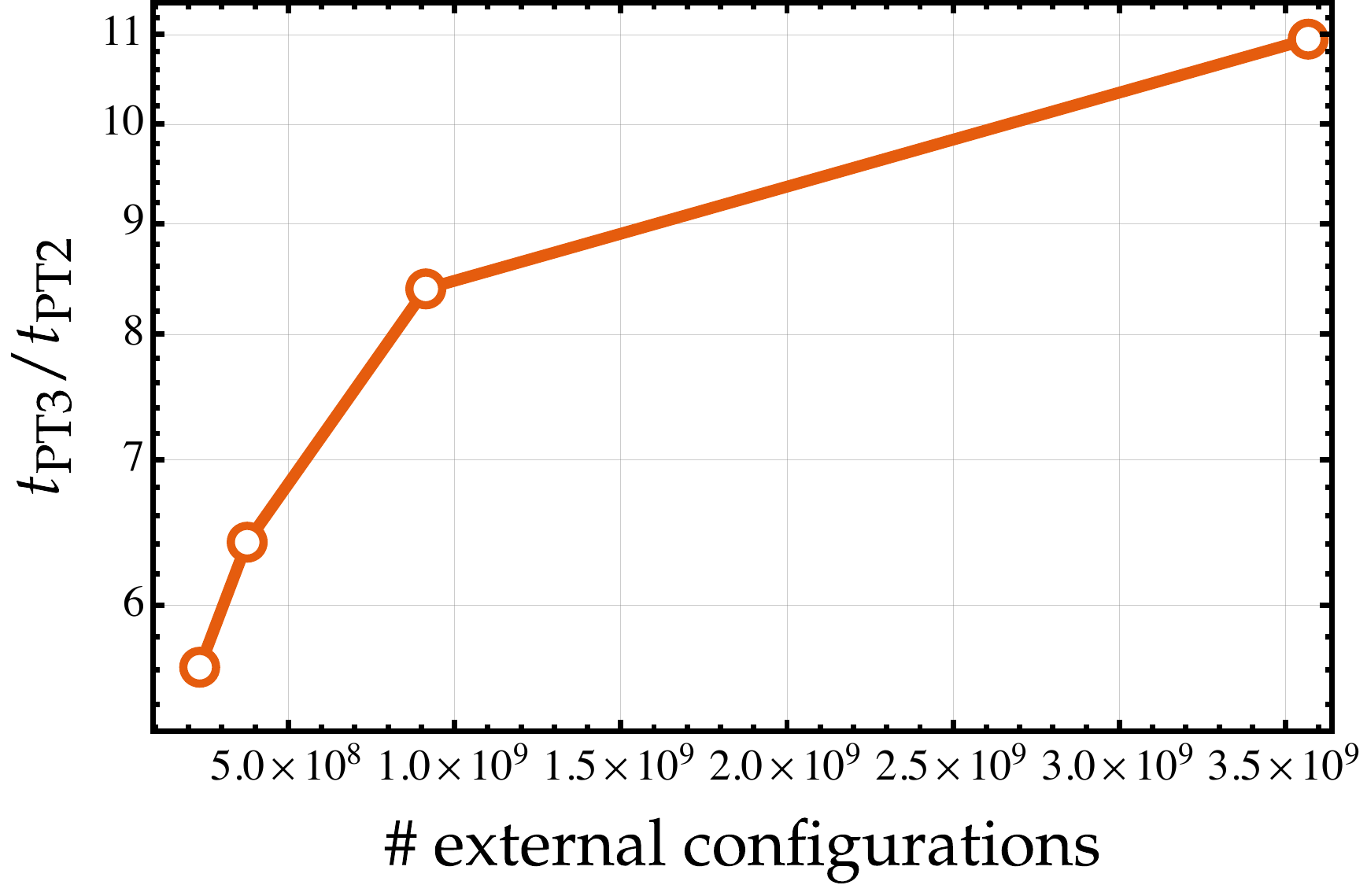}
	\caption{Ratio of the wall times associated with the computation of the third- and second-order energies as a function of the total number of contracted and uncontracted external configurations for benzene (see Table \ref{tab:timings} for raw data).
		Calculations have been performed in the frozen-core approximation and with the aug-cc-pVTZ basis set on an Intel Xeon node (see main text).}
	\label{fig:timings}
\end{figure}

Table \ref{tab:timings} reports the evolution of the wall times associated with the computation of the second- and third-order  energies in benzene with the aug-cc-pVTZ basis and the frozen-core approximation (42 electrons and 414 basis functions) for increasingly large active spaces. 
All these calculations have been performed on a single core of an Intel Xeon E5-2670 2.6 Ghz.
It is particularly instructive to study the wall time ratio as the number of (contracted and uncontracted) external configurations grows (see Fig.~\ref{fig:timings}).
Overall, the PT3 step takes between 5 and 10 times longer than the PT2 step for the active spaces that we have considered here, and remains thus typically affordable for these kinds of calculations.

\section{Conclusion}
\label{sec:ccl}
In the present study, we have benchmarked, using 280 highly-accurate electronic transitions extracted from the QUEST database, \cite{Veril_2021} the third-order multi-reference perturbation theory method, CASPT3, by computing vertical excitation energies with and without IPEA shift.
The two principal take-home messages of this study are that: 
(i) CASPT3 transition energies are almost independent of the IPEA shift; 
(ii) CASPT2(IPEA) and CASPT3 have a very similar accuracy.
These global trends are also true for specific sets of excitations and various system sizes.
Therefore, if one can afford the extra computation cost associated with the third-order energy (which is only several times more than its second-order counterpart), one can eschew the delicate choice of the IPEA value in CASPT2, and rely solely on the CASPT3(NOIPEA) excitation energies.
\alert{Of course, it is worth stressing that the present conclusions are only valid for small- and medium-sized molecules and that the present study does not cover transition metal complexes.}

\section*{Supplementary Material}
\label{sec:supmat}
Included in the {\SupMat} are the error distributions obtained for CASPT2 and CASPT3 with and without IPEA shift for various subsets of transitions, as well as the description and specification of the active space for each molecule.

\begin{acknowledgements}
This work was performed using HPC resources from CALMIP (Toulouse) under allocation 2021-18005.
DJ is indebted to the CCIPL computational center installed in Nantes for a generous allocation of computational time.
PFL thanks the European Research Council (ERC) under the European Union's Horizon 2020 research and innovation programme (Grant agreement No.~863481) for funding.
\end{acknowledgements}

\section*{Data availability statement}
The data that supports the findings of this study are available within the article and its supplementary material.

%


\begin{thebibliography}{131}%
\makeatletter
\providecommand \@ifxundefined [1]{%
 \@ifx{#1\undefined}
}%
\providecommand \@ifnum [1]{%
 \ifnum #1\expandafter \@firstoftwo
 \else \expandafter \@secondoftwo
 \fi
}%
\providecommand \@ifx [1]{%
 \ifx #1\expandafter \@firstoftwo
 \else \expandafter \@secondoftwo
 \fi
}%
\providecommand \natexlab [1]{#1}%
\providecommand \enquote  [1]{``#1''}%
\providecommand \bibnamefont  [1]{#1}%
\providecommand \bibfnamefont [1]{#1}%
\providecommand \citenamefont [1]{#1}%
\providecommand \href@noop [0]{\@secondoftwo}%
\providecommand \href [0]{\begingroup \@sanitize@url \@href}%
\providecommand \@href[1]{\@@startlink{#1}\@@href}%
\providecommand \@@href[1]{\endgroup#1\@@endlink}%
\providecommand \@sanitize@url [0]{\catcode `\\12\catcode `\$12\catcode
  `\&12\catcode `\#12\catcode `\^12\catcode `\_12\catcode `\%12\relax}%
\providecommand \@@startlink[1]{}%
\providecommand \@@endlink[0]{}%
\providecommand \url  [0]{\begingroup\@sanitize@url \@url }%
\providecommand \@url [1]{\endgroup\@href {#1}{\urlprefix }}%
\providecommand \urlprefix  [0]{URL }%
\providecommand \Eprint [0]{\href }%
\providecommand \doibase [0]{http://dx.doi.org/}%
\providecommand \selectlanguage [0]{\@gobble}%
\providecommand \bibinfo  [0]{\@secondoftwo}%
\providecommand \bibfield  [0]{\@secondoftwo}%
\providecommand \translation [1]{[#1]}%
\providecommand \BibitemOpen [0]{}%
\providecommand \bibitemStop [0]{}%
\providecommand \bibitemNoStop [0]{.\EOS\space}%
\providecommand \EOS [0]{\spacefactor3000\relax}%
\providecommand \BibitemShut  [1]{\csname bibitem#1\endcsname}%
\let\auto@bib@innerbib\@empty
\bibitem [{\citenamefont {Olsen}\ \emph {et~al.}(1996)\citenamefont {Olsen},
  \citenamefont {Christiansen}, \citenamefont {Koch},\ and\ \citenamefont
  {J{\o}rgensen}}]{Olsen_1996}%
  \BibitemOpen
  \bibfield  {author} {\bibinfo {author} {\bibfnamefont {J.}~\bibnamefont
  {Olsen}}, \bibinfo {author} {\bibfnamefont {O.}~\bibnamefont {Christiansen}},
  \bibinfo {author} {\bibfnamefont {H.}~\bibnamefont {Koch}}, \ and\ \bibinfo
  {author} {\bibfnamefont {P.}~\bibnamefont {J{\o}rgensen}},\ }\href {\doibase
  10.1063/1.472352} {\bibfield  {journal} {\bibinfo  {journal} {J. Chem.
  Phys.}\ }\textbf {\bibinfo {volume} {105}},\ \bibinfo {pages} {5082}
  (\bibinfo {year} {1996})}\BibitemShut {NoStop}%
\bibitem [{\citenamefont {Christiansen}\ \emph {et~al.}(1996)\citenamefont
  {Christiansen}, \citenamefont {Olsen}, \citenamefont {J{\o}rgensen},
  \citenamefont {Koch},\ and\ \citenamefont {Malmqvist}}]{Christiansen_1996}%
  \BibitemOpen
  \bibfield  {author} {\bibinfo {author} {\bibfnamefont {O.}~\bibnamefont
  {Christiansen}}, \bibinfo {author} {\bibfnamefont {J.}~\bibnamefont {Olsen}},
  \bibinfo {author} {\bibfnamefont {P.}~\bibnamefont {J{\o}rgensen}}, \bibinfo
  {author} {\bibfnamefont {H.}~\bibnamefont {Koch}}, \ and\ \bibinfo {author}
  {\bibfnamefont {P.-{\AA}.}\ \bibnamefont {Malmqvist}},\ }\href {\doibase
  10.1016/0009-2614(96)00974-8} {\bibfield  {journal} {\bibinfo  {journal}
  {Chem. Phys. Lett.}\ }\textbf {\bibinfo {volume} {261}},\ \bibinfo {pages}
  {369} (\bibinfo {year} {1996})}\BibitemShut {NoStop}%
\bibitem [{\citenamefont {Cremer}\ and\ \citenamefont
  {He}(1996)}]{Cremer_1996}%
  \BibitemOpen
  \bibfield  {author} {\bibinfo {author} {\bibfnamefont {D.}~\bibnamefont
  {Cremer}}\ and\ \bibinfo {author} {\bibfnamefont {Z.}~\bibnamefont {He}},\
  }\href {\doibase 10.1021/jp952815d} {\bibfield  {journal} {\bibinfo
  {journal} {J. Phys. Chem.}\ }\textbf {\bibinfo {volume} {100}},\ \bibinfo
  {pages} {6173} (\bibinfo {year} {1996})}\BibitemShut {NoStop}%
\bibitem [{\citenamefont {Olsen}\ \emph {et~al.}(2000)\citenamefont {Olsen},
  \citenamefont {J{\o}rgensen}, \citenamefont {Helgaker},\ and\ \citenamefont
  {Christiansen}}]{Olsen_2000}%
  \BibitemOpen
  \bibfield  {author} {\bibinfo {author} {\bibfnamefont {J.}~\bibnamefont
  {Olsen}}, \bibinfo {author} {\bibfnamefont {P.}~\bibnamefont {J{\o}rgensen}},
  \bibinfo {author} {\bibfnamefont {T.}~\bibnamefont {Helgaker}}, \ and\
  \bibinfo {author} {\bibfnamefont {O.}~\bibnamefont {Christiansen}},\ }\href
  {\doibase 10.1063/1.481611} {\bibfield  {journal} {\bibinfo  {journal} {J.
  Chem. Phys.}\ }\textbf {\bibinfo {volume} {112}},\ \bibinfo {pages} {9736}
  (\bibinfo {year} {2000})}\BibitemShut {NoStop}%
\bibitem [{\citenamefont {Olsen}\ and\ \citenamefont
  {J{\o}rgensen}(2019)}]{Olsen_2019}%
  \BibitemOpen
  \bibfield  {author} {\bibinfo {author} {\bibfnamefont {J.}~\bibnamefont
  {Olsen}}\ and\ \bibinfo {author} {\bibfnamefont {P.}~\bibnamefont
  {J{\o}rgensen}},\ }\href {\doibase 10.1063/1.5110554} {\bibfield  {journal}
  {\bibinfo  {journal} {J. Chem. Phys.}\ }\textbf {\bibinfo {volume} {151}},\
  \bibinfo {pages} {084108} (\bibinfo {year} {2019})}\BibitemShut {NoStop}%
\bibitem [{\citenamefont {Stillinger}(2000)}]{Stillinger_2000}%
  \BibitemOpen
  \bibfield  {author} {\bibinfo {author} {\bibfnamefont {F.~H.}\ \bibnamefont
  {Stillinger}},\ }\href {\doibase 10.1063/1.481608} {\bibfield  {journal}
  {\bibinfo  {journal} {J. Chem. Phys.}\ }\textbf {\bibinfo {volume} {112}},\
  \bibinfo {pages} {9711} (\bibinfo {year} {2000})}\BibitemShut {NoStop}%
\bibitem [{\citenamefont {Goodson}(2000{\natexlab{a}})}]{Goodson_2000a}%
  \BibitemOpen
  \bibfield  {author} {\bibinfo {author} {\bibfnamefont {D.~Z.}\ \bibnamefont
  {Goodson}},\ }\href {\doibase 10.1063/1.481044} {\bibfield  {journal}
  {\bibinfo  {journal} {J. Chem. Phys.}\ }\textbf {\bibinfo {volume} {112}},\
  \bibinfo {pages} {4901} (\bibinfo {year} {2000}{\natexlab{a}})}\BibitemShut
  {NoStop}%
\bibitem [{\citenamefont {Goodson}(2000{\natexlab{b}})}]{Goodson_2000b}%
  \BibitemOpen
  \bibfield  {author} {\bibinfo {author} {\bibfnamefont {D.~Z.}\ \bibnamefont
  {Goodson}},\ }\href {\doibase 10.1063/1.1318740} {\bibfield  {journal}
  {\bibinfo  {journal} {J. Chem. Phys.}\ }\textbf {\bibinfo {volume} {113}},\
  \bibinfo {pages} {6461} (\bibinfo {year} {2000}{\natexlab{b}})}\BibitemShut
  {NoStop}%
\bibitem [{\citenamefont {Goodson}\ and\ \citenamefont
  {Sergeev}(2004)}]{Goodson_2004}%
  \BibitemOpen
  \bibfield  {author} {\bibinfo {author} {\bibfnamefont {D.~Z.}\ \bibnamefont
  {Goodson}}\ and\ \bibinfo {author} {\bibfnamefont {A.~V.}\ \bibnamefont
  {Sergeev}},\ }in\ \href {\doibase 10.1016/S0065-3276(04)47011-7} {\emph
  {\bibinfo {booktitle} {Adv. Quantum Chem.}}},\ Vol.~\bibinfo {volume} {47}\
  (\bibinfo  {publisher} {Academic Press},\ \bibinfo {year} {2004})\ pp.\
  \bibinfo {pages} {193--208}\BibitemShut {NoStop}%
\bibitem [{\citenamefont {Sergeev}\ \emph {et~al.}(2005)\citenamefont
  {Sergeev}, \citenamefont {Goodson}, \citenamefont {Wheeler},\ and\
  \citenamefont {Allen}}]{Sergeev_2005}%
  \BibitemOpen
  \bibfield  {author} {\bibinfo {author} {\bibfnamefont {A.~V.}\ \bibnamefont
  {Sergeev}}, \bibinfo {author} {\bibfnamefont {D.~Z.}\ \bibnamefont
  {Goodson}}, \bibinfo {author} {\bibfnamefont {S.~E.}\ \bibnamefont
  {Wheeler}}, \ and\ \bibinfo {author} {\bibfnamefont {W.~D.}\ \bibnamefont
  {Allen}},\ }\href {\doibase 10.1063/1.1991854} {\bibfield  {journal}
  {\bibinfo  {journal} {J. Chem. Phys.}\ }\textbf {\bibinfo {volume} {123}},\
  \bibinfo {pages} {064105} (\bibinfo {year} {2005})}\BibitemShut {NoStop}%
\bibitem [{\citenamefont {Sergeev}\ and\ \citenamefont
  {Goodson}(2006)}]{Sergeev_2006}%
  \BibitemOpen
  \bibfield  {author} {\bibinfo {author} {\bibfnamefont {A.~V.}\ \bibnamefont
  {Sergeev}}\ and\ \bibinfo {author} {\bibfnamefont {D.~Z.}\ \bibnamefont
  {Goodson}},\ }\href {\doibase 10.1063/1.2173989} {\bibfield  {journal}
  {\bibinfo  {journal} {J. Chem. Phys.}\ }\textbf {\bibinfo {volume} {124}},\
  \bibinfo {pages} {094111} (\bibinfo {year} {2006})}\BibitemShut {NoStop}%
\bibitem [{\citenamefont {Goodson}(2011)}]{Goodson_2011}%
  \BibitemOpen
  \bibfield  {author} {\bibinfo {author} {\bibfnamefont {D.~Z.}\ \bibnamefont
  {Goodson}},\ }\href {\doibase 10.1002/wcms.92} {\bibfield  {journal}
  {\bibinfo  {journal} {{WIREs} Comput. Mol. Sci.}\ }\textbf {\bibinfo {volume}
  {2}},\ \bibinfo {pages} {743} (\bibinfo {year} {2011})}\BibitemShut {NoStop}%
\bibitem [{\citenamefont {M\o{}ller}\ and\ \citenamefont
  {Plesset}(1934)}]{Moller_1934}%
  \BibitemOpen
  \bibfield  {author} {\bibinfo {author} {\bibfnamefont {C.}~\bibnamefont
  {M\o{}ller}}\ and\ \bibinfo {author} {\bibfnamefont {M.~S.}\ \bibnamefont
  {Plesset}},\ }\href {\doibase 10.1103/PhysRev.46.618} {\bibfield  {journal}
  {\bibinfo  {journal} {Phys. Rev.}\ }\textbf {\bibinfo {volume} {46}},\
  \bibinfo {pages} {618} (\bibinfo {year} {1934})}\BibitemShut {NoStop}%
\bibitem [{\citenamefont {Laidig}, \citenamefont {Fitzgerald},\ and\
  \citenamefont {Bartlett}(1985)}]{Laidig_1985}%
  \BibitemOpen
  \bibfield  {author} {\bibinfo {author} {\bibfnamefont {W.~D.}\ \bibnamefont
  {Laidig}}, \bibinfo {author} {\bibfnamefont {G.}~\bibnamefont {Fitzgerald}},
  \ and\ \bibinfo {author} {\bibfnamefont {R.~J.}\ \bibnamefont {Bartlett}},\
  }\href {\doibase 10.1016/0009-2614(85)80934-9} {\bibfield  {journal}
  {\bibinfo  {journal} {Chem. Phys. Lett.}\ }\textbf {\bibinfo {volume}
  {113}},\ \bibinfo {pages} {151} (\bibinfo {year} {1985})}\BibitemShut
  {NoStop}%
\bibitem [{\citenamefont {Knowles}\ \emph {et~al.}(1985)\citenamefont
  {Knowles}, \citenamefont {Somasundram}, \citenamefont {Handy},\ and\
  \citenamefont {Hirao}}]{Knowles_1985}%
  \BibitemOpen
  \bibfield  {author} {\bibinfo {author} {\bibfnamefont {P.~J.}\ \bibnamefont
  {Knowles}}, \bibinfo {author} {\bibfnamefont {K.}~\bibnamefont
  {Somasundram}}, \bibinfo {author} {\bibfnamefont {N.~C.}\ \bibnamefont
  {Handy}}, \ and\ \bibinfo {author} {\bibfnamefont {K.}~\bibnamefont
  {Hirao}},\ }\href {\doibase 10.1016/0009-2614(85)85002-8} {\bibfield
  {journal} {\bibinfo  {journal} {Chem. Phys. Lett.}\ }\textbf {\bibinfo
  {volume} {113}},\ \bibinfo {pages} {8} (\bibinfo {year} {1985})}\BibitemShut
  {NoStop}%
\bibitem [{\citenamefont {Handy}, \citenamefont {Knowles},\ and\ \citenamefont
  {Somasundram}(1985)}]{Handy_1985}%
  \BibitemOpen
  \bibfield  {author} {\bibinfo {author} {\bibfnamefont {N.~C.}\ \bibnamefont
  {Handy}}, \bibinfo {author} {\bibfnamefont {P.~J.}\ \bibnamefont {Knowles}},
  \ and\ \bibinfo {author} {\bibfnamefont {K.}~\bibnamefont {Somasundram}},\
  }\href {\doibase 10.1007/BF00698753} {\bibfield  {journal} {\bibinfo
  {journal} {Theoret. Chim. Acta}\ }\textbf {\bibinfo {volume} {68}},\ \bibinfo
  {pages} {87} (\bibinfo {year} {1985})}\BibitemShut {NoStop}%
\bibitem [{\citenamefont {Gill}\ and\ \citenamefont {Radom}(1986)}]{Gill_1986}%
  \BibitemOpen
  \bibfield  {author} {\bibinfo {author} {\bibfnamefont {P.~M.~W.}\
  \bibnamefont {Gill}}\ and\ \bibinfo {author} {\bibfnamefont {L.}~\bibnamefont
  {Radom}},\ }\href {\doibase 10.1016/0009-2614(86)80686-8} {\bibfield
  {journal} {\bibinfo  {journal} {Chem. Phys. Lett.}\ }\textbf {\bibinfo
  {volume} {132}},\ \bibinfo {pages} {16} (\bibinfo {year} {1986})}\BibitemShut
  {NoStop}%
\bibitem [{\citenamefont {Laidig}, \citenamefont {Saxe},\ and\ \citenamefont
  {Bartlett}(1987)}]{Laidig_1987}%
  \BibitemOpen
  \bibfield  {author} {\bibinfo {author} {\bibfnamefont {W.~D.}\ \bibnamefont
  {Laidig}}, \bibinfo {author} {\bibfnamefont {P.}~\bibnamefont {Saxe}}, \ and\
  \bibinfo {author} {\bibfnamefont {R.~J.}\ \bibnamefont {Bartlett}},\ }\href
  {\doibase 10.1063/1.452291} {\bibfield  {journal} {\bibinfo  {journal} {J.
  Chem. Phys.}\ }\textbf {\bibinfo {volume} {86}},\ \bibinfo {pages} {887}
  (\bibinfo {year} {1987})}\BibitemShut {NoStop}%
\bibitem [{\citenamefont {Nobes}\ \emph {et~al.}(1987)\citenamefont {Nobes},
  \citenamefont {Pople}, \citenamefont {Radom}, \citenamefont {Handy},\ and\
  \citenamefont {Knowles}}]{Nobes_1987}%
  \BibitemOpen
  \bibfield  {author} {\bibinfo {author} {\bibfnamefont {R.~H.}\ \bibnamefont
  {Nobes}}, \bibinfo {author} {\bibfnamefont {J.~A.}\ \bibnamefont {Pople}},
  \bibinfo {author} {\bibfnamefont {L.}~\bibnamefont {Radom}}, \bibinfo
  {author} {\bibfnamefont {N.~C.}\ \bibnamefont {Handy}}, \ and\ \bibinfo
  {author} {\bibfnamefont {P.~J.}\ \bibnamefont {Knowles}},\ }\href {\doibase
  10.1016/0009-2614(87)80545-6} {\bibfield  {journal} {\bibinfo  {journal}
  {Chem. Phys. Lett.}\ }\textbf {\bibinfo {volume} {138}},\ \bibinfo {pages}
  {481} (\bibinfo {year} {1987})}\BibitemShut {NoStop}%
\bibitem [{\citenamefont {Gill}\ \emph
  {et~al.}(1988{\natexlab{a}})\citenamefont {Gill}, \citenamefont {Pople},
  \citenamefont {Radom},\ and\ \citenamefont {Nobes}}]{Gill_1988}%
  \BibitemOpen
  \bibfield  {author} {\bibinfo {author} {\bibfnamefont {P.~M.~W.}\
  \bibnamefont {Gill}}, \bibinfo {author} {\bibfnamefont {J.~A.}\ \bibnamefont
  {Pople}}, \bibinfo {author} {\bibfnamefont {L.}~\bibnamefont {Radom}}, \ and\
  \bibinfo {author} {\bibfnamefont {R.~H.}\ \bibnamefont {Nobes}},\ }\href
  {\doibase 10.1063/1.455312} {\bibfield  {journal} {\bibinfo  {journal} {J.
  Chem. Phys.}\ }\textbf {\bibinfo {volume} {89}},\ \bibinfo {pages} {7307}
  (\bibinfo {year} {1988}{\natexlab{a}})}\BibitemShut {NoStop}%
\bibitem [{\citenamefont {Gill}\ \emph
  {et~al.}(1988{\natexlab{b}})\citenamefont {Gill}, \citenamefont {Wong},
  \citenamefont {Nobes},\ and\ \citenamefont {Radom}}]{Gill_1988a}%
  \BibitemOpen
  \bibfield  {author} {\bibinfo {author} {\bibfnamefont {P.~M.~W.}\
  \bibnamefont {Gill}}, \bibinfo {author} {\bibfnamefont {M.~W.}\ \bibnamefont
  {Wong}}, \bibinfo {author} {\bibfnamefont {R.~H.}\ \bibnamefont {Nobes}}, \
  and\ \bibinfo {author} {\bibfnamefont {L.}~\bibnamefont {Radom}},\ }\href
  {\doibase 10.1016/0009-2614(88)80328-2} {\bibfield  {journal} {\bibinfo
  {journal} {Chem. Phys. Lett.}\ }\textbf {\bibinfo {volume} {148}},\ \bibinfo
  {pages} {541} (\bibinfo {year} {1988}{\natexlab{b}})}\BibitemShut {NoStop}%
\bibitem [{\citenamefont {Lepetit}, \citenamefont {P{\'e}lissier},\ and\
  \citenamefont {Malrieu}(1988)}]{Lepetit_1988}%
  \BibitemOpen
  \bibfield  {author} {\bibinfo {author} {\bibfnamefont {M.~B.}\ \bibnamefont
  {Lepetit}}, \bibinfo {author} {\bibfnamefont {M.}~\bibnamefont
  {P{\'e}lissier}}, \ and\ \bibinfo {author} {\bibfnamefont {J.~P.}\
  \bibnamefont {Malrieu}},\ }\href {\doibase 10.1063/1.455170} {\bibfield
  {journal} {\bibinfo  {journal} {J. Chem. Phys.}\ }\textbf {\bibinfo {volume}
  {89}},\ \bibinfo {pages} {998} (\bibinfo {year} {1988})}\BibitemShut
  {NoStop}%
\bibitem [{\citenamefont {Leininger}\ \emph {et~al.}(2000)\citenamefont
  {Leininger}, \citenamefont {Allen}, \citenamefont {Schaefer},\ and\
  \citenamefont {Sherrill}}]{Leininger_2000a}%
  \BibitemOpen
  \bibfield  {author} {\bibinfo {author} {\bibfnamefont {M.~L.}\ \bibnamefont
  {Leininger}}, \bibinfo {author} {\bibfnamefont {W.~D.}\ \bibnamefont
  {Allen}}, \bibinfo {author} {\bibfnamefont {H.~F.}\ \bibnamefont {Schaefer}},
  \ and\ \bibinfo {author} {\bibfnamefont {C.~D.}\ \bibnamefont {Sherrill}},\
  }\href {\doibase 10.1063/1.481764} {\bibfield  {journal} {\bibinfo  {journal}
  {J. Chem. Phys.}\ }\textbf {\bibinfo {volume} {112}},\ \bibinfo {pages}
  {9213} (\bibinfo {year} {2000})}\BibitemShut {NoStop}%
\bibitem [{\citenamefont {Malrieu}\ and\ \citenamefont
  {Angeli}(2013)}]{Malrieu_2003}%
  \BibitemOpen
  \bibfield  {author} {\bibinfo {author} {\bibfnamefont {J.-P.}\ \bibnamefont
  {Malrieu}}\ and\ \bibinfo {author} {\bibfnamefont {C.}~\bibnamefont
  {Angeli}},\ }\href {\doibase 10.1080/00268976.2013.788745} {\bibfield
  {journal} {\bibinfo  {journal} {Mol. Phys.}\ }\textbf {\bibinfo {volume}
  {111}},\ \bibinfo {pages} {1092} (\bibinfo {year} {2013})}\BibitemShut
  {NoStop}%
\bibitem [{\citenamefont {Damour}\ \emph {et~al.}(2021)\citenamefont {Damour},
  \citenamefont {V{\'e}ril}, \citenamefont {Kossoski}, \citenamefont
  {Caffarel}, \citenamefont {Jacquemin}, \citenamefont {Scemama},\ and\
  \citenamefont {Loos}}]{Damour_2021}%
  \BibitemOpen
  \bibfield  {author} {\bibinfo {author} {\bibfnamefont {Y.}~\bibnamefont
  {Damour}}, \bibinfo {author} {\bibfnamefont {M.}~\bibnamefont {V{\'e}ril}},
  \bibinfo {author} {\bibfnamefont {F.}~\bibnamefont {Kossoski}}, \bibinfo
  {author} {\bibfnamefont {M.}~\bibnamefont {Caffarel}}, \bibinfo {author}
  {\bibfnamefont {D.}~\bibnamefont {Jacquemin}}, \bibinfo {author}
  {\bibfnamefont {A.}~\bibnamefont {Scemama}}, \ and\ \bibinfo {author}
  {\bibfnamefont {P.-F.}\ \bibnamefont {Loos}},\ }\href {\doibase
  10.1063/5.0065314} {\bibfield  {journal} {\bibinfo  {journal} {J. Chem.
  Phys.}\ }\textbf {\bibinfo {volume} {155}},\ \bibinfo {pages} {134104}
  (\bibinfo {year} {2021})}\BibitemShut {NoStop}%
\bibitem [{\citenamefont {Marie}, \citenamefont {Burton},\ and\ \citenamefont
  {Loos}(2021)}]{Marie_2021a}%
  \BibitemOpen
  \bibfield  {author} {\bibinfo {author} {\bibfnamefont {A.}~\bibnamefont
  {Marie}}, \bibinfo {author} {\bibfnamefont {H.~G.~A.}\ \bibnamefont
  {Burton}}, \ and\ \bibinfo {author} {\bibfnamefont {P.-F.}\ \bibnamefont
  {Loos}},\ }\href {\doibase 10.1088/1361-648X/abe795} {\bibfield  {journal}
  {\bibinfo  {journal} {J. Phys.: Condens. Matter}\ }\textbf {\bibinfo {volume}
  {33}},\ \bibinfo {pages} {283001} (\bibinfo {year} {2021})}\BibitemShut
  {NoStop}%
\bibitem [{\citenamefont {Pito{\v n}{\'a}k}\ \emph {et~al.}(2009)\citenamefont
  {Pito{\v n}{\'a}k}, \citenamefont {Neogr{\'a}dy}, \citenamefont {{\v
  C}ern{\'y}}, \citenamefont {Grimme},\ and\ \citenamefont
  {Hobza}}]{Pitonak_2009}%
  \BibitemOpen
  \bibfield  {author} {\bibinfo {author} {\bibfnamefont {M.}~\bibnamefont
  {Pito{\v n}{\'a}k}}, \bibinfo {author} {\bibfnamefont {P.}~\bibnamefont
  {Neogr{\'a}dy}}, \bibinfo {author} {\bibfnamefont {J.}~\bibnamefont {{\v
  C}ern{\'y}}}, \bibinfo {author} {\bibfnamefont {S.}~\bibnamefont {Grimme}}, \
  and\ \bibinfo {author} {\bibfnamefont {P.}~\bibnamefont {Hobza}},\ }\href
  {\doibase 10.1002/cphc.200800718} {\bibfield  {journal} {\bibinfo  {journal}
  {ChemPhysChem}\ }\textbf {\bibinfo {volume} {10}},\ \bibinfo {pages} {282}
  (\bibinfo {year} {2009})}\BibitemShut {NoStop}%
\bibitem [{\citenamefont {Schirmer}(1982)}]{Schirmer_1982}%
  \BibitemOpen
  \bibfield  {author} {\bibinfo {author} {\bibfnamefont {J.}~\bibnamefont
  {Schirmer}},\ }\href {\doibase 10.1103/PhysRevA.26.2395} {\bibfield
  {journal} {\bibinfo  {journal} {Phys. Rev. A}\ }\textbf {\bibinfo {volume}
  {26}},\ \bibinfo {pages} {2395} (\bibinfo {year} {1982})}\BibitemShut
  {NoStop}%
\bibitem [{\citenamefont {Schirmer}(1991)}]{Schirmer_1991}%
  \BibitemOpen
  \bibfield  {author} {\bibinfo {author} {\bibfnamefont {J.}~\bibnamefont
  {Schirmer}},\ }\href {\doibase 10.1103/PhysRevA.43.4647} {\bibfield
  {journal} {\bibinfo  {journal} {Phys. Rev. A.}\ }\textbf {\bibinfo {volume}
  {43}},\ \bibinfo {pages} {4647} (\bibinfo {year} {1991})}\BibitemShut
  {NoStop}%
\bibitem [{\citenamefont {Barth}\ and\ \citenamefont
  {Schirmer}(1995)}]{Barth_1995}%
  \BibitemOpen
  \bibfield  {author} {\bibinfo {author} {\bibfnamefont {A.}~\bibnamefont
  {Barth}}\ and\ \bibinfo {author} {\bibfnamefont {J.}~\bibnamefont
  {Schirmer}},\ }\href {\doibase 10.1088/0022-3700/18/5/008} {\bibfield
  {journal} {\bibinfo  {journal} {J. Phys. B: At. Mol. Phys.}\ }\textbf
  {\bibinfo {volume} {18}},\ \bibinfo {pages} {867} (\bibinfo {year}
  {1995})}\BibitemShut {NoStop}%
\bibitem [{\citenamefont {Schirmer}\ and\ \citenamefont
  {Trofimov}(2004)}]{Schirmer_2004}%
  \BibitemOpen
  \bibfield  {author} {\bibinfo {author} {\bibfnamefont {J.}~\bibnamefont
  {Schirmer}}\ and\ \bibinfo {author} {\bibfnamefont {A.~B.}\ \bibnamefont
  {Trofimov}},\ }\href {\doibase 10.1063/1.1752875} {\bibfield  {journal}
  {\bibinfo  {journal} {J. Chem. Phys.}\ }\textbf {\bibinfo {volume} {120}},\
  \bibinfo {pages} {11449} (\bibinfo {year} {2004})}\BibitemShut {NoStop}%
\bibitem [{\citenamefont {Schirmer}(2018)}]{Schirmer_2018}%
  \BibitemOpen
  \bibfield  {author} {\bibinfo {author} {\bibfnamefont {J.}~\bibnamefont
  {Schirmer}},\ }\href@noop {} {\emph {\bibinfo {title} {Many-Body Methods for
  Atoms, Molecules and Clusters}}}\ (\bibinfo  {publisher} {Springer},\
  \bibinfo {year} {2018})\BibitemShut {NoStop}%
\bibitem [{\citenamefont {Trofimov}\ and\ \citenamefont
  {Schirmer}(1997{\natexlab{a}})}]{Trofimov_1997}%
  \BibitemOpen
  \bibfield  {author} {\bibinfo {author} {\bibfnamefont {A.}~\bibnamefont
  {Trofimov}}\ and\ \bibinfo {author} {\bibfnamefont {J.}~\bibnamefont
  {Schirmer}},\ }\href {\doibase https://doi.org/10.1016/S0301-0104(96)00303-5}
  {\bibfield  {journal} {\bibinfo  {journal} {Chem. Phys.}\ }\textbf {\bibinfo
  {volume} {214}},\ \bibinfo {pages} {153} (\bibinfo {year}
  {1997}{\natexlab{a}})}\BibitemShut {NoStop}%
\bibitem [{\citenamefont {Trofimov}\ and\ \citenamefont
  {Schirmer}(1997{\natexlab{b}})}]{Trofimov_1997b}%
  \BibitemOpen
  \bibfield  {author} {\bibinfo {author} {\bibfnamefont {A.}~\bibnamefont
  {Trofimov}}\ and\ \bibinfo {author} {\bibfnamefont {J.}~\bibnamefont
  {Schirmer}},\ }\href {\doibase https://doi.org/10.1016/S0301-0104(97)00256-5}
  {\bibfield  {journal} {\bibinfo  {journal} {Chem. Phys.}\ }\textbf {\bibinfo
  {volume} {224}},\ \bibinfo {pages} {175} (\bibinfo {year}
  {1997}{\natexlab{b}})}\BibitemShut {NoStop}%
\bibitem [{\citenamefont {Trofimov}, \citenamefont {Stelter},\ and\
  \citenamefont {Schirmer}(2002)}]{Trofimov_2002}%
  \BibitemOpen
  \bibfield  {author} {\bibinfo {author} {\bibfnamefont {A.~B.}\ \bibnamefont
  {Trofimov}}, \bibinfo {author} {\bibfnamefont {G.}~\bibnamefont {Stelter}}, \
  and\ \bibinfo {author} {\bibfnamefont {J.}~\bibnamefont {Schirmer}},\ }\href
  {\doibase 10.1063/1.1504708} {\bibfield  {journal} {\bibinfo  {journal} {J.
  Chem. Phys.}\ }\textbf {\bibinfo {volume} {117}},\ \bibinfo {pages} {6402}
  (\bibinfo {year} {2002})}\BibitemShut {NoStop}%
\bibitem [{\citenamefont {Trofimov}\ and\ \citenamefont
  {Schirmer}(2005)}]{Trofimov_2005}%
  \BibitemOpen
  \bibfield  {author} {\bibinfo {author} {\bibfnamefont {A.~B.}\ \bibnamefont
  {Trofimov}}\ and\ \bibinfo {author} {\bibfnamefont {J.}~\bibnamefont
  {Schirmer}},\ }\href {\doibase 10.1063/1.2047550} {\bibfield  {journal}
  {\bibinfo  {journal} {J. Chem. Phys.}\ }\textbf {\bibinfo {volume} {123}},\
  \bibinfo {pages} {144115} (\bibinfo {year} {2005})}\BibitemShut {NoStop}%
\bibitem [{\citenamefont {Trofimov}\ \emph {et~al.}(2006)\citenamefont
  {Trofimov}, \citenamefont {Krivdina}, \citenamefont {Weller},\ and\
  \citenamefont {Schirmer}}]{Trofimov_2006}%
  \BibitemOpen
  \bibfield  {author} {\bibinfo {author} {\bibfnamefont {A.}~\bibnamefont
  {Trofimov}}, \bibinfo {author} {\bibfnamefont {I.}~\bibnamefont {Krivdina}},
  \bibinfo {author} {\bibfnamefont {J.}~\bibnamefont {Weller}}, \ and\ \bibinfo
  {author} {\bibfnamefont {J.}~\bibnamefont {Schirmer}},\ }\href {\doibase
  10.1016/j.chemphys.2006.07.015} {\bibfield  {journal} {\bibinfo  {journal}
  {Chem. Phys.}\ }\textbf {\bibinfo {volume} {329}},\ \bibinfo {pages} {1}
  (\bibinfo {year} {2006})}\BibitemShut {NoStop}%
\bibitem [{\citenamefont {Harbach}, \citenamefont {Wormit},\ and\ \citenamefont
  {Dreuw}(2014)}]{Harbach_2014}%
  \BibitemOpen
  \bibfield  {author} {\bibinfo {author} {\bibfnamefont {P.~H.~P.}\
  \bibnamefont {Harbach}}, \bibinfo {author} {\bibfnamefont {M.}~\bibnamefont
  {Wormit}}, \ and\ \bibinfo {author} {\bibfnamefont {A.}~\bibnamefont
  {Dreuw}},\ }\href {\doibase 10.1063/1.4892418} {\bibfield  {journal}
  {\bibinfo  {journal} {J. Chem. Phys.}\ }\textbf {\bibinfo {volume} {141}},\
  \bibinfo {pages} {064113} (\bibinfo {year} {2014})}\BibitemShut {NoStop}%
\bibitem [{\citenamefont {Dreuw}\ and\ \citenamefont
  {Wormit}(2015)}]{Dreuw_2015}%
  \BibitemOpen
  \bibfield  {author} {\bibinfo {author} {\bibfnamefont {A.}~\bibnamefont
  {Dreuw}}\ and\ \bibinfo {author} {\bibfnamefont {M.}~\bibnamefont {Wormit}},\
  }\href {\doibase 10.1002/wcms.1206} {\bibfield  {journal} {\bibinfo
  {journal} {Wiley Interdiscip. Rev. Comput. Mol. Sci.}\ }\textbf {\bibinfo
  {volume} {5}},\ \bibinfo {pages} {82} (\bibinfo {year} {2015})}\BibitemShut
  {NoStop}%
\bibitem [{\citenamefont {Loos}\ \emph {et~al.}(2018)\citenamefont {Loos},
  \citenamefont {Scemama}, \citenamefont {Blondel}, \citenamefont {Garniron},
  \citenamefont {Caffarel},\ and\ \citenamefont {Jacquemin}}]{Loos_2018a}%
  \BibitemOpen
  \bibfield  {author} {\bibinfo {author} {\bibfnamefont {P.~F.}\ \bibnamefont
  {Loos}}, \bibinfo {author} {\bibfnamefont {A.}~\bibnamefont {Scemama}},
  \bibinfo {author} {\bibfnamefont {A.}~\bibnamefont {Blondel}}, \bibinfo
  {author} {\bibfnamefont {Y.}~\bibnamefont {Garniron}}, \bibinfo {author}
  {\bibfnamefont {M.}~\bibnamefont {Caffarel}}, \ and\ \bibinfo {author}
  {\bibfnamefont {D.}~\bibnamefont {Jacquemin}},\ }\href {\doibase
  10.1021/acs.jctc.8b00406} {\bibfield  {journal} {\bibinfo  {journal} {J.
  Chem. Theory Comput.}\ }\textbf {\bibinfo {volume} {14}},\ \bibinfo {pages}
  {4360} (\bibinfo {year} {2018})}\BibitemShut {NoStop}%
\bibitem [{\citenamefont {Loos}, \citenamefont {Scemama},\ and\ \citenamefont
  {Jacquemin}(2020)}]{Loos_2020a}%
  \BibitemOpen
  \bibfield  {author} {\bibinfo {author} {\bibfnamefont {P.~F.}\ \bibnamefont
  {Loos}}, \bibinfo {author} {\bibfnamefont {A.}~\bibnamefont {Scemama}}, \
  and\ \bibinfo {author} {\bibfnamefont {D.}~\bibnamefont {Jacquemin}},\ }\href
  {\doibase 10.1021/acs.jpclett.0c00014} {\bibfield  {journal} {\bibinfo
  {journal} {J. Phys. Chem. Lett.}\ }\textbf {\bibinfo {volume} {11}},\
  \bibinfo {pages} {2374} (\bibinfo {year} {2020})}\BibitemShut {NoStop}%
\bibitem [{\citenamefont {V{\'e}ril}\ \emph {et~al.}()\citenamefont
  {V{\'e}ril}, \citenamefont {Scemama}, \citenamefont {Caffarel}, \citenamefont
  {Lipparini}, \citenamefont {Boggio-Pasqua}, \citenamefont {Jacquemin},\ and\
  \citenamefont {Loos}}]{Veril_2021}%
  \BibitemOpen
  \bibfield  {author} {\bibinfo {author} {\bibfnamefont {M.}~\bibnamefont
  {V{\'e}ril}}, \bibinfo {author} {\bibfnamefont {A.}~\bibnamefont {Scemama}},
  \bibinfo {author} {\bibfnamefont {M.}~\bibnamefont {Caffarel}}, \bibinfo
  {author} {\bibfnamefont {F.}~\bibnamefont {Lipparini}}, \bibinfo {author}
  {\bibfnamefont {M.}~\bibnamefont {Boggio-Pasqua}}, \bibinfo {author}
  {\bibfnamefont {D.}~\bibnamefont {Jacquemin}}, \ and\ \bibinfo {author}
  {\bibfnamefont {P.-F.}\ \bibnamefont {Loos}},\ }\href {\doibase
  https://doi.org/10.1002/wcms.1517} {\bibfield  {journal} {\bibinfo  {journal}
  {WIREs Comput. Mol. Sci.}\ }\textbf {\bibinfo {volume} {11}},\ \bibinfo
  {pages} {e1517}}\BibitemShut {NoStop}%
\bibitem [{\citenamefont {Loos}\ and\ \citenamefont
  {Jacquemin}(2020)}]{Loos_2020d}%
  \BibitemOpen
  \bibfield  {author} {\bibinfo {author} {\bibfnamefont {P.-F.}\ \bibnamefont
  {Loos}}\ and\ \bibinfo {author} {\bibfnamefont {D.}~\bibnamefont
  {Jacquemin}},\ }\href {\doibase 10.1021/acs.jpclett.9b03652} {\bibfield
  {journal} {\bibinfo  {journal} {J. Phys. Chem. Lett.}\ }\textbf {\bibinfo
  {volume} {11}},\ \bibinfo {pages} {974} (\bibinfo {year} {2020})}\BibitemShut
  {NoStop}%
\bibitem [{\citenamefont {Andersson}\ \emph {et~al.}(1990)\citenamefont
  {Andersson}, \citenamefont {Malmqvist}, \citenamefont {Roos}, \citenamefont
  {Sadlej},\ and\ \citenamefont {Wolinski}}]{Andersson_1990}%
  \BibitemOpen
  \bibfield  {author} {\bibinfo {author} {\bibfnamefont {K.}~\bibnamefont
  {Andersson}}, \bibinfo {author} {\bibfnamefont {P.~A.}\ \bibnamefont
  {Malmqvist}}, \bibinfo {author} {\bibfnamefont {B.~O.}\ \bibnamefont {Roos}},
  \bibinfo {author} {\bibfnamefont {A.~J.}\ \bibnamefont {Sadlej}}, \ and\
  \bibinfo {author} {\bibfnamefont {K.}~\bibnamefont {Wolinski}},\ }\href
  {\doibase 10.1021/j100377a012} {\bibfield  {journal} {\bibinfo  {journal} {J.
  Phys. Chem.}\ }\textbf {\bibinfo {volume} {94}},\ \bibinfo {pages} {5483}
  (\bibinfo {year} {1990})}\BibitemShut {NoStop}%
\bibitem [{\citenamefont {Andersson}, \citenamefont {Malmqvist},\ and\
  \citenamefont {Roos}(1992)}]{Andersson_1992}%
  \BibitemOpen
  \bibfield  {author} {\bibinfo {author} {\bibfnamefont {K.}~\bibnamefont
  {Andersson}}, \bibinfo {author} {\bibfnamefont {P.-A.}\ \bibnamefont
  {Malmqvist}}, \ and\ \bibinfo {author} {\bibfnamefont {B.~O.}\ \bibnamefont
  {Roos}},\ }\href {\doibase 10.1063/1.462209} {\bibfield  {journal} {\bibinfo
  {journal} {J. Chem. Phys.}\ }\textbf {\bibinfo {volume} {96}},\ \bibinfo
  {pages} {1218} (\bibinfo {year} {1992})}\BibitemShut {NoStop}%
\bibitem [{\citenamefont {Roos}\ \emph {et~al.}(1995)\citenamefont {Roos},
  \citenamefont {F{\"u}lscher}, \citenamefont {Malmqvist}, \citenamefont
  {Merch{\'a}n},\ and\ \citenamefont {Serrano-Andr{\'e}s}}]{Roos_1995a}%
  \BibitemOpen
  \bibfield  {author} {\bibinfo {author} {\bibfnamefont {B.~O.}\ \bibnamefont
  {Roos}}, \bibinfo {author} {\bibfnamefont {M.}~\bibnamefont {F{\"u}lscher}},
  \bibinfo {author} {\bibfnamefont {P.-{\AA}.}\ \bibnamefont {Malmqvist}},
  \bibinfo {author} {\bibfnamefont {M.}~\bibnamefont {Merch{\'a}n}}, \ and\
  \bibinfo {author} {\bibfnamefont {L.}~\bibnamefont {Serrano-Andr{\'e}s}},\
  }\href {\doibase 10.1007/978-94-011-0193-6_8} {\emph {\bibinfo {title}
  {Quantum Mechanical Electronic Structure Calculations with Chemical
  Accuracy}}}\ (\bibinfo  {publisher} {Springer Netherlands},\ \bibinfo
  {address} {Dordrecht},\ \bibinfo {year} {1995})\ pp.\ \bibinfo {pages}
  {357--438}\BibitemShut {NoStop}%
\bibitem [{\citenamefont {Hirao}(1992)}]{Hirao_1992}%
  \BibitemOpen
  \bibfield  {author} {\bibinfo {author} {\bibfnamefont {K.}~\bibnamefont
  {Hirao}},\ }\href {\doibase https://doi.org/10.1016/0009-2614(92)85354-D}
  {\bibfield  {journal} {\bibinfo  {journal} {Chem. Phys. Lett.}\ }\textbf
  {\bibinfo {volume} {190}},\ \bibinfo {pages} {374} (\bibinfo {year}
  {1992})}\BibitemShut {NoStop}%
\bibitem [{\citenamefont {Angeli}, \citenamefont {Cimiraglia},\ and\
  \citenamefont {Malrieu}(2001)}]{Angeli_2001a}%
  \BibitemOpen
  \bibfield  {author} {\bibinfo {author} {\bibfnamefont {C.}~\bibnamefont
  {Angeli}}, \bibinfo {author} {\bibfnamefont {R.}~\bibnamefont {Cimiraglia}},
  \ and\ \bibinfo {author} {\bibfnamefont {J.-P.}\ \bibnamefont {Malrieu}},\
  }\href {\doibase 10.1016/S0009-2614(01)01303-3} {\bibfield  {journal}
  {\bibinfo  {journal} {Chem. Phys. Lett.}\ }\textbf {\bibinfo {volume}
  {350}},\ \bibinfo {pages} {297} (\bibinfo {year} {2001})}\BibitemShut
  {NoStop}%
\bibitem [{\citenamefont {Angeli}\ \emph {et~al.}(2001)\citenamefont {Angeli},
  \citenamefont {Cimiraglia}, \citenamefont {Evangelisti}, \citenamefont
  {Leininger},\ and\ \citenamefont {Malrieu}}]{Angeli_2001b}%
  \BibitemOpen
  \bibfield  {author} {\bibinfo {author} {\bibfnamefont {C.}~\bibnamefont
  {Angeli}}, \bibinfo {author} {\bibfnamefont {R.}~\bibnamefont {Cimiraglia}},
  \bibinfo {author} {\bibfnamefont {S.}~\bibnamefont {Evangelisti}}, \bibinfo
  {author} {\bibfnamefont {T.}~\bibnamefont {Leininger}}, \ and\ \bibinfo
  {author} {\bibfnamefont {J.-P.}\ \bibnamefont {Malrieu}},\ }\href {\doibase
  10.1063/1.1361246} {\bibfield  {journal} {\bibinfo  {journal} {J. Chem.
  Phys.}\ }\textbf {\bibinfo {volume} {114}},\ \bibinfo {pages} {10252}
  (\bibinfo {year} {2001})}\BibitemShut {NoStop}%
\bibitem [{\citenamefont {Angeli}, \citenamefont {Cimiraglia},\ and\
  \citenamefont {Malrieu}(2002)}]{Angeli_2002}%
  \BibitemOpen
  \bibfield  {author} {\bibinfo {author} {\bibfnamefont {C.}~\bibnamefont
  {Angeli}}, \bibinfo {author} {\bibfnamefont {R.}~\bibnamefont {Cimiraglia}},
  \ and\ \bibinfo {author} {\bibfnamefont {J.-P.}\ \bibnamefont {Malrieu}},\
  }\href {\doibase 10.1063/1.1515317} {\bibfield  {journal} {\bibinfo
  {journal} {J. Chem. Phys.}\ }\textbf {\bibinfo {volume} {117}},\ \bibinfo
  {pages} {9138} (\bibinfo {year} {2002})}\BibitemShut {NoStop}%
\bibitem [{\citenamefont {Angeli}\ \emph {et~al.}(2006)\citenamefont {Angeli},
  \citenamefont {Bories}, \citenamefont {Cavallini},\ and\ \citenamefont
  {Cimiraglia}}]{Angeli_2006}%
  \BibitemOpen
  \bibfield  {author} {\bibinfo {author} {\bibfnamefont {C.}~\bibnamefont
  {Angeli}}, \bibinfo {author} {\bibfnamefont {B.}~\bibnamefont {Bories}},
  \bibinfo {author} {\bibfnamefont {A.}~\bibnamefont {Cavallini}}, \ and\
  \bibinfo {author} {\bibfnamefont {R.}~\bibnamefont {Cimiraglia}},\ }\href
  {\doibase 10.1063/1.2148946} {\bibfield  {journal} {\bibinfo  {journal} {J.
  Chem. Phys.}\ }\textbf {\bibinfo {volume} {124}},\ \bibinfo {pages} {054108}
  (\bibinfo {year} {2006})}\BibitemShut {NoStop}%
\bibitem [{\citenamefont {Roos}\ and\ \citenamefont
  {Andersson}(1995)}]{Roos_1995b}%
  \BibitemOpen
  \bibfield  {author} {\bibinfo {author} {\bibfnamefont {B.~O.}\ \bibnamefont
  {Roos}}\ and\ \bibinfo {author} {\bibfnamefont {K.}~\bibnamefont
  {Andersson}},\ }\href {\doibase https://doi.org/10.1016/0009-2614(95)01010-7}
  {\bibfield  {journal} {\bibinfo  {journal} {Chem. Phys. Lett.}\ }\textbf
  {\bibinfo {volume} {245}},\ \bibinfo {pages} {215} (\bibinfo {year}
  {1995})}\BibitemShut {NoStop}%
\bibitem [{\citenamefont {Roos}\ \emph {et~al.}(1996)\citenamefont {Roos},
  \citenamefont {Andersson}, \citenamefont {F{\"u}lscher}, \citenamefont
  {Serrano-Andr{\'e}s}, \citenamefont {Pierloot}, \citenamefont {Merch{\'a}n},\
  and\ \citenamefont {Molina}}]{Roos_1996}%
  \BibitemOpen
  \bibfield  {author} {\bibinfo {author} {\bibfnamefont {B.~O.}\ \bibnamefont
  {Roos}}, \bibinfo {author} {\bibfnamefont {K.}~\bibnamefont {Andersson}},
  \bibinfo {author} {\bibfnamefont {M.~P.}\ \bibnamefont {F{\"u}lscher}},
  \bibinfo {author} {\bibfnamefont {L.}~\bibnamefont {Serrano-Andr{\'e}s}},
  \bibinfo {author} {\bibfnamefont {K.}~\bibnamefont {Pierloot}}, \bibinfo
  {author} {\bibfnamefont {M.}~\bibnamefont {Merch{\'a}n}}, \ and\ \bibinfo
  {author} {\bibfnamefont {V.}~\bibnamefont {Molina}},\ }\href {\doibase
  10.1016/S0166-1280(96)80039-X} {\bibfield  {journal} {\bibinfo  {journal} {J.
  Mol. Struct. (THEOCHEM)}\ }\textbf {\bibinfo {volume} {388}},\ \bibinfo
  {pages} {257} (\bibinfo {year} {1996})}\BibitemShut {NoStop}%
\bibitem [{\citenamefont {Forsberg}\ and\ \citenamefont
  {Malmqvist.}(1997)}]{Forsberg_1997}%
  \BibitemOpen
  \bibfield  {author} {\bibinfo {author} {\bibfnamefont {N.}~\bibnamefont
  {Forsberg}}\ and\ \bibinfo {author} {\bibfnamefont {P.-A.}\ \bibnamefont
  {Malmqvist.}},\ }\href {\doibase 10.1016/S0009-2614(97)00669-6} {\bibfield
  {journal} {\bibinfo  {journal} {Chem. Phys. Lett.}\ }\textbf {\bibinfo
  {volume} {274}},\ \bibinfo {pages} {196} (\bibinfo {year}
  {1997})}\BibitemShut {NoStop}%
\bibitem [{\citenamefont {Schapiro}, \citenamefont {Sivalingam},\ and\
  \citenamefont {Neese}(2013)}]{Schapiro_2013}%
  \BibitemOpen
  \bibfield  {author} {\bibinfo {author} {\bibfnamefont {I.}~\bibnamefont
  {Schapiro}}, \bibinfo {author} {\bibfnamefont {K.}~\bibnamefont
  {Sivalingam}}, \ and\ \bibinfo {author} {\bibfnamefont {F.}~\bibnamefont
  {Neese}},\ }\href {\doibase 10.1021/ct400136y} {\bibfield  {journal}
  {\bibinfo  {journal} {J. Chem. Theory Comput.}\ }\textbf {\bibinfo {volume}
  {9}},\ \bibinfo {pages} {3567} (\bibinfo {year} {2013})}\BibitemShut
  {NoStop}%
\bibitem [{\citenamefont {Zobel}, \citenamefont {Nogueira},\ and\ \citenamefont
  {Gonzalez}(2017)}]{Zobel_2017}%
  \BibitemOpen
  \bibfield  {author} {\bibinfo {author} {\bibfnamefont {J.~P.}\ \bibnamefont
  {Zobel}}, \bibinfo {author} {\bibfnamefont {J.~J.}\ \bibnamefont {Nogueira}},
  \ and\ \bibinfo {author} {\bibfnamefont {L.}~\bibnamefont {Gonzalez}},\
  }\href {\doibase 10.1039/C6SC03759C} {\bibfield  {journal} {\bibinfo
  {journal} {Chem. Sci.}\ }\textbf {\bibinfo {volume} {8}},\ \bibinfo {pages}
  {1482} (\bibinfo {year} {2017})}\BibitemShut {NoStop}%
\bibitem [{\citenamefont {Sarkar}\ \emph {et~al.}(2022)\citenamefont {Sarkar},
  \citenamefont {Loos}, \citenamefont {Boggio-Pasqua},\ and\ \citenamefont
  {Jacquemin.}}]{Sarkar_2022}%
  \BibitemOpen
  \bibfield  {author} {\bibinfo {author} {\bibfnamefont {R.}~\bibnamefont
  {Sarkar}}, \bibinfo {author} {\bibfnamefont {P.~F.}\ \bibnamefont {Loos}},
  \bibinfo {author} {\bibfnamefont {M.}~\bibnamefont {Boggio-Pasqua}}, \ and\
  \bibinfo {author} {\bibfnamefont {D.}~\bibnamefont {Jacquemin.}},\ }\href
  {\doibase 10.1021/acs.jctc.1c01197} {\bibfield  {journal} {\bibinfo
  {journal} {J. Chem. Theory Comput.}\ }\textbf {\bibinfo {volume} {18}},\
  \bibinfo {pages} {2418} (\bibinfo {year} {2022})}\BibitemShut {NoStop}%
\bibitem [{\citenamefont {Andersson}\ and\ \citenamefont
  {Roos}(1993)}]{Andersson_1993}%
  \BibitemOpen
  \bibfield  {author} {\bibinfo {author} {\bibfnamefont {K.}~\bibnamefont
  {Andersson}}\ and\ \bibinfo {author} {\bibfnamefont {B.~O.}\ \bibnamefont
  {Roos}},\ }\href {\doibase https://doi.org/10.1002/qua.560450610} {\bibfield
  {journal} {\bibinfo  {journal} {Int. J. Quantum Chem.}\ }\textbf {\bibinfo
  {volume} {45}},\ \bibinfo {pages} {591} (\bibinfo {year} {1993})}\BibitemShut
  {NoStop}%
\bibitem [{\citenamefont {Andersson}(1995)}]{Andersson_1995}%
  \BibitemOpen
  \bibfield  {author} {\bibinfo {author} {\bibfnamefont {K.}~\bibnamefont
  {Andersson}},\ }\href {\doibase 10.1007/BF01113860} {\bibfield  {journal}
  {\bibinfo  {journal} {Theor. Chim. Acta}\ }\textbf {\bibinfo {volume} {91}},\
  \bibinfo {pages} {31} (\bibinfo {year} {1995})}\BibitemShut {NoStop}%
\bibitem [{\citenamefont {Ghigo}, \citenamefont {Roos},\ and\ \citenamefont
  {Malmqvist}(2004)}]{Ghigo_2004}%
  \BibitemOpen
  \bibfield  {author} {\bibinfo {author} {\bibfnamefont {G.}~\bibnamefont
  {Ghigo}}, \bibinfo {author} {\bibfnamefont {B.~O.}\ \bibnamefont {Roos}}, \
  and\ \bibinfo {author} {\bibfnamefont {P.-{\AA}.}\ \bibnamefont
  {Malmqvist}},\ }\href {\doibase https://doi.org/10.1016/j.cplett.2004.08.032}
  {\bibfield  {journal} {\bibinfo  {journal} {Chem. Phys. Lett.}\ }\textbf
  {\bibinfo {volume} {396}},\ \bibinfo {pages} {142} (\bibinfo {year}
  {2004})}\BibitemShut {NoStop}%
\bibitem [{\citenamefont {Pierloot}\ and\ \citenamefont
  {Vancoillie}(2006)}]{Pierloot_2006}%
  \BibitemOpen
  \bibfield  {author} {\bibinfo {author} {\bibfnamefont {K.}~\bibnamefont
  {Pierloot}}\ and\ \bibinfo {author} {\bibfnamefont {S.}~\bibnamefont
  {Vancoillie}},\ }\href {\doibase 10.1063/1.2353829} {\bibfield  {journal}
  {\bibinfo  {journal} {J. Chem. Phys.}\ }\textbf {\bibinfo {volume} {125}},\
  \bibinfo {pages} {124303} (\bibinfo {year} {2006})}\BibitemShut {NoStop}%
\bibitem [{\citenamefont {Pierloot}\ and\ \citenamefont
  {Vancoillie}(2008)}]{Pierloot_2008}%
  \BibitemOpen
  \bibfield  {author} {\bibinfo {author} {\bibfnamefont {K.}~\bibnamefont
  {Pierloot}}\ and\ \bibinfo {author} {\bibfnamefont {S.}~\bibnamefont
  {Vancoillie}},\ }\href {\doibase 10.1063/1.2820786} {\bibfield  {journal}
  {\bibinfo  {journal} {J. Chem. Phys.}\ }\textbf {\bibinfo {volume} {128}},\
  \bibinfo {pages} {034104} (\bibinfo {year} {2008})}\BibitemShut {NoStop}%
\bibitem [{\citenamefont {Suaud}\ \emph {et~al.}(2009)\citenamefont {Suaud},
  \citenamefont {Bonnet}, \citenamefont {Boilleau}, \citenamefont
  {Lab{\`e}guerie},\ and\ \citenamefont {Guih{\'e}ry}}]{Suaud_2009}%
  \BibitemOpen
  \bibfield  {author} {\bibinfo {author} {\bibfnamefont {N.}~\bibnamefont
  {Suaud}}, \bibinfo {author} {\bibfnamefont {M.-L.}\ \bibnamefont {Bonnet}},
  \bibinfo {author} {\bibfnamefont {C.}~\bibnamefont {Boilleau}}, \bibinfo
  {author} {\bibfnamefont {P.}~\bibnamefont {Lab{\`e}guerie}}, \ and\ \bibinfo
  {author} {\bibfnamefont {N.}~\bibnamefont {Guih{\'e}ry}},\ }\href {\doibase
  10.1021/ja805626s} {\bibfield  {journal} {\bibinfo  {journal} {J. Am. Chem.
  Soc.}\ }\textbf {\bibinfo {volume} {131}},\ \bibinfo {pages} {715} (\bibinfo
  {year} {2009})}\BibitemShut {NoStop}%
\bibitem [{\citenamefont {Kepenekian}, \citenamefont {Robert},\ and\
  \citenamefont {Le~Guennic}(2009)}]{Kepenekian_2009}%
  \BibitemOpen
  \bibfield  {author} {\bibinfo {author} {\bibfnamefont {M.}~\bibnamefont
  {Kepenekian}}, \bibinfo {author} {\bibfnamefont {V.}~\bibnamefont {Robert}},
  \ and\ \bibinfo {author} {\bibfnamefont {B.}~\bibnamefont {Le~Guennic}},\
  }\href {\doibase 10.1063/1.3211020} {\bibfield  {journal} {\bibinfo
  {journal} {J. Chem. Phys.}\ }\textbf {\bibinfo {volume} {131}},\ \bibinfo
  {pages} {114702} (\bibinfo {year} {2009})}\BibitemShut {NoStop}%
\bibitem [{\citenamefont {Lawson~Daku}\ \emph {et~al.}(2012)\citenamefont
  {Lawson~Daku}, \citenamefont {Aquilante}, \citenamefont {Robinson},\ and\
  \citenamefont {Hauser}}]{Daku_2012}%
  \BibitemOpen
  \bibfield  {author} {\bibinfo {author} {\bibfnamefont {L.~M.}\ \bibnamefont
  {Lawson~Daku}}, \bibinfo {author} {\bibfnamefont {F.}~\bibnamefont
  {Aquilante}}, \bibinfo {author} {\bibfnamefont {T.~W.}\ \bibnamefont
  {Robinson}}, \ and\ \bibinfo {author} {\bibfnamefont {A.}~\bibnamefont
  {Hauser}},\ }\href {\doibase 10.1021/ct300592w} {\bibfield  {journal}
  {\bibinfo  {journal} {J. Chem. Theory Comput.}\ }\textbf {\bibinfo {volume}
  {8}},\ \bibinfo {pages} {4216} (\bibinfo {year} {2012})}\BibitemShut
  {NoStop}%
\bibitem [{\citenamefont {Rudavskyi}\ \emph {et~al.}(2014)\citenamefont
  {Rudavskyi}, \citenamefont {Sousa}, \citenamefont {de~Graaf}, \citenamefont
  {Havenith},\ and\ \citenamefont {Broer}}]{Rudavskyi_2014}%
  \BibitemOpen
  \bibfield  {author} {\bibinfo {author} {\bibfnamefont {A.}~\bibnamefont
  {Rudavskyi}}, \bibinfo {author} {\bibfnamefont {C.}~\bibnamefont {Sousa}},
  \bibinfo {author} {\bibfnamefont {C.}~\bibnamefont {de~Graaf}}, \bibinfo
  {author} {\bibfnamefont {R.~W.~A.}\ \bibnamefont {Havenith}}, \ and\ \bibinfo
  {author} {\bibfnamefont {R.}~\bibnamefont {Broer}},\ }\href {\doibase
  10.1063/1.4875695} {\bibfield  {journal} {\bibinfo  {journal} {J. Chem.
  Phys.}\ }\textbf {\bibinfo {volume} {140}},\ \bibinfo {pages} {184318}
  (\bibinfo {year} {2014})}\BibitemShut {NoStop}%
\bibitem [{\citenamefont {Vela}\ \emph {et~al.}(2016)\citenamefont {Vela},
  \citenamefont {Fumanal}, \citenamefont {Ribas-Ari{\~n}o},\ and\ \citenamefont
  {Robert}}]{Vela_2016}%
  \BibitemOpen
  \bibfield  {author} {\bibinfo {author} {\bibfnamefont {S.}~\bibnamefont
  {Vela}}, \bibinfo {author} {\bibfnamefont {M.}~\bibnamefont {Fumanal}},
  \bibinfo {author} {\bibfnamefont {J.}~\bibnamefont {Ribas-Ari{\~n}o}}, \ and\
  \bibinfo {author} {\bibfnamefont {V.}~\bibnamefont {Robert}},\ }\href
  {\doibase https://doi.org/10.1002/jcc.24283} {\bibfield  {journal} {\bibinfo
  {journal} {J. Comput. Chem.}\ }\textbf {\bibinfo {volume} {37}},\ \bibinfo
  {pages} {947} (\bibinfo {year} {2016})}\BibitemShut {NoStop}%
\bibitem [{\citenamefont {Wen}\ \emph {et~al.}(2018)\citenamefont {Wen},
  \citenamefont {Han}, \citenamefont {Havlas},\ and\ \citenamefont
  {Michl}}]{Wen_2018}%
  \BibitemOpen
  \bibfield  {author} {\bibinfo {author} {\bibfnamefont {J.}~\bibnamefont
  {Wen}}, \bibinfo {author} {\bibfnamefont {B.}~\bibnamefont {Han}}, \bibinfo
  {author} {\bibfnamefont {Z.}~\bibnamefont {Havlas}}, \ and\ \bibinfo {author}
  {\bibfnamefont {J.}~\bibnamefont {Michl}},\ }\href {\doibase
  10.1021/acs.jctc.8b00136} {\bibfield  {journal} {\bibinfo  {journal} {J.
  Chem. Theory Comput.}\ }\textbf {\bibinfo {volume} {14}},\ \bibinfo {pages}
  {4291} (\bibinfo {year} {2018})}\BibitemShut {NoStop}%
\bibitem [{\citenamefont {Loos}\ \emph {et~al.}(2019)\citenamefont {Loos},
  \citenamefont {Boggio-Pasqua}, \citenamefont {Scemama}, \citenamefont
  {Caffarel},\ and\ \citenamefont {Jacquemin}}]{Loos_2019}%
  \BibitemOpen
  \bibfield  {author} {\bibinfo {author} {\bibfnamefont {P.-F.}\ \bibnamefont
  {Loos}}, \bibinfo {author} {\bibfnamefont {M.}~\bibnamefont {Boggio-Pasqua}},
  \bibinfo {author} {\bibfnamefont {A.}~\bibnamefont {Scemama}}, \bibinfo
  {author} {\bibfnamefont {M.}~\bibnamefont {Caffarel}}, \ and\ \bibinfo
  {author} {\bibfnamefont {D.}~\bibnamefont {Jacquemin}},\ }\href {\doibase
  10.1021/acs.jctc.8b01205} {\bibfield  {journal} {\bibinfo  {journal} {J.
  Chem. Theory Comput.}\ }\textbf {\bibinfo {volume} {15}},\ \bibinfo {pages}
  {1939} (\bibinfo {year} {2019})}\BibitemShut {NoStop}%
\bibitem [{\citenamefont {Loos}\ \emph
  {et~al.}(2020{\natexlab{a}})\citenamefont {Loos}, \citenamefont {Lipparini},
  \citenamefont {Boggio-Pasqua}, \citenamefont {Scemama},\ and\ \citenamefont
  {Jacquemin}}]{Loos_2020b}%
  \BibitemOpen
  \bibfield  {author} {\bibinfo {author} {\bibfnamefont {P.~F.}\ \bibnamefont
  {Loos}}, \bibinfo {author} {\bibfnamefont {F.}~\bibnamefont {Lipparini}},
  \bibinfo {author} {\bibfnamefont {M.}~\bibnamefont {Boggio-Pasqua}}, \bibinfo
  {author} {\bibfnamefont {A.}~\bibnamefont {Scemama}}, \ and\ \bibinfo
  {author} {\bibfnamefont {D.}~\bibnamefont {Jacquemin}},\ }\href {\doibase
  10.1021/acs.jctc.9b01216} {\bibfield  {journal} {\bibinfo  {journal} {J.
  Chem. Theory Comput.}\ }\textbf {\bibinfo {volume} {16}},\ \bibinfo {pages}
  {1711} (\bibinfo {year} {2020}{\natexlab{a}})}\BibitemShut {NoStop}%
\bibitem [{\citenamefont {Loos}\ \emph
  {et~al.}(2020{\natexlab{b}})\citenamefont {Loos}, \citenamefont {Scemama},
  \citenamefont {Boggio-Pasqua},\ and\ \citenamefont {Jacquemin}}]{Loos_2020c}%
  \BibitemOpen
  \bibfield  {author} {\bibinfo {author} {\bibfnamefont {P.~F.}\ \bibnamefont
  {Loos}}, \bibinfo {author} {\bibfnamefont {A.}~\bibnamefont {Scemama}},
  \bibinfo {author} {\bibfnamefont {M.}~\bibnamefont {Boggio-Pasqua}}, \ and\
  \bibinfo {author} {\bibfnamefont {D.}~\bibnamefont {Jacquemin}},\ }\href
  {\doibase 10.1021/acs.jctc.0c00227} {\bibfield  {journal} {\bibinfo
  {journal} {J. Chem. Theory Comput.}\ }\textbf {\bibinfo {volume} {16}},\
  \bibinfo {pages} {3720} (\bibinfo {year} {2020}{\natexlab{b}})}\BibitemShut
  {NoStop}%
\bibitem [{\citenamefont {Loos}\ \emph {et~al.}(2021)\citenamefont {Loos},
  \citenamefont {Comin}, \citenamefont {Blase},\ and\ \citenamefont
  {Jacquemin}}]{Loos_2021c}%
  \BibitemOpen
  \bibfield  {author} {\bibinfo {author} {\bibfnamefont {P.-F.}\ \bibnamefont
  {Loos}}, \bibinfo {author} {\bibfnamefont {M.}~\bibnamefont {Comin}},
  \bibinfo {author} {\bibfnamefont {X.}~\bibnamefont {Blase}}, \ and\ \bibinfo
  {author} {\bibfnamefont {D.}~\bibnamefont {Jacquemin}},\ }\href {\doibase
  10.1021/acs.jctc.1c00226} {\bibfield  {journal} {\bibinfo  {journal} {J.
  Chem. Theory Comput.}\ }\textbf {\bibinfo {volume} {17}},\ \bibinfo {pages}
  {3666} (\bibinfo {year} {2021})}\BibitemShut {NoStop}%
\bibitem [{\citenamefont {Loos}\ and\ \citenamefont
  {Jacquemin}(2021)}]{Loos_2021b}%
  \BibitemOpen
  \bibfield  {author} {\bibinfo {author} {\bibfnamefont {P.-F.}\ \bibnamefont
  {Loos}}\ and\ \bibinfo {author} {\bibfnamefont {D.}~\bibnamefont
  {Jacquemin}},\ }\href {\doibase 10.1021/acs.jpca.1c08524} {\bibfield
  {journal} {\bibinfo  {journal} {J. Phys. Chem. A}\ }\textbf {\bibinfo
  {volume} {125}},\ \bibinfo {pages} {10174} (\bibinfo {year}
  {2021})}\BibitemShut {NoStop}%
\bibitem [{\citenamefont {Rettig}\ \emph {et~al.}(2020)\citenamefont {Rettig},
  \citenamefont {Hait}, \citenamefont {Bertels},\ and\ \citenamefont
  {Head-Gordon}}]{Rettig_2020}%
  \BibitemOpen
  \bibfield  {author} {\bibinfo {author} {\bibfnamefont {A.}~\bibnamefont
  {Rettig}}, \bibinfo {author} {\bibfnamefont {D.}~\bibnamefont {Hait}},
  \bibinfo {author} {\bibfnamefont {L.~W.}\ \bibnamefont {Bertels}}, \ and\
  \bibinfo {author} {\bibfnamefont {M.}~\bibnamefont {Head-Gordon}},\ }\href
  {\doibase 10.1021/acs.jctc.0c00986} {\bibfield  {journal} {\bibinfo
  {journal} {J. Chem. Theory Comput.}\ }\textbf {\bibinfo {volume} {16}},\
  \bibinfo {pages} {7473} (\bibinfo {year} {2020})}\BibitemShut {NoStop}%
\bibitem [{\citenamefont {Werner}(1996)}]{Werner_1996}%
  \BibitemOpen
  \bibfield  {author} {\bibinfo {author} {\bibfnamefont {H.-J.}\ \bibnamefont
  {Werner}},\ }\href {\doibase 10.1080/002689796173967} {\bibfield  {journal}
  {\bibinfo  {journal} {Mol. Phys.}\ }\textbf {\bibinfo {volume} {89}},\
  \bibinfo {pages} {645} (\bibinfo {year} {1996})}\BibitemShut {NoStop}%
\bibitem [{\citenamefont {Werner}\ \emph {et~al.}(2020)\citenamefont {Werner},
  \citenamefont {Knowles}, \citenamefont {Manby}, \citenamefont {Black},
  \citenamefont {Doll}, \citenamefont {He{\ss}elmann}, \citenamefont {Kats},
  \citenamefont {K{\"o}hn}, \citenamefont {Korona}, \citenamefont {Kreplin},
  \citenamefont {Ma}, \citenamefont {Miller}, \citenamefont {Mitrushchenkov},
  \citenamefont {Peterson}, \citenamefont {Polyak}, \citenamefont {Rauhut},\
  and\ \citenamefont {Sibaev}}]{Werner_2020}%
  \BibitemOpen
  \bibfield  {author} {\bibinfo {author} {\bibfnamefont {H.-J.}\ \bibnamefont
  {Werner}}, \bibinfo {author} {\bibfnamefont {P.~J.}\ \bibnamefont {Knowles}},
  \bibinfo {author} {\bibfnamefont {F.~R.}\ \bibnamefont {Manby}}, \bibinfo
  {author} {\bibfnamefont {J.~A.}\ \bibnamefont {Black}}, \bibinfo {author}
  {\bibfnamefont {K.}~\bibnamefont {Doll}}, \bibinfo {author} {\bibfnamefont
  {A.}~\bibnamefont {He{\ss}elmann}}, \bibinfo {author} {\bibfnamefont
  {D.}~\bibnamefont {Kats}}, \bibinfo {author} {\bibfnamefont {A.}~\bibnamefont
  {K{\"o}hn}}, \bibinfo {author} {\bibfnamefont {T.}~\bibnamefont {Korona}},
  \bibinfo {author} {\bibfnamefont {D.~A.}\ \bibnamefont {Kreplin}}, \bibinfo
  {author} {\bibfnamefont {Q.}~\bibnamefont {Ma}}, \bibinfo {author}
  {\bibfnamefont {T.~F.}\ \bibnamefont {Miller}}, \bibinfo {author}
  {\bibfnamefont {A.}~\bibnamefont {Mitrushchenkov}}, \bibinfo {author}
  {\bibfnamefont {K.~A.}\ \bibnamefont {Peterson}}, \bibinfo {author}
  {\bibfnamefont {I.}~\bibnamefont {Polyak}}, \bibinfo {author} {\bibfnamefont
  {G.}~\bibnamefont {Rauhut}}, \ and\ \bibinfo {author} {\bibfnamefont
  {M.}~\bibnamefont {Sibaev}},\ }\href {\doibase 10.1063/5.0005081} {\bibfield
  {journal} {\bibinfo  {journal} {J. Chem. Phys.}\ }\textbf {\bibinfo {volume}
  {152}},\ \bibinfo {pages} {144107} (\bibinfo {year} {2020})}\BibitemShut
  {NoStop}%
\bibitem [{\citenamefont {Yanai}\ and\ \citenamefont
  {Chan}(2007)}]{Yanai_2007}%
  \BibitemOpen
  \bibfield  {author} {\bibinfo {author} {\bibfnamefont {T.}~\bibnamefont
  {Yanai}}\ and\ \bibinfo {author} {\bibfnamefont {G.~K.-L.}\ \bibnamefont
  {Chan}},\ }\href {\doibase 10.1063/1.2761870} {\bibfield  {journal} {\bibinfo
   {journal} {J. Chem. Phys.}\ }\textbf {\bibinfo {volume} {127}},\ \bibinfo
  {pages} {104107} (\bibinfo {year} {2007})}\BibitemShut {NoStop}%
\bibitem [{\citenamefont {Grabarek}, \citenamefont {Walczak},\ and\
  \citenamefont {Andruni{\'o}w}(2016)}]{Grabarek_2016}%
  \BibitemOpen
  \bibfield  {author} {\bibinfo {author} {\bibfnamefont {D.}~\bibnamefont
  {Grabarek}}, \bibinfo {author} {\bibfnamefont {E.}~\bibnamefont {Walczak}}, \
  and\ \bibinfo {author} {\bibfnamefont {T.}~\bibnamefont {Andruni{\'o}w}},\
  }\href {\doibase 10.1021/acs.jctc.6b00108} {\bibfield  {journal} {\bibinfo
  {journal} {J. Chem. Theory Comput.}\ }\textbf {\bibinfo {volume} {12}},\
  \bibinfo {pages} {2346} (\bibinfo {year} {2016})}\BibitemShut {NoStop}%
\bibitem [{\citenamefont {Li}\ and\ \citenamefont
  {Evangelista}(2017)}]{Li_2017}%
  \BibitemOpen
  \bibfield  {author} {\bibinfo {author} {\bibfnamefont {C.}~\bibnamefont
  {Li}}\ and\ \bibinfo {author} {\bibfnamefont {F.~A.}\ \bibnamefont
  {Evangelista}},\ }\href {\doibase 10.1063/1.4979016} {\bibfield  {journal}
  {\bibinfo  {journal} {J. Chem. Phys.}\ }\textbf {\bibinfo {volume} {146}},\
  \bibinfo {pages} {124132} (\bibinfo {year} {2017})}\BibitemShut {NoStop}%
\bibitem [{\citenamefont {Li}\ and\ \citenamefont
  {Evangelista}(2018)}]{Li_2018}%
  \BibitemOpen
  \bibfield  {author} {\bibinfo {author} {\bibfnamefont {C.}~\bibnamefont
  {Li}}\ and\ \bibinfo {author} {\bibfnamefont {F.~A.}\ \bibnamefont
  {Evangelista}},\ }\href {\doibase 10.1063/1.5019793} {\bibfield  {journal}
  {\bibinfo  {journal} {J. Chem. Phys.}\ }\textbf {\bibinfo {volume} {148}},\
  \bibinfo {pages} {124106} (\bibinfo {year} {2018})}\BibitemShut {NoStop}%
\bibitem [{\citenamefont {Li}\ and\ \citenamefont
  {Evangelista}(2021)}]{Li_2021}%
  \BibitemOpen
  \bibfield  {author} {\bibinfo {author} {\bibfnamefont {C.}~\bibnamefont
  {Li}}\ and\ \bibinfo {author} {\bibfnamefont {F.~A.}\ \bibnamefont
  {Evangelista}},\ }\href {\doibase 10.1063/5.0059362} {\bibfield  {journal}
  {\bibinfo  {journal} {J. Chem. Phys.}\ }\textbf {\bibinfo {volume} {155}},\
  \bibinfo {pages} {114111} (\bibinfo {year} {2021})}\BibitemShut {NoStop}%
\bibitem [{\citenamefont {Bittererova}, \citenamefont {Brinck},\ and\
  \citenamefont {Ostmark}(2001)}]{Bittererova_2001}%
  \BibitemOpen
  \bibfield  {author} {\bibinfo {author} {\bibfnamefont {M.}~\bibnamefont
  {Bittererova}}, \bibinfo {author} {\bibfnamefont {T.}~\bibnamefont {Brinck}},
  \ and\ \bibinfo {author} {\bibfnamefont {H.}~\bibnamefont {Ostmark}},\ }\href
  {\doibase 10.1016/S0009-2614(01)00454-7} {\bibfield  {journal} {\bibinfo
  {journal} {Chem. Phys. Lett.}\ }\textbf {\bibinfo {volume} {340}},\ \bibinfo
  {pages} {597} (\bibinfo {year} {2001})}\BibitemShut {NoStop}%
\bibitem [{\citenamefont {Bokarev}\ \emph {et~al.}(2009)\citenamefont
  {Bokarev}, \citenamefont {Dolgov}, \citenamefont {Bataev},\ and\
  \citenamefont {Godunov}}]{Bokarev_2009}%
  \BibitemOpen
  \bibfield  {author} {\bibinfo {author} {\bibfnamefont {S.~I.}\ \bibnamefont
  {Bokarev}}, \bibinfo {author} {\bibfnamefont {E.~K.}\ \bibnamefont {Dolgov}},
  \bibinfo {author} {\bibfnamefont {V.~A.}\ \bibnamefont {Bataev}}, \ and\
  \bibinfo {author} {\bibfnamefont {I.~A.}\ \bibnamefont {Godunov}},\ }\href
  {\doibase 10.1002/qua.21838} {\bibfield  {journal} {\bibinfo  {journal} {Int.
  J. Quantum Chem.}\ }\textbf {\bibinfo {volume} {109}},\ \bibinfo {pages}
  {569} (\bibinfo {year} {2009})}\BibitemShut {NoStop}%
\bibitem [{\citenamefont {Frankcombe}, \citenamefont {McNeil},\ and\
  \citenamefont {Nyman}(2011)}]{Frankcombe_2011}%
  \BibitemOpen
  \bibfield  {author} {\bibinfo {author} {\bibfnamefont {T.~J.}\ \bibnamefont
  {Frankcombe}}, \bibinfo {author} {\bibfnamefont {S.~D.}\ \bibnamefont
  {McNeil}}, \ and\ \bibinfo {author} {\bibfnamefont {G.}~\bibnamefont
  {Nyman}},\ }\href {\doibase 10.1016/j.cplett.2011.08.047} {\bibfield
  {journal} {\bibinfo  {journal} {Chem. Phys. Lett.}\ }\textbf {\bibinfo
  {volume} {514}},\ \bibinfo {pages} {40} (\bibinfo {year} {2011})}\BibitemShut
  {NoStop}%
\bibitem [{\citenamefont {Gu}\ \emph {et~al.}(2008)\citenamefont {Gu},
  \citenamefont {Lin}, \citenamefont {Ma}, \citenamefont {Wu},\ and\
  \citenamefont {Shaik}}]{Gu_2008}%
  \BibitemOpen
  \bibfield  {author} {\bibinfo {author} {\bibfnamefont {J.}~\bibnamefont
  {Gu}}, \bibinfo {author} {\bibfnamefont {Y.}~\bibnamefont {Lin}}, \bibinfo
  {author} {\bibfnamefont {B.}~\bibnamefont {Ma}}, \bibinfo {author}
  {\bibfnamefont {W.}~\bibnamefont {Wu}}, \ and\ \bibinfo {author}
  {\bibfnamefont {S.}~\bibnamefont {Shaik}},\ }\href {\doibase
  10.1021/ct800341z} {\bibfield  {journal} {\bibinfo  {journal} {J. Chem.
  Theory Comput.}\ }\textbf {\bibinfo {volume} {4}},\ \bibinfo {pages} {2101}
  (\bibinfo {year} {2008})}\BibitemShut {NoStop}%
\bibitem [{\citenamefont {Kerkines}, \citenamefont {Carsky},\ and\
  \citenamefont {Mavridis}(2005)}]{Kerkines_2005}%
  \BibitemOpen
  \bibfield  {author} {\bibinfo {author} {\bibfnamefont {I.}~\bibnamefont
  {Kerkines}}, \bibinfo {author} {\bibfnamefont {P.}~\bibnamefont {Carsky}}, \
  and\ \bibinfo {author} {\bibfnamefont {A.}~\bibnamefont {Mavridis}},\ }\href
  {\doibase 10.1021/jp054530z} {\bibfield  {journal} {\bibinfo  {journal} {J.
  Phys. Chem. A}\ }\textbf {\bibinfo {volume} {109}},\ \bibinfo {pages} {10148}
  (\bibinfo {year} {2005})}\BibitemShut {NoStop}%
\bibitem [{\citenamefont {Lampart}\ \emph {et~al.}(2008)\citenamefont
  {Lampart}, \citenamefont {Schofield}, \citenamefont {Christie},\ and\
  \citenamefont {Jordan}}]{Lampart_2008}%
  \BibitemOpen
  \bibfield  {author} {\bibinfo {author} {\bibfnamefont {W.~M.}\ \bibnamefont
  {Lampart}}, \bibinfo {author} {\bibfnamefont {D.~P.}\ \bibnamefont
  {Schofield}}, \bibinfo {author} {\bibfnamefont {R.~A.}\ \bibnamefont
  {Christie}}, \ and\ \bibinfo {author} {\bibfnamefont {K.~D.}\ \bibnamefont
  {Jordan}},\ }\href {\doibase 10.1080/00268970802317504} {\bibfield  {journal}
  {\bibinfo  {journal} {Mol. Phys.}\ }\textbf {\bibinfo {volume} {106}},\
  \bibinfo {pages} {1697} (\bibinfo {year} {2008})}\BibitemShut {NoStop}%
\bibitem [{\citenamefont {Leininger}\ and\ \citenamefont
  {Gadea}(2000)}]{Leininger_2000}%
  \BibitemOpen
  \bibfield  {author} {\bibinfo {author} {\bibfnamefont {T.}~\bibnamefont
  {Leininger}}\ and\ \bibinfo {author} {\bibfnamefont {F.}~\bibnamefont
  {Gadea}},\ }\href {\doibase 10.1088/0953-4075/33/4/311} {\bibfield  {journal}
  {\bibinfo  {journal} {J. Phys. B At. Mol. Opt .}\ }\textbf {\bibinfo {volume}
  {33}},\ \bibinfo {pages} {735} (\bibinfo {year} {2000})}\BibitemShut
  {NoStop}%
\bibitem [{\citenamefont {Maranzana}\ and\ \citenamefont
  {Tonachini}(2020)}]{Maranzana_2020}%
  \BibitemOpen
  \bibfield  {author} {\bibinfo {author} {\bibfnamefont {A.}~\bibnamefont
  {Maranzana}}\ and\ \bibinfo {author} {\bibfnamefont {G.}~\bibnamefont
  {Tonachini}},\ }\href {\doibase 10.1021/acs.jpca.9b11430} {\bibfield
  {journal} {\bibinfo  {journal} {J. Phys. Chem. A}\ }\textbf {\bibinfo
  {volume} {124}},\ \bibinfo {pages} {1112} (\bibinfo {year}
  {2020})}\BibitemShut {NoStop}%
\bibitem [{\citenamefont {Papakondylis}\ and\ \citenamefont
  {Mavridis}(1999)}]{Papakondylis_1999}%
  \BibitemOpen
  \bibfield  {author} {\bibinfo {author} {\bibfnamefont {A.}~\bibnamefont
  {Papakondylis}}\ and\ \bibinfo {author} {\bibfnamefont {A.}~\bibnamefont
  {Mavridis}},\ }\href {\doibase 10.1021/jp983403i} {\bibfield  {journal}
  {\bibinfo  {journal} {J. Phys. Chem. A}\ }\textbf {\bibinfo {volume} {103}},\
  \bibinfo {pages} {1255} (\bibinfo {year} {1999})}\BibitemShut {NoStop}%
\bibitem [{\citenamefont {Schild}\ and\ \citenamefont
  {Paulus}(2013)}]{Schild_2013}%
  \BibitemOpen
  \bibfield  {author} {\bibinfo {author} {\bibfnamefont {A.}~\bibnamefont
  {Schild}}\ and\ \bibinfo {author} {\bibfnamefont {B.}~\bibnamefont
  {Paulus}},\ }\href {\doibase 10.1002/jcc.23273} {\bibfield  {journal}
  {\bibinfo  {journal} {J. Comput. Chem.}\ }\textbf {\bibinfo {volume} {34}},\
  \bibinfo {pages} {1393} (\bibinfo {year} {2013})}\BibitemShut {NoStop}%
\bibitem [{\citenamefont {Sun}\ and\ \citenamefont
  {Schaefer}(2018)}]{Sun_2018}%
  \BibitemOpen
  \bibfield  {author} {\bibinfo {author} {\bibfnamefont {Z.}~\bibnamefont
  {Sun}}\ and\ \bibinfo {author} {\bibfnamefont {H.~F.}\ \bibnamefont
  {Schaefer}, \bibfnamefont {III}},\ }\href {\doibase 10.1039/c8cp03434f}
  {\bibfield  {journal} {\bibinfo  {journal} {Phys. Chem. Chem. Phys.}\
  }\textbf {\bibinfo {volume} {20}},\ \bibinfo {pages} {18986} (\bibinfo {year}
  {2018})}\BibitemShut {NoStop}%
\bibitem [{\citenamefont {Takatani}, \citenamefont {Sears},\ and\ \citenamefont
  {Sherrill}(2009)}]{Takatani_2009}%
  \BibitemOpen
  \bibfield  {author} {\bibinfo {author} {\bibfnamefont {T.}~\bibnamefont
  {Takatani}}, \bibinfo {author} {\bibfnamefont {J.~S.}\ \bibnamefont {Sears}},
  \ and\ \bibinfo {author} {\bibfnamefont {C.~D.}\ \bibnamefont {Sherrill}},\
  }\href {\doibase 10.1021/jp903865t} {\bibfield  {journal} {\bibinfo
  {journal} {J. Phys. Chem. A}\ }\textbf {\bibinfo {volume} {113}},\ \bibinfo
  {pages} {9231} (\bibinfo {year} {2009})}\BibitemShut {NoStop}%
\bibitem [{\citenamefont {Takatani}, \citenamefont {Sears},\ and\ \citenamefont
  {Sherrill}(2010)}]{Takatani_2010}%
  \BibitemOpen
  \bibfield  {author} {\bibinfo {author} {\bibfnamefont {T.}~\bibnamefont
  {Takatani}}, \bibinfo {author} {\bibfnamefont {J.~S.}\ \bibnamefont {Sears}},
  \ and\ \bibinfo {author} {\bibfnamefont {C.~D.}\ \bibnamefont {Sherrill}},\
  }\href {\doibase 10.1021/jp1046084} {\bibfield  {journal} {\bibinfo
  {journal} {J. Phys. Chem. A}\ }\textbf {\bibinfo {volume} {114}},\ \bibinfo
  {pages} {11714} (\bibinfo {year} {2010})}\BibitemShut {NoStop}%
\bibitem [{\citenamefont {Verma}, \citenamefont {Varga},\ and\ \citenamefont
  {Truhlar}(2018)}]{Verma_2018}%
  \BibitemOpen
  \bibfield  {author} {\bibinfo {author} {\bibfnamefont {P.}~\bibnamefont
  {Verma}}, \bibinfo {author} {\bibfnamefont {Z.}~\bibnamefont {Varga}}, \ and\
  \bibinfo {author} {\bibfnamefont {D.~G.}\ \bibnamefont {Truhlar}},\ }\href
  {\doibase 10.1021/acs.jpca.7b12652} {\bibfield  {journal} {\bibinfo
  {journal} {J. Phys. Chem. A}\ }\textbf {\bibinfo {volume} {122}},\ \bibinfo
  {pages} {2563} (\bibinfo {year} {2018})}\BibitemShut {NoStop}%
\bibitem [{\citenamefont {Woywod}\ \emph {et~al.}(2010)\citenamefont {Woywod},
  \citenamefont {Papp}, \citenamefont {Halasz},\ and\ \citenamefont
  {Vibok}}]{Woywod_2010}%
  \BibitemOpen
  \bibfield  {author} {\bibinfo {author} {\bibfnamefont {C.}~\bibnamefont
  {Woywod}}, \bibinfo {author} {\bibfnamefont {A.}~\bibnamefont {Papp}},
  \bibinfo {author} {\bibfnamefont {G.~J.}\ \bibnamefont {Halasz}}, \ and\
  \bibinfo {author} {\bibfnamefont {A.}~\bibnamefont {Vibok}},\ }\href
  {\doibase 10.1007/s00214-009-0678-x} {\bibfield  {journal} {\bibinfo
  {journal} {Theor. Chem. Acc.}\ }\textbf {\bibinfo {volume} {125}},\ \bibinfo
  {pages} {521} (\bibinfo {year} {2010})}\BibitemShut {NoStop}%
\bibitem [{\citenamefont {Yan}, \citenamefont {Hase},\ and\ \citenamefont
  {Doubleday}(2004)}]{Yan_2004}%
  \BibitemOpen
  \bibfield  {author} {\bibinfo {author} {\bibfnamefont {T.}~\bibnamefont
  {Yan}}, \bibinfo {author} {\bibfnamefont {W.}~\bibnamefont {Hase}}, \ and\
  \bibinfo {author} {\bibfnamefont {C.}~\bibnamefont {Doubleday}},\ }\href
  {\doibase 10.1063/1.1705574} {\bibfield  {journal} {\bibinfo  {journal} {J.
  Chem. Phys.}\ }\textbf {\bibinfo {volume} {120}},\ \bibinfo {pages} {9253}
  (\bibinfo {year} {2004})}\BibitemShut {NoStop}%
\bibitem [{\citenamefont {Zhang}\ and\ \citenamefont
  {Truhlar}(2020)}]{Zhang_2020}%
  \BibitemOpen
  \bibfield  {author} {\bibinfo {author} {\bibfnamefont {D.}~\bibnamefont
  {Zhang}}\ and\ \bibinfo {author} {\bibfnamefont {D.~G.}\ \bibnamefont
  {Truhlar}},\ }\href {\doibase 10.1021/acs.jctc.0c00518} {\bibfield  {journal}
  {\bibinfo  {journal} {J. Chem. Theory Comput.}\ }\textbf {\bibinfo {volume}
  {16}},\ \bibinfo {pages} {4416} (\bibinfo {year} {2020})}\BibitemShut
  {NoStop}%
\bibitem [{\citenamefont {Zhu}\ and\ \citenamefont {Lin}(2005)}]{Zhu_2005}%
  \BibitemOpen
  \bibfield  {author} {\bibinfo {author} {\bibfnamefont {R.}~\bibnamefont
  {Zhu}}\ and\ \bibinfo {author} {\bibfnamefont {M.}~\bibnamefont {Lin}},\
  }\href {\doibase 10.1002/kin.20066} {\bibfield  {journal} {\bibinfo
  {journal} {Int. J. Quantum Chem.}\ }\textbf {\bibinfo {volume} {37}},\
  \bibinfo {pages} {593} (\bibinfo {year} {2005})}\BibitemShut {NoStop}%
\bibitem [{\citenamefont {Zhu}\ and\ \citenamefont {Lin}(2007)}]{Zhu_2007}%
  \BibitemOpen
  \bibfield  {author} {\bibinfo {author} {\bibfnamefont {R.~S.}\ \bibnamefont
  {Zhu}}\ and\ \bibinfo {author} {\bibfnamefont {M.~C.}\ \bibnamefont {Lin}},\
  }\href {\doibase 10.1021/jp068991b} {\bibfield  {journal} {\bibinfo
  {journal} {J. Phys. Chem. A}\ }\textbf {\bibinfo {volume} {111}},\ \bibinfo
  {pages} {6766} (\bibinfo {year} {2007})}\BibitemShut {NoStop}%
\bibitem [{\citenamefont {Zhu}, \citenamefont {Raghunath},\ and\ \citenamefont
  {Lin}(2013)}]{Zhu_2013}%
  \BibitemOpen
  \bibfield  {author} {\bibinfo {author} {\bibfnamefont {R.~S.}\ \bibnamefont
  {Zhu}}, \bibinfo {author} {\bibfnamefont {P.}~\bibnamefont {Raghunath}}, \
  and\ \bibinfo {author} {\bibfnamefont {M.~C.}\ \bibnamefont {Lin}},\ }\href
  {\doibase 10.1021/jp401148q} {\bibfield  {journal} {\bibinfo  {journal} {J.
  Phys. Chem. A}\ }\textbf {\bibinfo {volume} {117}},\ \bibinfo {pages} {7308}
  (\bibinfo {year} {2013})}\BibitemShut {NoStop}%
\bibitem [{\citenamefont {Zou}\ and\ \citenamefont {Liu}(2009)}]{Zou_2009}%
  \BibitemOpen
  \bibfield  {author} {\bibinfo {author} {\bibfnamefont {W.}~\bibnamefont
  {Zou}}\ and\ \bibinfo {author} {\bibfnamefont {W.}~\bibnamefont {Liu}},\
  }\href {\doibase 10.1002/jcc.21080} {\bibfield  {journal} {\bibinfo
  {journal} {J. Comput. Chem.}\ }\textbf {\bibinfo {volume} {30}},\ \bibinfo
  {pages} {524} (\bibinfo {year} {2009})}\BibitemShut {NoStop}%
\bibitem [{\citenamefont {Pastore}, \citenamefont {Angeli},\ and\ \citenamefont
  {Cimiraglia}(2006{\natexlab{a}})}]{Pastore_2006a}%
  \BibitemOpen
  \bibfield  {author} {\bibinfo {author} {\bibfnamefont {M.}~\bibnamefont
  {Pastore}}, \bibinfo {author} {\bibfnamefont {C.}~\bibnamefont {Angeli}}, \
  and\ \bibinfo {author} {\bibfnamefont {R.}~\bibnamefont {Cimiraglia}},\
  }\href {\doibase 10.1016/j.cplett.2006.03.011} {\bibfield  {journal}
  {\bibinfo  {journal} {Chem. Phys. Lett.}\ }\textbf {\bibinfo {volume}
  {422}},\ \bibinfo {pages} {522} (\bibinfo {year}
  {2006}{\natexlab{a}})}\BibitemShut {NoStop}%
\bibitem [{\citenamefont {Pastore}, \citenamefont {Angeli},\ and\ \citenamefont
  {Cimiraglia}(2006{\natexlab{b}})}]{Pastore_2006b}%
  \BibitemOpen
  \bibfield  {author} {\bibinfo {author} {\bibfnamefont {M.}~\bibnamefont
  {Pastore}}, \bibinfo {author} {\bibfnamefont {C.}~\bibnamefont {Angeli}}, \
  and\ \bibinfo {author} {\bibfnamefont {R.}~\bibnamefont {Cimiraglia}},\
  }\href {\doibase 10.1016/j.cplett.2006.06.009} {\bibfield  {journal}
  {\bibinfo  {journal} {Chem. Phys. Lett.}\ }\textbf {\bibinfo {volume}
  {426}},\ \bibinfo {pages} {445} (\bibinfo {year}
  {2006}{\natexlab{b}})}\BibitemShut {NoStop}%
\bibitem [{\citenamefont {Pastore}, \citenamefont {Angeli},\ and\ \citenamefont
  {Cimiraglia}(2007)}]{Pastore_2007}%
  \BibitemOpen
  \bibfield  {author} {\bibinfo {author} {\bibfnamefont {M.}~\bibnamefont
  {Pastore}}, \bibinfo {author} {\bibfnamefont {C.}~\bibnamefont {Angeli}}, \
  and\ \bibinfo {author} {\bibfnamefont {R.}~\bibnamefont {Cimiraglia}},\
  }\href {\doibase 10.1007/s00214-006-0239-5} {\bibfield  {journal} {\bibinfo
  {journal} {Theor. Chem. Acc.}\ }\textbf {\bibinfo {volume} {118}},\ \bibinfo
  {pages} {35} (\bibinfo {year} {2007})}\BibitemShut {NoStop}%
\bibitem [{\citenamefont {Angeli}, \citenamefont {Cavallini},\ and\
  \citenamefont {Cimiraglia}(2007)}]{Angeli_2007}%
  \BibitemOpen
  \bibfield  {author} {\bibinfo {author} {\bibfnamefont {C.}~\bibnamefont
  {Angeli}}, \bibinfo {author} {\bibfnamefont {A.}~\bibnamefont {Cavallini}}, \
  and\ \bibinfo {author} {\bibfnamefont {R.}~\bibnamefont {Cimiraglia}},\
  }\href {\doibase 10.1063/1.2768529} {\bibfield  {journal} {\bibinfo
  {journal} {J. Chem. Phys.}\ }\textbf {\bibinfo {volume} {127}},\ \bibinfo
  {pages} {074306} (\bibinfo {year} {2007})}\BibitemShut {NoStop}%
\bibitem [{\citenamefont {Camacho}, \citenamefont {Witek},\ and\ \citenamefont
  {Cimiraglia}(2010)}]{Camacho_2010}%
  \BibitemOpen
  \bibfield  {author} {\bibinfo {author} {\bibfnamefont {C.}~\bibnamefont
  {Camacho}}, \bibinfo {author} {\bibfnamefont {H.~A.}\ \bibnamefont {Witek}},
  \ and\ \bibinfo {author} {\bibfnamefont {R.}~\bibnamefont {Cimiraglia}},\
  }\href {\doibase 10.1063/1.3442374} {\bibfield  {journal} {\bibinfo
  {journal} {J. Chem. Phys.}\ }\textbf {\bibinfo {volume} {132}},\ \bibinfo
  {pages} {244306} (\bibinfo {year} {2010})}\BibitemShut {NoStop}%
\bibitem [{\citenamefont {Angeli}\ and\ \citenamefont
  {Cimiraglia}(2011)}]{Angeli_2011}%
  \BibitemOpen
  \bibfield  {author} {\bibinfo {author} {\bibfnamefont {C.}~\bibnamefont
  {Angeli}}\ and\ \bibinfo {author} {\bibfnamefont {R.}~\bibnamefont
  {Cimiraglia}},\ }\href {\doibase 10.1080/00268976.2011.566586} {\bibfield
  {journal} {\bibinfo  {journal} {Mol. Phys.}\ }\textbf {\bibinfo {volume}
  {109}},\ \bibinfo {pages} {1503} (\bibinfo {year} {2011})}\BibitemShut
  {NoStop}%
\bibitem [{\citenamefont {Angeli}, \citenamefont {Cimiraglia},\ and\
  \citenamefont {Pastore}(2012)}]{Angeli_2012}%
  \BibitemOpen
  \bibfield  {author} {\bibinfo {author} {\bibfnamefont {C.}~\bibnamefont
  {Angeli}}, \bibinfo {author} {\bibfnamefont {R.}~\bibnamefont {Cimiraglia}},
  \ and\ \bibinfo {author} {\bibfnamefont {M.}~\bibnamefont {Pastore}},\ }\href
  {\doibase 10.1080/00268976.2012.689872} {\bibfield  {journal} {\bibinfo
  {journal} {Mol. Phys.}\ }\textbf {\bibinfo {volume} {110}},\ \bibinfo {pages}
  {2963} (\bibinfo {year} {2012})}\BibitemShut {NoStop}%
\bibitem [{\citenamefont {Loos}, \citenamefont {Galland},\ and\ \citenamefont
  {Jacquemin}(2018)}]{Loos_2018b}%
  \BibitemOpen
  \bibfield  {author} {\bibinfo {author} {\bibfnamefont {P.-F.}\ \bibnamefont
  {Loos}}, \bibinfo {author} {\bibfnamefont {N.}~\bibnamefont {Galland}}, \
  and\ \bibinfo {author} {\bibfnamefont {D.}~\bibnamefont {Jacquemin}},\ }\href
  {\doibase 10.1021/acs.jpclett.8b02058} {\bibfield  {journal} {\bibinfo
  {journal} {J. Phys. Chem. Lett.}\ }\textbf {\bibinfo {volume} {9}},\ \bibinfo
  {pages} {4646} (\bibinfo {year} {2018})}\BibitemShut {NoStop}%
\bibitem [{\citenamefont {Loos}\ and\ \citenamefont
  {Jacquemin}(2019{\natexlab{a}})}]{Loos_2019a}%
  \BibitemOpen
  \bibfield  {author} {\bibinfo {author} {\bibfnamefont {P.-F.}\ \bibnamefont
  {Loos}}\ and\ \bibinfo {author} {\bibfnamefont {D.}~\bibnamefont
  {Jacquemin}},\ }\href {\doibase 10.1021/acs.jctc.8b01103} {\bibfield
  {journal} {\bibinfo  {journal} {J. Chem. Theory Comput.}\ }\textbf {\bibinfo
  {volume} {15}},\ \bibinfo {pages} {2481} (\bibinfo {year}
  {2019}{\natexlab{a}})}\BibitemShut {NoStop}%
\bibitem [{\citenamefont {Loos}\ and\ \citenamefont
  {Jacquemin}(2019{\natexlab{b}})}]{Loos_2019b}%
  \BibitemOpen
  \bibfield  {author} {\bibinfo {author} {\bibfnamefont {P.-F.}\ \bibnamefont
  {Loos}}\ and\ \bibinfo {author} {\bibfnamefont {D.}~\bibnamefont
  {Jacquemin}},\ }\href@noop {} {\bibfield  {journal} {\bibinfo  {journal}
  {ChemPhotoChem}\ }\textbf {\bibinfo {volume} {3}},\ \bibinfo {pages} {684}
  (\bibinfo {year} {2019}{\natexlab{b}})}\BibitemShut {NoStop}%
\bibitem [{\citenamefont {Kendall}, \citenamefont {Dunning},\ and\
  \citenamefont {Harisson}(1992)}]{Kendall_1992}%
  \BibitemOpen
  \bibfield  {author} {\bibinfo {author} {\bibfnamefont {R.~A.}\ \bibnamefont
  {Kendall}}, \bibinfo {author} {\bibfnamefont {T.~H.}\ \bibnamefont
  {Dunning}}, \ and\ \bibinfo {author} {\bibfnamefont {R.~J.}\ \bibnamefont
  {Harisson}},\ }\href {\doibase 10.1063/1.462569} {\bibfield  {journal}
  {\bibinfo  {journal} {J. Chem. Phys.}\ }\textbf {\bibinfo {volume} {96}},\
  \bibinfo {pages} {6796} (\bibinfo {year} {1992})}\BibitemShut {NoStop}%
\bibitem [{\citenamefont {Werner}\ and\ \citenamefont
  {Knowles}(1988)}]{Werner_1988}%
  \BibitemOpen
  \bibfield  {author} {\bibinfo {author} {\bibfnamefont {H.}~\bibnamefont
  {Werner}}\ and\ \bibinfo {author} {\bibfnamefont {P.~J.}\ \bibnamefont
  {Knowles}},\ }\href {\doibase 10.1063/1.455556} {\bibfield  {journal}
  {\bibinfo  {journal} {J. Chem. Phys.}\ }\textbf {\bibinfo {volume} {89}},\
  \bibinfo {pages} {5803} (\bibinfo {year} {1988})}\BibitemShut {NoStop}%
\bibitem [{\citenamefont {Knowles}\ and\ \citenamefont
  {Werner}(1988)}]{Knowles_1988}%
  \BibitemOpen
  \bibfield  {author} {\bibinfo {author} {\bibfnamefont {P.~J.}\ \bibnamefont
  {Knowles}}\ and\ \bibinfo {author} {\bibfnamefont {H.-J.}\ \bibnamefont
  {Werner}},\ }\href {\doibase https://doi.org/10.1016/0009-2614(88)87412-8}
  {\bibfield  {journal} {\bibinfo  {journal} {Chem. Phys. Lett.}\ }\textbf
  {\bibinfo {volume} {145}},\ \bibinfo {pages} {514} (\bibinfo {year}
  {1988})}\BibitemShut {NoStop}%
\bibitem [{\citenamefont {Davidson}(1996)}]{Davidson_1996}%
  \BibitemOpen
  \bibfield  {author} {\bibinfo {author} {\bibfnamefont {E.~R.}\ \bibnamefont
  {Davidson}},\ }\href {\doibase 10.1021/jp952794n} {\bibfield  {journal}
  {\bibinfo  {journal} {J. Phys. Chem.}\ }\textbf {\bibinfo {volume} {100}},\
  \bibinfo {pages} {6161} (\bibinfo {year} {1996})}\BibitemShut {NoStop}%
\bibitem [{\citenamefont {Borden}\ and\ \citenamefont
  {Davidson}(1996)}]{Borden_1996}%
  \BibitemOpen
  \bibfield  {author} {\bibinfo {author} {\bibfnamefont {W.~T.}\ \bibnamefont
  {Borden}}\ and\ \bibinfo {author} {\bibfnamefont {E.~R.}\ \bibnamefont
  {Davidson}},\ }\href {\doibase 10.1021/ar950134v} {\bibfield  {journal}
  {\bibinfo  {journal} {Acc. Chem. Res.}\ }\textbf {\bibinfo {volume} {29}},\
  \bibinfo {pages} {67} (\bibinfo {year} {1996})}\BibitemShut {NoStop}%
\bibitem [{\citenamefont {Boggio-Pasqua}\ \emph {et~al.}(2004)\citenamefont
  {Boggio-Pasqua}, \citenamefont {Bearpark}, \citenamefont {Klene},\ and\
  \citenamefont {Robb}}]{Boggio-Pasqua_2004}%
  \BibitemOpen
  \bibfield  {author} {\bibinfo {author} {\bibfnamefont {M.}~\bibnamefont
  {Boggio-Pasqua}}, \bibinfo {author} {\bibfnamefont {M.~J.}\ \bibnamefont
  {Bearpark}}, \bibinfo {author} {\bibfnamefont {M.}~\bibnamefont {Klene}}, \
  and\ \bibinfo {author} {\bibfnamefont {M.~A.}\ \bibnamefont {Robb}},\ }\href
  {\doibase 10.1063/1.1690756} {\bibfield  {journal} {\bibinfo  {journal} {J.
  Chem. Phys.}\ }\textbf {\bibinfo {volume} {120}},\ \bibinfo {pages} {7849}
  (\bibinfo {year} {2004})}\BibitemShut {NoStop}%
\bibitem [{\citenamefont {Angeli}(2009)}]{Angeli_2009}%
  \BibitemOpen
  \bibfield  {author} {\bibinfo {author} {\bibfnamefont {C.}~\bibnamefont
  {Angeli}},\ }\href {\doibase 10.1002/jcc.21155} {\bibfield  {journal}
  {\bibinfo  {journal} {J. Comput. Chem.}\ }\textbf {\bibinfo {volume} {30}},\
  \bibinfo {pages} {1319} (\bibinfo {year} {2009})}\BibitemShut {NoStop}%
\bibitem [{\citenamefont {Garniron}\ \emph {et~al.}(2018)\citenamefont
  {Garniron}, \citenamefont {Scemama}, \citenamefont {Giner}, \citenamefont
  {Caffarel},\ and\ \citenamefont {Loos}}]{Garniron_2018}%
  \BibitemOpen
  \bibfield  {author} {\bibinfo {author} {\bibfnamefont {Y.}~\bibnamefont
  {Garniron}}, \bibinfo {author} {\bibfnamefont {A.}~\bibnamefont {Scemama}},
  \bibinfo {author} {\bibfnamefont {E.}~\bibnamefont {Giner}}, \bibinfo
  {author} {\bibfnamefont {M.}~\bibnamefont {Caffarel}}, \ and\ \bibinfo
  {author} {\bibfnamefont {P.~F.}\ \bibnamefont {Loos}},\ }\href {\doibase
  10.1063/1.5044503} {\bibfield  {journal} {\bibinfo  {journal} {J. Chem.
  Phys.}\ }\textbf {\bibinfo {volume} {149}},\ \bibinfo {pages} {064103}
  (\bibinfo {year} {2018})}\BibitemShut {NoStop}%
\bibitem [{\citenamefont {Tran}\ \emph {et~al.}(2019)\citenamefont {Tran},
  \citenamefont {Segarra-Marti}, \citenamefont {Bearpark},\ and\ \citenamefont
  {Robb}}]{Tran_2019}%
  \BibitemOpen
  \bibfield  {author} {\bibinfo {author} {\bibfnamefont {T.}~\bibnamefont
  {Tran}}, \bibinfo {author} {\bibfnamefont {J.}~\bibnamefont {Segarra-Marti}},
  \bibinfo {author} {\bibfnamefont {M.~J.}\ \bibnamefont {Bearpark}}, \ and\
  \bibinfo {author} {\bibfnamefont {M.~A.}\ \bibnamefont {Robb}},\ }\href
  {\doibase 10.1021/acs.jpca.9b03715} {\bibfield  {journal} {\bibinfo
  {journal} {J. Phys. Chem. A}\ }\textbf {\bibinfo {volume} {123}},\ \bibinfo
  {pages} {5223} (\bibinfo {year} {2019})}\BibitemShut {NoStop}%
\bibitem [{\citenamefont {Ben~Amor}\ \emph {et~al.}(2020)\citenamefont
  {Ben~Amor}, \citenamefont {No{\^u}s}, \citenamefont {Trinquier},\ and\
  \citenamefont {Malrieu}}]{BenAmor_2020}%
  \BibitemOpen
  \bibfield  {author} {\bibinfo {author} {\bibfnamefont {N.}~\bibnamefont
  {Ben~Amor}}, \bibinfo {author} {\bibfnamefont {C.}~\bibnamefont {No{\^u}s}},
  \bibinfo {author} {\bibfnamefont {G.}~\bibnamefont {Trinquier}}, \ and\
  \bibinfo {author} {\bibfnamefont {J.-P.}\ \bibnamefont {Malrieu}},\ }\href
  {\doibase 10.1063/5.0011582} {\bibfield  {journal} {\bibinfo  {journal} {J.
  Chem. Phys.}\ }\textbf {\bibinfo {volume} {153}},\ \bibinfo {pages} {044118}
  (\bibinfo {year} {2020})}\BibitemShut {NoStop}%
\bibitem [{\citenamefont {Mok}, \citenamefont {Neumann},\ and\ \citenamefont
  {Handy}(1996)}]{Mok_1996}%
  \BibitemOpen
  \bibfield  {author} {\bibinfo {author} {\bibfnamefont {D.~K.~W.}\
  \bibnamefont {Mok}}, \bibinfo {author} {\bibfnamefont {R.}~\bibnamefont
  {Neumann}}, \ and\ \bibinfo {author} {\bibfnamefont {N.~C.}\ \bibnamefont
  {Handy}},\ }\href {\doibase 10.1021/jp9528020} {\bibfield  {journal}
  {\bibinfo  {journal} {J. Phys. Chem.}\ }\textbf {\bibinfo {volume} {100}},\
  \bibinfo {pages} {6225} (\bibinfo {year} {1996})}\BibitemShut {NoStop}%
\bibitem [{\citenamefont {Fulscher}\ and\ \citenamefont
  {Roos}(1994)}]{Fulscher_1994}%
  \BibitemOpen
  \bibfield  {author} {\bibinfo {author} {\bibfnamefont {M.~P.}\ \bibnamefont
  {Fulscher}}\ and\ \bibinfo {author} {\bibfnamefont {B.~O.}\ \bibnamefont
  {Roos}},\ }\href {\doibase 10.1007/BF01113393} {\bibfield  {journal}
  {\bibinfo  {journal} {Theor. Chim. Acta}\ }\textbf {\bibinfo {volume} {87}},\
  \bibinfo {pages} {403} (\bibinfo {year} {1994})}\BibitemShut {NoStop}%
\bibitem [{\citenamefont {Olsen}\ \emph {et~al.}(1988)\citenamefont {Olsen},
  \citenamefont {Roos}, \citenamefont {Jorgensen},\ and\ \citenamefont
  {Jensen}}]{Olsen_1988}%
  \BibitemOpen
  \bibfield  {author} {\bibinfo {author} {\bibfnamefont {J.}~\bibnamefont
  {Olsen}}, \bibinfo {author} {\bibfnamefont {B.~O.}\ \bibnamefont {Roos}},
  \bibinfo {author} {\bibfnamefont {P.}~\bibnamefont {Jorgensen}}, \ and\
  \bibinfo {author} {\bibfnamefont {H.~J.~A.}\ \bibnamefont {Jensen}},\ }\href
  {\doibase 10.1063/1.455063} {\bibfield  {journal} {\bibinfo  {journal} {J.
  Chem. Phys.}\ }\textbf {\bibinfo {volume} {89}},\ \bibinfo {pages} {2185}
  (\bibinfo {year} {1988})}\BibitemShut {NoStop}%
\bibitem [{\citenamefont {Liang}\ \emph {et~al.}(ress)\citenamefont {Liang},
  \citenamefont {Feng}, \citenamefont {Hait},\ and\ \citenamefont
  {Head-Gordon}}]{Liang_2022}%
  \BibitemOpen
  \bibfield  {author} {\bibinfo {author} {\bibfnamefont {J.}~\bibnamefont
  {Liang}}, \bibinfo {author} {\bibfnamefont {X.}~\bibnamefont {Feng}},
  \bibinfo {author} {\bibfnamefont {D.}~\bibnamefont {Hait}}, \ and\ \bibinfo
  {author} {\bibfnamefont {M.}~\bibnamefont {Head-Gordon}},\ }\href {\doibase
  10.1021/acs.jctc.2c00160} {\bibfield  {journal} {\bibinfo  {journal} {J.
  Chem. Theory Comput.}\ } (\bibinfo {year} {in press}),\
  10.1021/acs.jctc.2c00160}\BibitemShut {NoStop}%
\bibitem [{\citenamefont {Head-Gordon}\ \emph {et~al.}(1994)\citenamefont
  {Head-Gordon}, \citenamefont {Rico}, \citenamefont {Oumi},\ and\
  \citenamefont {Lee}}]{Head-Gordon_1994}%
  \BibitemOpen
  \bibfield  {author} {\bibinfo {author} {\bibfnamefont {M.}~\bibnamefont
  {Head-Gordon}}, \bibinfo {author} {\bibfnamefont {R.~J.}\ \bibnamefont
  {Rico}}, \bibinfo {author} {\bibfnamefont {M.}~\bibnamefont {Oumi}}, \ and\
  \bibinfo {author} {\bibfnamefont {T.~J.}\ \bibnamefont {Lee}},\ }\href
  {\doibase 10.1016/0009-2614(94)00070-0} {\bibfield  {journal} {\bibinfo
  {journal} {Chem. Phys. Lett.}\ }\textbf {\bibinfo {volume} {219}},\ \bibinfo
  {pages} {21} (\bibinfo {year} {1994})}\BibitemShut {NoStop}%
\bibitem [{\citenamefont {Head-Gordon}, \citenamefont {Maurice},\ and\
  \citenamefont {Oumi}(1995)}]{Head-Gordon_1995}%
  \BibitemOpen
  \bibfield  {author} {\bibinfo {author} {\bibfnamefont {M.}~\bibnamefont
  {Head-Gordon}}, \bibinfo {author} {\bibfnamefont {D.}~\bibnamefont
  {Maurice}}, \ and\ \bibinfo {author} {\bibfnamefont {M.}~\bibnamefont
  {Oumi}},\ }\href {\doibase 10.1016/0009-2614(95)01111-L} {\bibfield
  {journal} {\bibinfo  {journal} {Chem. Phys. Lett.}\ }\textbf {\bibinfo
  {volume} {246}},\ \bibinfo {pages} {114} (\bibinfo {year}
  {1995})}\BibitemShut {NoStop}%
\bibitem [{\citenamefont {Christiansen}, \citenamefont {Koch},\ and\
  \citenamefont {J{\o}rgensen}(1995)}]{Christiansen_1995b}%
  \BibitemOpen
  \bibfield  {author} {\bibinfo {author} {\bibfnamefont {O.}~\bibnamefont
  {Christiansen}}, \bibinfo {author} {\bibfnamefont {H.}~\bibnamefont {Koch}},
  \ and\ \bibinfo {author} {\bibfnamefont {P.}~\bibnamefont {J{\o}rgensen}},\
  }\href {\doibase http://dx.doi.org/10.1063/1.470315} {\bibfield  {journal}
  {\bibinfo  {journal} {J. Chem. Phys.}\ }\textbf {\bibinfo {volume} {103}},\
  \bibinfo {pages} {7429} (\bibinfo {year} {1995})}\BibitemShut {NoStop}%
\bibitem [{\citenamefont {Koch}\ \emph {et~al.}(1997)\citenamefont {Koch},
  \citenamefont {Christiansen}, \citenamefont {Jorgensen}, \citenamefont
  {Sanchez~de Mer{\'a}s},\ and\ \citenamefont {Helgaker}}]{Koch_1997}%
  \BibitemOpen
  \bibfield  {author} {\bibinfo {author} {\bibfnamefont {H.}~\bibnamefont
  {Koch}}, \bibinfo {author} {\bibfnamefont {O.}~\bibnamefont {Christiansen}},
  \bibinfo {author} {\bibfnamefont {P.}~\bibnamefont {Jorgensen}}, \bibinfo
  {author} {\bibfnamefont {A.~M.}\ \bibnamefont {Sanchez~de Mer{\'a}s}}, \ and\
  \bibinfo {author} {\bibfnamefont {T.}~\bibnamefont {Helgaker}},\ }\href
  {\doibase http://dx.doi.org/10.1063/1.473322} {\bibfield  {journal} {\bibinfo
   {journal} {J. Chem. Phys.}\ }\textbf {\bibinfo {volume} {106}},\ \bibinfo
  {pages} {1808} (\bibinfo {year} {1997})}\BibitemShut {NoStop}%
\bibitem [{\citenamefont {{Boggio-Pasqua}}, \citenamefont {Bearpark},\ and\
  \citenamefont {Robb}(2007)}]{Boggio-Pasqua_2007}%
  \BibitemOpen
  \bibfield  {author} {\bibinfo {author} {\bibfnamefont {M.}~\bibnamefont
  {{Boggio-Pasqua}}}, \bibinfo {author} {\bibfnamefont {M.~J.}\ \bibnamefont
  {Bearpark}}, \ and\ \bibinfo {author} {\bibfnamefont {M.~A.}\ \bibnamefont
  {Robb}},\ }\href {\doibase 10.1021/jo070452v} {\bibfield  {journal} {\bibinfo
   {journal} {J. Org. Chem.}\ }\textbf {\bibinfo {volume} {72}},\ \bibinfo
  {pages} {4497} (\bibinfo {year} {2007})}\BibitemShut {NoStop}%
\end{thebibliography}
\end{document}